\documentclass[a4paper,fleqn,usenatbib,useAMS,onecolumn]{mnras}
\usepackage{color}
\usepackage{amsmath} 
\usepackage{amssymb} 
\usepackage{multicol} 
\usepackage{bm} 

\usepackage{textcomp,amssymb}
\usepackage{leftidx}
\usepackage{hyperref} 
\hypersetup{dvips, colorlinks=true, linkcolor=black, citecolor=black, filecolor=blue, urlcolor=black}
\usepackage{float}
\usepackage{epsfig}
\usepackage{array}
\usepackage{supertabular}
\usepackage{hhline}
\usepackage{multirow}
\usepackage{balance}
\usepackage{times}
\usepackage{graphicx}
\usepackage{times}
\usepackage{soul}
\usepackage{bm}

\allowdisplaybreaks

\newcommand{\ri}{\mathrm{i}}
\newcommand{\re}{\mathrm{e}}
\newcommand{\ra}{\mathrm{a}}
\newcommand{\rb}{\mathrm{b}}
\newcommand{\rB}{\mathrm{B}}
\newcommand{\rc}{\mathrm{c}}
\newcommand{\rd}{\mathrm{d}}
\newcommand{\rD}{\mathrm{D}}
\newcommand{\rp}{\mathrm{p}}

\newcommand{\rt}{\mathrm{t}}

\begin{document}

\setcounter{tocdepth}{3}

\title[Dressed dynamical friction]
{Dressed diffusion and friction coefficients
\\
in inhomogeneous multicomponent self-gravitating systems}

\author[J. Heyvaerts, J-B. Fouvry, P-H. Chavanis \& C. Pichon]{Jean Heyvaerts$^{1}$, Jean-Baptiste Fouvry$^{2}$\thanks{Hubble Fellow.}, Pierre-Henri Chavanis$^{3}$ and Christophe Pichon$^{4,5}$
\vspace*{6pt}\\
\noindent
$^{1}$ Observatoire Astronomique de Strasbourg, 11 rue de l'Universit\'e, 67000 Strasbourg, France\\
$^{2}$ Institute for Advanced Study, Einstein Drive, Princeton, NJ 08540, USA\\
$^{3}$ Laboratoire de Physique Th\'eorique (IRSAMC), CNRS and UPS, Univ. de Toulouse, F-31062 Toulouse, France\\
$^{4}$ Institut d'Astrophysique de Paris, and UPMC Univ. Paris 06, (UMR7095), 98 bis Boulevard Arago, 75014 Paris, France\\
$^{5}$ Korea Institute for Advanced Study (KIAS), 85 Hoegiro, Dongdaemun-gu, Seoul, 02455, Republic of Korea
}

\date{\today}
\label{firstpage}
\pagerange{\pageref{firstpage}--\pageref{lastpage}}

\maketitle

\begin{abstract}
General self-consistent expressions for the coefficients of diffusion and dynamical friction in
a stable, bound, multicomponent self-gravitating and inhomogeneous system are
derived. They account for the detailed dynamics of the colliding particles and their self-consistent dressing by
collective gravitational interactions. 
The associated Fokker-Planck equation is shown to be fully
consistent with the corresponding inhomogeneous Balescu-Lenard equation and, in the weak
self-gravitating limit, to the inhomogeneous Landau equation. Hence it provides
an alternative derivation to both and demonstrates their equivalence. 
The corresponding stochastic
Langevin equations are presented: they can be a practical alternative to numerically
solving the inhomogeneous Fokker-Planck and Balescu-Lenard equations. 
The present formalism allows for a self-consistent description of the secular evolution of 
different populations covering a spectrum of masses,
with a proper accounting of the induced secular mass segregation, which should be of interest
to various astrophysical contexts, from galactic centers to protostellar discs. 
\end{abstract}

\begin{keywords}
Galaxies: kinematics and dynamics - Galaxies: nuclei - Diffusion - Gravitation
\end{keywords}

\section{Introduction and motivation}
\label{sec:introduction}

The kinetic theory of stellar systems was initiated by~\cite{Chandrasekhar1942}.
He first described the motion of a star in a stellar system semi-heuristically
by using an analogy with Brownian motion~\citep{Chandrasekhar1943I}. He argued
that the force acting on a star has two components: a mean field component due
to the smooth distribution of the system and a fluctuating component arising
from discreteness effects. Discreteness effects (also called finite${-N}$
effects, granularities, graininess...) account for gravitational encounters. For
a spatially homogeneous system, the mean field force vanishes so that only
gravitational encounters (``collisions'') can produce an evolution. The
fluctuating force acting on a star gives rise to diffusive motion in velocity
space. However, a purely diffusive motion would lead to a divergence of the
kinetic energy of the star and would not establish a statistical equilibrium state 
at late time. As a result, Chandrasekhar realised that something was
``missing'' in his description and that the diffusive process must be
accompanied by a dissipative process. He concluded that the star must also experience
a dynamical friction. Using a phenomenological Langevin equation incorporating a
friction force proportional and opposite to the velocity of the star and a
random force modeled as a white noise, he derived a Fokker-Planck equation
describing the evolution of the velocity distribution function ${ f (\bm{v} , t)
}$ of the star. Then, requiring that the Maxwell-Boltzmann distribution must be
a stationary state of this Fokker-Planck equation, he showed that the
coefficients of diffusion and friction are related to each other by an Einstein
relation. This is the manifestation of the fluctuation-dissipation theorem. In
parallel, using a more rigorous kinetic theory, he directly computed the
coefficients of diffusion and friction in the approximation of close binary
encounters and explicitly checked that, for a Maxwellian distribution, the
Einstein relation is indeed satisfied. He then used his kinetic theory to
estimate the rate of escape of stars from clusters~\citep{Chandrasekhar1943II},
and found observational evidence for the operation of dynamical friction.

In the approach of Chandrasekhar, further developed
by~\cite{Cohen1950,Rosenbluth1957,King1960,BinneyTremaine2008}, the diffusion and friction
coefficients are obtained independently, from two different calculations, and
are then injected into the Fokker-Planck equation. When the distribution function
is allowed to change self-consistently, one obtains an integrodifferential
equation describing the evolution of the distribution function of the stellar
system as a whole. This integrodifferential equation is formally equivalent to
the Landau equation that was introduced earlier in the different context of plasma
physics~\citep{Landau1936}.\footnote{The Landau equation was sometimes misunderstood. For example, the paper of~\cite{Cohen1950} comments: ``A similar but incomplete
approach [to the work of Chandrasekhar] was made somewhat earlier by Landau. In
this reference, the important terms representing dynamical friction, which
should appear in the diffusion equation, are set equal to zero as a result of
certain approximations.'' This claim is misleading since the Landau
equation includes both terms of diffusion and friction.
}
Indeed, a few years before
Chandrasekhar's seminal papers on stellar dynamics, Landau derived an
integrodifferential kinetic equation for homogeneous Coulombian plasmas
interacting via inverse-square forces. He obtained this equation as a weak
deflection approximation of the Boltzmann equation. In this equation, the terms
of diffusion and friction arise simultaneously from a unique formalism. The
Landau equation has a form similar to the Fokker-Planck equation except that the
diffusion tensor is placed between the two velocity derivatives, ${ \partial_{t}
f \!=\! \partial_{v_{i}} (D_{ij}\partial_{v_{j}} f) \!-\! \partial_{v_{i}}
(F^{\rm pol}_{i} \! f ) }$, while in the standard Fokker-Planck equation it is
placed after the two velocity derivatives, ${ \partial_{t} f \!=\!
\partial_{v_{i}} \partial_{v_{j}} (D_{ij} f) \!-\! \partial_{v_{i}} (F^{\rm
fric}_{i} \! f ) }$. As a result, the friction force that appears in the Landau
equation differs from the friction force that appears in the Fokker-Planck
equation of Chandrasekhar. We shall call it the ``friction by polarisation''
$\bm{F}_{\rm pol}$, to distinguish it from the ``true friction'' $\bm{F}_{\rm
fric}$. Of course, one can immediately transform the Landau equation into the
standard Fokker-Planck equation and find the relation ${ F^{\rm fric}_{i} \!=\!
F^{\rm pol}_{i} \!+\! \partial_{v_{j}}D_{ij} }$ between the two friction forces.
One can then check~\cite[see, e.g.][]{Chavanis2013} that the Landau equation is
fully equivalent to the Fokker-Planck equation of Chandrasekhar even if the
equations do not appear in the same form.\footnote{A virtue of the Landau
equation resides in its symmetric structure from which we can immediately derive
the conservation laws (mass, energy, impulse, angular momentum) and the
${H-}$theorem for the Boltzmann entropy. 
}

The force of dynamical friction was calculated
by~\cite{Marochnik1968,Kalnajs1971,Kandrup1983,BekensteinZamir1990,Chavanis2008}
from a linear
response theory based on the Liouville or on the Klimontovich equation. In these
calculations, it results from a polarisation
process. A test star perturbs the distribution of the field stars and the
retroaction of the field stars on the test star leads to a friction force that
decelerates the test star. The force of dynamical friction calculated by these
authors differs from that calculated by Chandrasekhar
by a factor of two (or by a factor ${ (m \!+\! m_{\rm f}) / m }$ if the test
star and the field stars have different masses). This is because, as noted
in~\cite{Chavanis2013}, they actually
calculated the ``friction by polarisation'' (the one that arises in the Landau
equation), not the ``true friction'' (the one that arises in the Fokker-Planck
equation of Chandrasekhar).

The kinetic theory of Chandrasekhar is based on two simplifying assumptions. He
assumed that the system is spatially homogeneous and neglected collective
effects (the fact that a star is surrounded by a cloud of other stars that tends
to enhance the gravitational attraction).
Various generalisations and improvements of the theory
of~\cite{Chandrasekhar1943I} were subsequently proposed.
Some authors tried to deal with spatial inhomogeneity.
\cite{Kandrup1983,BekensteinMaoz1992,Maoz1993,NelsonTremaine1999,Chavanis2008}
reconsidered the interaction of a test particle with a background stochastic
force in the context of the fluctuation-dissipation theorem,
and showed that the friction force depends on the global structure of the system.
This was also investigated
in~\cite{DelPopoloGambera1999,DelPopolo2003} which
extended~\cite{ChandrasekharVonNeumann1943}'s
analysis to the
case where the background particles are inhomogeneously 
distributed with a density profile decaying as $\rho\sim
r^{-p}$.
Kinetic theories for spatially inhomogeneous systems were developed
in~\cite{SeverneHaggerty1976,ParisotSeverne1979,Kandrup1981,Chavanis2008,
Chavanis2013} using position-velocity variables.
Collective effects were first taken
into account in plasma physics, where the system is spatially homogeneous because
of electroneutrality.~\cite{Balescu1960} and~\cite{Lenard1960} derived a
generalisation of the Landau equation by accounting properly for collective
effects (the fact that a charge is surrounded by a cloud of opposite charges
that tends to shield the electrostatic interaction). In the Balescu-Lenard
equation, the bare potential of interaction is replaced by a dressed potential
of interaction that takes into account the dressing of the particles by their
polarisation cloud. As a result, the Debye length appears naturally without any
ad hoc assumptions and regularises the logarithmic divergence at large scales that
occurs in the Landau equation when collective effects are
neglected.~\cite{Balescu1960} started from the diagram technique introduced
by~\cite{PrigogineBalescu1959}, and~\cite{Lenard1960} started from
the~\cite{Bogoliubov1946} equations (now known as the BBGKY hierarchy).
Independently,~\cite{Hubbard1961} used a Fokker-Planck approach and directly
calculated the coefficients of diffusion and friction by taking collective
effects into account.\footnote{If one substitutes these expressions into the
Fokker-Planck
equation and performs minor transformations (a substitution that Hubbard did not
explicitly make), one obtains the Balescu-Lenard equation. Inversely, from the
Balescu-Lenard equation, one can recover the expressions of the diffusion and
friction coefficients obtained by Hubbard. These considerations show that the
self-consistent Fokker-Planck equation of Hubbard is fully equivalent to the
Balescu-Lenard equation~\cite[see, e.g.][]{Chavanis2012epjp}.}

In order to take collective effects into account 
in stellar systems, one must simultaneously account for their spatial
inhomogeneity otherwise the kinetic equation presents a strong (algebraic)
divergence at large-scales related to the Jeans
instability~\citep{Weinberg1993,Chavanis2013}.
Kinetic theories for spatially inhomogeneous systems with position-velocity variables taking
collective effects into account were developed
by~\cite{Miller1966,Thorne1968,Gilbert1968,Gilbert1970,Lerche1971}.
Their kinetic equations are very
complicated but they managed to show that collective effects are equivalent to
increasing the effective mass of the stars, hence reducing the relaxation time.
Spatial inhomogeneity can
conveniently be dealt with by using angle-action
variables for integrable systems~\citep{Goldstein1950,BinneyTremaine2008}.
On the other hand, collective effects can be dealt
with by introducing a biorthogonal basis of potentials and densities and
using~\cite{Kalnajs1976II}'s matrix method.
The friction force
and the energy exchange rate for spatially inhomogeneous stellar systems with
angle-action variables were
calculated by
\cite{LyndenBellKalnajs1972,TremaineWeinberg1984,PalmerPapaloizou1985,
Weinberg1986,Weinberg1989,RauchTremaine1996}.
The corresponding inhomogeneous
Balescu-Lenard equation has been derived by~\cite{Heyvaerts2010} from the BBGKY
hierarchy and by~\cite{Chavanis2012} from the Klimontovich
approach.\footnote{The inhomogeneous Balescu-Lenard equation with
angle-action variables was previously derived by~\cite{LucianiPellat1987},
where the response matrix is introduced at a formal level without
an explicit representation.}
\cite{Chavanis2012} developed a Fokker-Planck approach and
directly calculated the diffusion and friction coefficients with angle-action variables
by taking collective effects into account, thereby generalising the results
of~\cite{Hubbard1961} to inhomogeneous systems. 
He also developed a test particle approach relying on a
(thermal) bath approximation in which the integrodifferential Balescu-Lenard
equation is transformed into a differential Fokker-Planck equation. Neglecting
collective
effects, one recovers the results obtained from the inhomogeneous Landau
equation~\citep{Chavanis2007,Chavanis2013}.
Making a local
approximation, one recovers the original results of~\cite{Landau1936}
and~\cite{Chandrasekhar1942,Chandrasekhar1943I}.

The inhomogeneous Balescu-Lenard equation was 
recently implemented for the first time in astrophysics.
In~\cite{FouvryPichonChavanis2015,FouvryPichonMagorrianChavanis2015,
FouvryPichonChavanisMonk2017}, it was applied to razor-thin
and thickened stellar discs, and proved useful to probe complex secular regimes
of diffusion. These works showed in particular how collective effects cause cool
discs to have two-body relaxation times much shorter than naively expected. In
addition, they showed how this relaxation introduces small-scale
structures in the disc, which secularly destabilise it at the collisionless
level. The inhomogeneous Balescu-Lenard equation was also recently applied in the context of the ${1D}$ inhomogeneous Hamiltonian Mean Field (HMF) model~\citep{BenettiMarcos2017}.
Finally, it was specialised
to dynamically degenerate systems, such as quasi-Keplerian systems (galactic centers, protostellar discs)
in~\cite{FouvryPichonMagorrian2017} and~\cite{SridharTouma2016III} (without collective effects). Following these recent successes, the
inhomogeneous Balescu-Lenard equation now appears as a powerful and predictive
framework. It could also be used as a valuable check of the accuracy of
${N-}$body integrators on secular timescales.

In this paper, we present a derivation of the coefficients of
diffusion and dynamical friction for
self-gravitating stellar systems, while taking into account both inhomogeneity
and collective effects. We also present the set of stochastic Langevin equations, dual to the
Fokker-Planck equation, that can be a powerful alternative to numerically
solving the inhomogeneous Balescu-Lenard equation
through stochastic ${N-}$body techniques.
Compared to the derivation of~\cite{Chavanis2012}, the
present formalism presents several advantages: it shows more clearly the
dressing of particles by their
polarisation cloud, as we calculate from the start an explicit expression
of the dressed potential of
a moving particle. Collective effects are obtained by solving the
linearised Klimontovich equation. As a result, the approximation where
collective effects
are neglected is straightforwardly recovered. In addition,
this calculation also
takes into account the possibility for particles to have different masses, a
generalisation that is particularly important for astrophysical applications,
where mass segregation is deemed to play a significant role on secular timescales.
In the single component case, the results of~\cite{Chavanis2012} are recovered.

The paper is organised as follows.
Section~\ref{sec:dressedpotential} presents the dressed potential of a test particle.
Section~\ref{sec:diffusioncoefficients} computes the diffusion coefficient in
action space.
Section~\ref{sec:coeffdynamicalfriction} focuses on the coefficient of dynamical
friction.
Section~\ref{sec:fromFPtoBL} relates the corresponding Fokker-Planck equation to
the Balescu-Lenard equation. Section~\ref{sec:langevin} provides the
equivalent stochastic Langevin equations and stresses their practical interest.
Section~\ref{sec:stages} discusses the different stages in the evolution of
stellar systems in the light of kinetic theory, and finally, section~\ref{sec:conclusion}
wraps up. 
Appendix~\ref{sec:appendixdrift} computes the drift vector.
Appendix~\ref{sec:appendixFpol} gives a direct derivation of the friction force by polarisation.
Appendix~\ref{sec:prop} derives the main properties of the
multicomponent Balescu-Lenard equation (steady states, energy conservation, ${H-}$theorem).
Appendix~\ref{sec:tpa} develops the corresponding test
particle approach in a bath formalism and considers the thermal bath
and the sinking satellite problems.
Appendices~\ref{sec:appendixformula} and~\ref{sec:appendixsymmetries}
provide mathematical results needed in the calculations.

\section{Dressed potential of a moving particle}
\label{sec:dressedpotential}

When considering self-gravitating systems, two main difficulties have to be overcome. First, self-gravitating systems are spatially inhomogeneous, which makes the trajectories of individual particles intricate. Assuming the mean system to be integrable, one can deal with inhomogeneity thanks to the use of angle-action coordinates. The second difficulty arises from the system's self-gravity, i.e. its ability to amplify perturbations. Dealing with these collective effects requires to study the dressing of fluctuations on dynamical times. Let us therefore first compute the gravitational polarisation induced by a moving particle in a self-gravitating system.

\subsection{Notations}
\label{sec:notations}

We consider a test particle moving in a self-gravitating inhomogeneous system,
composed of various components denoted with ``$\ra$", ``$\rb$", etc. We note as
${ F^{\ra}_{\rm tot} (\bm{x} , \bm{v}) }$ the distribution function (DF) of
particles of component ``$\ra$" with individual mass $\mu_{\ra}$. The DFs are
normalised such that ${ \! \int \! \rd \bm{x} \rd \bm{v} \, F^{\ra}_{\rm tot} \!=\!
M_{\rm tot}^{\ra} }$, where ${ M_{\rm tot}^{\ra} \!=\! N_{\ra} \mu_{\ra} }$ is
the total active mass of the component ``$\ra$" composed of $N_{\ra}$ particles.
The test particle is denoted by a subscript ``$\rt$", which
should not be mixed up with the time $t$. The system's total gravitational
potential is written as ${ U (\bm{x} , t) }$. The DFs and the potential can
be decomposed as
\begin{equation}
F^{\ra}_{\rm tot} (\bm{x} , \bm{v} , t) = F^{\ra} (\bm{x}) + \delta F^{\ra}
(\bm{x} , t) \;\;\; ; \;\;\; U (\bm{x} , t) = U_{0} (\bm{x}) + \delta
U_{\rt} (\bm{x} , t) + \delta U_{\rp} (\bm{x} , t ) \, ,
\label{decomposition_potentials}
\end{equation}
where $F^{\ra}$ only varies on secular times and stands 
for the unperturbed part of the DF, while ${ \delta F^{\ra} }$ depends on time
and is associated with the polarisation cloud surrounding the test particle.
Here, $U_{0}$ stands for the system's mean potential and carries on secular
timescales, ${ \delta U_{\rt} }$ is the potential created by the test
particle, and finally ${ \delta U_{\rp} }$ is the polarisation cloud induced
by the test particle. We assume that all perturbations are small, so that
\begin{equation}
\delta F^{\ra} \ll F^{\ra} \;\;\; ; \;\;\; \delta U_{\rt} , \, \delta
U_{\rp} \ll U_{0} \, .
\label{small_perturbations}
\end{equation}
The dynamics of a particle of component ``$\ra$'' is given by the total Hamiltonian $H_{\ra}$ reading
\begin{equation}
H_{\ra} = \frac{\mu_{\ra}}{2} \bm{v}^{2} \!+\! \mu_{\ra} \big[ U_{0} (\bm{x})
\!+\! \delta U_{\rt} (\bm{x} , t) \!+\! \delta U_{\rp} (\bm{x} , t) \big]
= H_{0} \!+\! \delta H_{\rt} \!+\! \delta H_{\rp} \, .
\label{Hamiltonian_a}
\end{equation}
For simplicity, the phase coordinates are denoted by ${ \Gamma \!=\! (\bm{x} ,
\bm{v}) }$. We assume the system to be quasi-stationary and integrable. As a
consequence, one can introduce angle-action coordinates ${ (\bm{\theta} ,
\bm{J}) }$~\citep{BinneyTremaine2008}. Within these coordinates, the system's mean potential becomes ${
U_{0} \!=\! U_{0} (\bm{J}) }$ while, following Jeans' theorem, the mean
quasi-stationary DFs $F^{\ra}$ depend also only on the actions $\bm{J}$. These
coordinates are canonical and conserve infinitesimal volumes, so that
\begin{equation}
\rd \Gamma = \rd \bm{x} \rd \bm{v} = \rd \bm{\theta} \rd \bm{J} \, .
\label{volume_canonical}
\end{equation}
The gradients w.r.t. to a vectorial variable such as $\bm{x}$ are denoted by
$\partial_{\bm{x}}$, while the derivative w.r.t. the time $t$ will sometimes be
written as $\partial_{t}$ for brevity. We also rely on the matrix
method~\citep{Kalnajs1976II} and introduce a biorthogonal basis of 
potentials and densities ${ (\psi^{(\alpha)} , \rho^{(\alpha)}) }$ satisfying
\begin{equation}
\psi^{(\alpha)} (\bm{x}) = \!\! \int \!\! \rd \bm{x} ' \, \rho^{(\alpha)}
(\bm{x}') \, u (|\bm{x} \!-\! \bm{x}'|) \;\;\; ; \;\;\; \!\! \int \!\! \rd
\bm{x} \, \psi^{(\alpha)} (\bm{x}) \, \rho^{(\beta) *} (\bm{x}) = -
\delta_{\alpha}^{\beta} \, ,
\label{definition_basis}
\end{equation}
where ${ u (\bm{x}) }$ corresponds to the interaction potential, i.e. ${ u
(|\bm{x}|) \!=\! - G / |\bm{x}| }$ in the gravitational case. Any perturbation
in the system can then be represented using these basis elements. For example,
a density field and the corresponding potential can be written as ${ \delta
\rho (\bm{x} , t) \!=\! a^{\alpha} (t) \,
\rho^{(\alpha)} (\bm{x}) }$ and ${ \delta
U (\bm{x} , t) \!=\! a^{\alpha} (t) \,
\psi^{(\alpha)} (\bm{x}) }$, where the sum over $\alpha$ is implied.

\subsection{Dressing of a test particle}
\label{sec:dressingtestparticle}

The Klimontovich equation~\citep{Klimontovich1967} reads symbolically for the one particle DF, ${ F_{\rm tot}^{\ra} \!=\! \sum_{i} \mu_{\ra} \delta_{\rD} (\bm{x} \!-\! \bm{x}_{i} (t)) \delta_{\rD} (\bm{v} \!-\! \bm{v}_{i} (t) ) }$:
\begin{equation}
\partial_{t} F^{\ra}_{\rm tot} + \dot{\bm{x}}_{\ra} \!\cdot\! \frac{\partial F^{\ra}_{\rm tot}}{\partial \bm{x}_{\ra}} \!+\! \dot{\bm{v}}_{\ra} \!\cdot\! \frac{\partial F^{\ra}_{\rm tot}}{\partial \bm{v}_{\ra}} = 0 \, ,
\label{eq:vlasov}
\end{equation}
where individual motions are given by Hamilton's equations reading ${ \mu_{\ra}
\dot{\bm{x}}_{\ra} \!=\! \partial H_{\ra} / \partial \bm{v}_{\ra} }$ and ${
\mu_{\ra} \dot{\bm{v}}_{\ra} \!=\! - \partial H_{\ra} / \partial \bm{x}_{\ra}
}$. When the motion of the stars is described by the angle-action variables ${ (
\bm{J}, \bm{\theta} ) }$, associated with the mean potential $U_{0}$, the
components of the Hamiltonian $H_{\ra}$ from equation~\eqref{Hamiltonian_a} read
\begin{equation}
H_{0} = \mu_{\ra} \, \bm{\Omega} \!\cdot\! \bm{J} \;\;\; ; \;\;\; \delta H_{\rt} = \mu_{\ra} \, \delta U_{\rt} ( \bm{x}_{\ra}(\bm{\theta} , \bm{J}) , t ) \;\;\; ; \;\;\; \delta H_{\rp} = \mu_{\ra} \, \delta U_{\rp} ( \bm{x}_{\ra}(\bm{\theta} , \bm{J}) , t ) \, .
\end{equation}
Following equation~\eqref{eq:vlasov}, the Klimontovich equation for the total DF
${ F^{\ra}_{\rm tot} \!=\! F^{\ra} \!+\! \delta F^{\ra} }$ becomes
\begin{equation}
\mu_{\ra} \, \partial_{t} F^{\ra} + \mu_{\ra} \, \partial_{t} \delta F^{\ra} + \partial_{\bm{\theta}} (F^{\ra} \!+\! \delta F^{\ra}) \!\cdot\! \partial_{\bm{J}} (H_{0} \!+\! \delta H_{\rt} \!+\! \delta H_{\rp}) - \partial_{\bm{J}} (F^{\ra} \!+\! \delta F^{\ra}) \!\cdot\! \partial_{\bm{\theta}} (H_{0} \!+\! \delta H_{\rt} \!+\! \delta H_{\rp}) = 0 \, .
\label{eq:vlasov-linear1}
\end{equation}
Taking into account that ${ \partial_{\bm{\theta}} H_{0} \!=\! 0 }$, ${
\partial_{\bm{\theta}} F^{\ra} \!=\! 0 }$, and ${ \partial_{\bm{J}} H_{0} \!=\!
\mu_{\ra} \bm{\Omega} }$, and retaining only linear terms (quasilinear
approximation) (here ${ \partial_{t} F^{\ra} }$ is a second order term that can
be neglected), equation~\eqref{eq:vlasov-linear1} becomes
\begin{equation}
\partial_{t} \delta F^{\ra} + \bm{\Omega} \!\cdot\! \partial_{\bm{\theta}} \delta F^{\ra} - \partial_{\bm{J} } F^{\ra} \!\cdot\! \partial_{\bm{\theta}} (\delta U_{\rt} + \delta U_{\rp}) = 0 \, .
\label{eq:vlasov-linear2}
\end{equation}
Equation~\eqref{eq:vlasov-linear2} is the linearised Klimontovich equation and
describes the amplification of perturbations on dynamical timescales.

\subsection{Dressed potential of a test particle}
\label{sec:dressedpotentialtestparticle}

In this section, let us compute the dressed potential generated by a given test particle.
The density ${ \delta \rho_{\rt} }$ generated by the test particle is
straightforwardly expressed as
\begin{equation}
\delta \rho_{\rt} (\bm{x} , t) = \mu_{\rt} \delta_{\rD} (\bm{x} \!-\! \bm{x}_{\rt} (t)) \, ,
\end{equation}
where $\mu_{\rt}$ stands for the mass of the test particle, and ${ \bm{x}_{\rt}
(t) }$ for its position at time $t$. Here $\delta_{\rD}$ is the Dirac function.
One may then expand this density on the basis of densities so
that ${ \delta \rho_{\rt} (\bm{x} , t) \!=\! a_{\rt}^{\alpha} (t) \,
\rho^{(\alpha)} (\bm{x}) }$. One has
\begin{equation}
a^{\alpha}_{\rt} (t) = - \!\! \int \!\! \rd \bm{x} \, \delta \rho_{\rt} (\bm{x} , t) \, \psi^{(\alpha) *} (\bm{x}) = - \mu_{\rt} \psi^{(\alpha) *} ( \bm{x}_{\rt} (t) ) \, .
\label{expression_at}
\end{equation}
The polarisation potential ${ \delta U_{\rp} }$ is self-consistently
generated by the contributions of the perturbation DFs, ${ \delta F^{\rb} }$, so
that the associated density perturbation ${ \delta \rho_{\rp} }$ reads 
\begin{equation}
\delta \rho_{\rp} (\bm{x} , t) = \sum_{\rb} \!\! \int \!\! \rd \bm{v} \, \delta F^{\rb} (\bm{x} , \bm{v} , t) \, .
\end{equation}
The associated coefficients in the potential-density expansion ${ \delta
\rho_{\rp} (\bm{x} , t) \!=\! a_{\rp}^{\alpha} (t) \, \rho^{(\alpha)} (\bm{x})
}$ read
\begin{equation}
a^{\alpha}_{\rp} (t) = - \!\! \int \!\! \rd \bm{x} \, \delta \rho_{\rp} (\bm{x} , t) \, \psi^{(\alpha) *} (\bm{x}) = - \sum_{\rb} \!\! \int \!\! \rd \bm{x} \rd \bm{v} \, \delta F^{\rb} (\bm{x} , \bm{v} , t) \, \psi^{(\alpha) *} (\bm{x}) \, .
\label{eq:aexpt1}
\end{equation}
Relying on the ${ 2 \pi-}$periodicity of the angles $\bm{\theta}$, one can
perform harmonic (Fourier) expansions of these expressions. Let us first
decompose the basis elements as
\begin{equation}
\psi^{(\alpha)} (\bm{x}) = \sum_{\bm{m}} \psi_{\bm{m}}^{(\alpha)} (\bm{J}) \, \re^{\ri \bm{m} \cdot \bm{\theta}} \;\;\; \text{with} \;\;\; \psi_{\bm{m}}^{(\alpha)} (\bm{J}) = \frac{1}{(2 \pi)^{d}} \!\! \int \!\! \rd \bm{\theta} \, \psi^{(\alpha)} (\bm{x} (\bm{\theta} , \bm{J})) \, \re^{- \ri \bm{m} \cdot \bm{\theta}} \, ,
\end{equation}
where we introduced $d$ as the dimension of the physical space. In the angle-action coordinates, the unperturbed motion of the test particle is given by ${ \bm{\theta}_{\rt} \!=\! \bm{\theta}_{\rt}^{0} \!+\! \bm{\Omega}_{\rt} t }$ and ${ \bm{J}_{\rt} \!=\! \text{cst.} }$, so that equation~\eqref{expression_at} becomes
\begin{equation}
a^{\alpha}_{\rt} (t) = - \mu_{\rt} \sum_{\bm{m}} \psi_{\bm{m}}^{(\alpha) *} (\bm{J}_{\rt}) \, \re^{- \ri \bm{m} \cdot (\bm{\theta}_{\rt}^{0} + \bm{\Omega}_{\rt} t)} \, .
\label{eq:aexpt2}
\end{equation}
Let us now introduce the temporal Fourier transform with the convention
\begin{equation}
\widehat{f} (\omega) = \!\! \int_{- \infty}^{+ \infty} \!\!\!\!\!\! \rd t \, f (t) \, \re^{\ri \omega t} \;\;\; ; \;\;\; f (t) = \frac{1}{2 \pi} \!\! \int_{- \infty}^{+ \infty} \!\!\!\!\!\! \rd \omega \, \widehat{f} (\omega) \, \re^{- \ri \omega t} \, .
\label{definition_Fourier_time}
\end{equation}
Starting from equation~\eqref{eq:aexpt2}, the temporal Fourier transform of ${
a^{\alpha}_{\rt} (t) }$ reads
\begin{equation}
\widehat{a}^{\alpha}_{\rt} (\omega) = - 2 \pi \mu_{\rt} \sum_{\bm{m}} \psi_{\bm{m}}^{(\alpha) *} (\bm{J}_{\rt}) \, \re^{-\ri \bm{m} \cdot \bm{\theta}_{\rt}^{0}} \, \delta_{\rD} (\omega \!-\! \bm{m} \!\cdot\! \bm{\Omega}_{\rt} ) \, .
\label{eq:aexpt3}
\end{equation}
The coefficients produced by the polarisation cloud of the test particle,
$a^{\alpha}_{\rp}$, can similarly be expressed via the harmonic Fourier
transforms of the DFs, ${ \delta F^{\rb}_{\bm{m}} (\bm{J} , t) }$, so that
equation~\eqref{eq:aexpt1} gives
\begin{equation}
a^{\alpha}_{\rp} (t) = - \sum_{\rb} \!\!\sum_{\bm{m} , \bm{m}'} \! \int \!\! \rd
\bm{\theta} \rd \bm{J} \, \delta F^{\rb}_{\bm{m}} (\bm{J} , t) \, \psi^{(\alpha)
*}_{\bm{m}'} (\bm{J}) \, \re^{\ri ( \bm{m} - \bm{m}') \cdot \bm{\theta}} \, .
\label{a_alpha_p_time}
\end{equation}
The integral over ${ \rd \bm{\theta} }$ yields ${ (2 \pi)^{d}
\delta_{\bm{m}}^{\bm{m}'} }$. The Fourier transform
of equation~\eqref{a_alpha_p_time} w.r.t. time then reads
\begin{equation}
\widehat{a}^{\alpha}_{\rp} (\omega) = - (2 \pi)^{d} \sum_{\rb} \sum_{\bm{m}} \!\! \int \!\! \rd \bm{J} \, \delta \widehat{F}^{\rb}_{\bm{m}} (\bm{J} , \omega) \, \psi^{(\alpha) *}_{\bm{m}}(\bm{J}) \, .
\label{eq:aalphaomega}
\end{equation}
Finally, the potentials ${ \delta U_{\rt} }$ and ${ \delta U_{\rp} }$ can straightforwardly be expressed in terms of the $a_{\rt}^{\alpha}$ and $a_{\rp}^{\alpha}$ coefficients as
\begin{equation}
\delta U_{\rt} (\bm{x} , t) = a^{\alpha}_{\rt} (t) \,
\psi^{(\alpha)} (\bm{x}) \;\;\; ; \;\;\; \delta U_{\rp} (\bm{x} , t) =
 a^{\alpha}_{\rp} (t) \, \psi^{(\alpha)} (\bm{x}) \, .
\label{eq:defU}
\end{equation}
The next step involves computing the coefficients $\widehat{a}^{\alpha}_{\rp}$ as
a function of the parameters of the test particle. This is possible by solving
equation~\eqref{eq:vlasov-linear2}, which describes the self-gravitating amplification of perturbations on dynamical timescales.
When Fourier transformed w.r.t. the angles and time, it yields
\begin{equation}
\delta \widehat{F}^{\ra}_{\bm{m}} (\bm{J} , \omega) = - \frac{\bm{m} \!\cdot\! \partial_{\bm{J}} F^{\ra}}{\omega \!-\! \bm{m} \!\cdot\! \bm{\Omega}} \bigg[ \delta \widehat{U}^{\rt}_{\bm{m}} (\bm{J} , \omega) \!+\! \delta \widehat{U}^{\rp}_{\bm{m}} (\bm{J} , \omega) \bigg] \, .
\label{eq:fsolved}
\end{equation}
Substituting equation~\eqref{eq:defU} into equation~\eqref{eq:fsolved}, ${
\delta \widehat{F}^{\ra}_{\bm{m}} (\bm{J} , \omega) }$ can be expressed as a
function of the $\widehat{a}_{\rp}^{\alpha}$ coefficients.
Equation~\eqref{eq:aalphaomega} then becomes a self-consistent equation for
these coefficients. After a few transformations, it can be recast as
\begin{equation}
\varepsilon_{\alpha \beta} (\omega) \, \widehat{a}^{\beta}_{\rp} (\omega) = \big[ \delta_{\alpha \beta} \!-\! \varepsilon_{\alpha \beta} (\omega) \big] \, \widehat{a}^{\beta}_{\rt} (\omega) = \widehat{\mathbf{M}}_{\alpha \beta} (\omega) \, \widehat{a}^{\beta}_{\rt} (\omega) \, ,
\label{eq:epsilon}
\end{equation}
where the sum over $\beta$ is implied. In equation~\eqref{eq:epsilon},
we introduced the usual Kronecker symbol $\delta_{\alpha \beta}$, and the
system's susceptibility ${ \varepsilon_{\alpha \beta} }$ as
\begin{equation}
\varepsilon_{\alpha \beta} (\omega) = \delta_{\alpha \beta} - \sum_{\bm{m}} (2 \pi)^{d} \!\! \int \!\! \rd \bm{J} \, \frac{\bm{m} \!\cdot\! \partial_{\bm{J}} \big[ \sum_{\rb} F^{\rb} (\bm{J}) \big]}{\omega \!-\! \bm{m} \!\cdot\! \bm{\Omega}} \, \psi^{(\alpha) *}_{\bm{m}} (\bm{J}) \, \psi^{(\beta)}_{\bm{m}} (\bm{J}) \, .
\end{equation}
Finally, in equation~\eqref{eq:epsilon}, we also introduced the system's
response matrix ${ \widehat{\mathbf{M}} (\omega)}$,
which controls the strength of the self-gravitating amplification in the system.
The total dressed potential, ${ \delta U_{\rd} }$, defined as ${ \delta U_{\rd}
\!=\! \delta U_{\rp} \!+\! \delta U_{\rt} }$ immediately follows. Its
coefficients ${ a_{\rd} \!=\! a_{\rt} \!+\! a_{\rp} }$ read
\begin{equation}
\widehat{a}^{\alpha}_{\rd} (\omega) = \widehat{a}^{\alpha}_{\rt} (\omega) + \widehat{a}^{\alpha}_{\rp}(\omega) = \varepsilon_{\alpha \beta}^{-1} (\omega) \, \widehat{a}^{\beta}_{\rt} (\omega) \, ,
\label{eq:calc_ad}
\end{equation}
where we assumed that the system is linearly stable so that ${ \varepsilon_{\alpha \beta} (\omega) }$ can be inverted.
In conclusion, thanks to the coefficients
${ \widehat{a}^{\beta}_{\rt} (\omega) }$
from equation~\eqref{eq:aexpt3}, the total dressed potential perturbation
${ \delta U_{\rd} }$
reads
\begin{equation}
\delta \widehat{U}_{\rd} (\bm{x} , \omega) = \psi^{(\alpha)} (\bm{x}) \,
\varepsilon_{\alpha \beta}^{-1} (\omega) \, \widehat{a}^{\beta}_{\rt} (\omega)=-
2\pi \mu_{\rt} \sum_{\bm{m}}
\psi^{(\alpha)} (\bm{x})
\, \varepsilon_{\alpha \beta}^{-1} (\omega) \,
\psi^{(\beta) *}_{\bm{m}} (\bm{J}_{\rt}) \, \re^{- \ri \bm{m} \cdot
\bm{\theta}_{\rt}^{0}} \, \delta_{D}(\omega-\bm{m} \!\cdot\!
\bm{\Omega}_{\rt})
\, .
\label{eq:defUdomega}
\end{equation}
Written as a function of time,
equation~\eqref{eq:defUdomega} becomes
\begin{equation}
\delta U_{\rd} (\bm{x} , t) = - \mu_{\rt} \sum_{\bm{m}} \psi^{(\alpha)} (\bm{x}) \, \varepsilon_{\alpha \beta}^{-1} (\bm{m} \!\cdot\! \bm{\Omega}_{\rt}) \, \psi^{(\beta) *}_{\bm{m}} (\bm{J}_{\rt}) \, \re^{- \ri \bm{m} \cdot (\bm{\theta}_{\rt}^{0} + \bm{\Omega}_{\rt} t)} \, .
\label{eq:defUd}
\end{equation}
Note that the bare potential perturbation ${ \delta U_{\rm bare} \!=\! \delta
U_{\rt} }$ has the same functional form with $\varepsilon$ being replaced
by the identity, so that
\begin{equation}
\delta U_{\rm bare} (\bm{x} , t) = - \mu_{\rt} \sum_{\bm{m}} \psi^{(\alpha)} (\bm{x}) \, \psi^{(\alpha) *}_{\bm{m}} (\bm{J}_{\rt}) \, \re^{- \ri \bm{m} \cdot (\bm{\theta}_{\rt}^{0} + \bm{\Omega}_{\rt} t)} \, .
\label{eq:potbare}
\end{equation}

The dressed potential, ${ \delta U_{2|1}^{\rd} }$, created by particle 2 and felt by particle 1 with action $\bm{J}_{1}$ and angle $\bm{\theta}_{1}$, is given by equation~\eqref{eq:defUd} when taking for ${ \psi^{(\alpha)} (\bm{x}_{1}) }$ the sum
\begin{equation}
\psi^{(\alpha)} (\bm{x}_{1}) = \sum_{\bm{m}_{1}} \psi^{(\alpha)}_{\bm{m}_{1}} (\bm{J}_{1}) \, \re^{ \ri \bm{m}_{1} \cdot \bm{\theta}_{1}} \, ,
\end{equation}
and replacing the test particle ``$\rt$" by the particle~2. After some
reordering, equation~\eqref{eq:defUd} gives
\begin{equation}
\delta U^{\rd}_{2 | 1} (1 , t) = - \!\!\sum_{\bm{m}_{1} , \bm{m}_{2}}\!\!
\mu_{2} \, \re^{\ri (\bm{m}_{1} \cdot \bm{\theta}_{1} - \bm{m}_{2} \cdot
\bm{\theta}_{2})} \, \psi^{(\alpha)}_{\bm{m}_{1}} (\bm{J}_{1}) \,
\varepsilon_{\alpha \beta}^{-1} (\bm{m}_{2} \!\cdot\! \bm{\Omega}_{2}) \,
\psi^{(\beta) *}_{\bm{m}_{2}} (\bm{J}_{2}) \, .
\end{equation}
Let us finally introduce the system's dressed susceptibility coefficients ${ 1 /
\mathcal{D}_{\bm{m}_{1} , \bm{m}_{2}} (\bm{J}_{1} , \bm{J}_{2} , \omega) }$ as
\begin{equation}
\frac{1}{\mathcal{D}_{\bm{m}_{1} , \bm{m}_{2}} (\bm{J}_{1} , \bm{J}_{2} , \omega)} = \psi^{(\alpha)}_{\bm{m}_{1}} (\bm{J}_{1}) \, \varepsilon_{\alpha \beta}^{-1} (\omega) \, \psi^{(\beta) *}_{\bm{m}_{2}} (\bm{J}_{2}) \, ,
\label{definition_1/D}
\end{equation}
so that one can write
\begin{equation}
\delta U^{\rd}_{2|1} (1 , t) = - \!\!\sum_{\bm{m}_{1} , \bm{m}_{2}}\!\! \mu_{2} \, \frac{\re^{\ri ( \bm{m}_{1} \cdot \bm{\theta}_{1} - \bm{m}_{2} \cdot \bm{\theta}_{2})} }{ \mathcal{D}_{\bm{m}_{1} , \bm{m}_{2}} (\bm{J}_{1} , \bm{J}_{2} , \bm{m}_{2} \!\cdot\! \bm{\Omega}_{2})} \, .
\label{eq:Ud21}
\end{equation}
With the shorthand notations
\begin{equation}
\omega_{1} = \bm{m}_{1} \!\cdot\! \bm{\Omega}_{1} \;\;\; ; \;\;\; \omega_{2} = \bm{m}_{2} \!\cdot\! \bm{\Omega}_{2} \;\;\; ; \;\;\; z_{12} = z_{\bm{m}_{1} , \bm{m}_{2}} (1 , 2) = \bm{m}_{1} \!\cdot\! \bm{\theta}_{1} \!-\! \bm{m}_{2} \!\cdot\! \bm{\theta}_{2} \;\;\; ; \;\;\; \Lambda_{\bm{m}_{1} , \bm{m}_{2}} (1 , 2 , \omega) = \frac{1}{\mathcal{D}_{\bm{m}_{1} , \bm{m}_{2}} (\bm{J}_{1} , \bm{J}_{2} , \omega)} \, ,
\label{eq:shorthand}
\end{equation}
equation~\eqref{eq:Ud21} becomes
\begin{equation}
\delta U^{\rd}_{2|1} (1 , t) = - \mu_{2} \!\!\sum_{\bm{m}_{1} , \bm{m}_{2}}\!\! \re^{\ri z_{12}} \, \Lambda_{\bm{m}_{1} , \bm{m}_{2}} (\bm{J}_{1} , \bm{J}_{2} , \omega_{2}) \, .
\label{eq:Ud21short}
\end{equation}
The total potential created by the sum over all discrete particles in the system, each of them having its own angle-action variables, is then simply the sum over all the individual contributions given by equation~\eqref{eq:Ud21short}, so that
\begin{equation}
U^{\rm di}_{1} (1 , t) = - \sum_{2} \mu_{2} \!\!\sum_{\bm{m}_{1} , \bm{m}_{2}}\!\! \re^{\ri z_{12}} \, \Lambda_{\bm{m}_{1} , \bm{m}_{2}} (\bm{J}_{1} , \bm{J}_{2} , \omega_{2}) \, .
\label{eq:Ud1short}
\end{equation}
Under the assumptions made in the present derivation, 
the discrete noise is small which means that the fluctuating part in $U^{\rm di}_{1}$
is small (compared to the typical kinetic energy of a given particle).
This implies that ${ \sum_{\bm{m}_{1}} \sum_{\bm{m}_{2} \neq 0}
\Lambda_{\bm{m}_{1} , \bm{m}_{2}} (\bm{J}_{1} , \bm{J}_{2} , \omega_{2}) }$
should be treated like a small contribution. The smallness in the noise level
should therefore be measured as a function of the number of $\Lambda$ in an
expression. For example, $U^{\rm di}_{1}$ in equation~\eqref{eq:Ud1short}, is
expressed to first order in the noise level.

\section{Diffusion coefficients in action space}
\label{sec:diffusioncoefficients}

The evolution of the action of particle 1 under the influence of the force applied by all other discrete particles is captured by the Hamiltonian ${ H_{1} \!=\! \mu_{1} U_{1}^{\rm di} }$. The associated Hamilton's equation for the action reads
\begin{equation}
\frac{\rd \bm{J}_{1}}{\rd t} = - \frac{1}{\mu_{1}} \frac{\partial H_{1}}{\partial \bm{\theta}_{1}} = - \partial_{\bm{\theta}_{1}} U^{\rm di}_{1} (1 , t) \, .
\end{equation}
Given equation~\eqref{eq:Ud1short}, it becomes
\begin{equation}
\frac{\rd \bm{J}_{1}}{\rd t} = \sum_{2} \mu_{2} \!\!\sum_{\bm{m}_{1} , \bm{m}_{2}}\!\! \ri \bm{m}_{1} \, \Lambda_{\bm{m}_{1} , \bm{m}_{2}} (\bm{J}_{1} , \bm{J}_{2} , \omega_{2}) \, \re^{\ri z_{12}} \, ,
\label{eq:rateofchangeJ1}
\end{equation}
which can be integrated for a time ${ \Delta t }$ to give
\begin{equation}
\Delta \bm{J}_{1} = \sum_{2} \mu_{2} \!\!\sum_{\bm{m}_{1} , \bm{m}_{2}}\!\! \ri \bm{m}_{1} \!\! \int_{0}^{\Delta t} \!\!\!\!\!\! \rd t \, \Lambda_{\bm{m}_{1} , \bm{m}_{2}} (\bm{J}_{1} , \bm{J}_{2} , \omega_{2} ; t) \, \re^{\ri z_{12} (t)} \, .
\label{eq:deltaJ1}
\end{equation}
Equation~\eqref{eq:deltaJ1} requires the full knowledge of the particles' motion
in order to account for the explicit time dependencies in the integral term.

The frequency spectrum of fluctuations generated by the particles 2 is
associated will all harmonics ${ \bm{m}_{2} \!\cdot\! \bm{\Omega}_{2} }$.
Provided the system is not dynamically degenerate, its zero frequency component
corresponds to ${ \bm{m}_{2} \!=\! \bm{0} }$. Its fluctuating part corresponds to
non-zero values of $\bm{m}_{2}$ in equation~\eqref{eq:deltaJ1}. Summation over
all but null vectors will be represented with a dash. The action diffusion
tensor is an average which will be written as ${ \big< \displaystyle \Delta
\bm{J}_{1} \!\otimes\! \Delta \bm{J}_{1} / \Delta t \big> }$. This average is
carried on the dynamical variables of the particles 2 and on the angles of
particle $1$. It enters the Fokker-Planck equation for the DF in action space
for particle 1 and component ``$\ra$''. As always in Fokker-Planck theory, this
diffusion coefficient must be evaluated at second order in the noise
level~\citep{Risken1996}. Since equation~\eqref{eq:deltaJ1} is by construction
first order, and, since the diffusion tensor is quadratic in ${ \Delta \bm{J}
}$, one should compute other involved expression, e.g., ${ \re^{z_{12} (t)}
\!=\! \re^{\ri (\bm{m}_{1} \cdot \bm{\theta}_{1} (t) - \bm{m}_{2} \cdot
\bm{\theta}_{2} (t) )} }$ to zeroth order. For this term, this corresponds to
the uniform angular motion, for which one can write
\begin{equation}
\bm{m}_{1} \!\cdot\! \bm{\theta}_{1} (t) - \bm{m}_{2} \!\cdot\! \bm{\theta}_{2} (t) \approx \bm{m}_{1} \!\cdot\! \bm{\theta}_{1}^{0} - \bm{m}_{2} \!\cdot\! \bm{\theta}_{2}^{0} + \bm{m}_{1} \!\cdot\! \bm{\Omega}_{1} (0) \, t - \bm{m}_{2} \!\cdot\! \bm{\Omega}_{2} (0) \, t \, .
\end{equation}
Similarly, the susceptibility coefficient ${ \Lambda (... ; t) }$ can be evaluated at ${ t \!=\! 0 }$. At the level of this approximation, one therefore gets
\begin{equation}
\Delta \bm{J}_{1} = \sum_{2} \mu_{2} \!\!\sum_{\bm{m}_{1} , \bm{m}_{2}}' \!\! \bm{m}_{1} \, \Lambda_{\bm{m}_{1} , \bm{m}_{2}} (\bm{J}_{1} , \bm{J}_{2} , \omega_{2} ; 0) \, \re^{\ri (\bm{m}_{1} \cdot \bm{\theta}_{1}^{0} - \bm{m}_{2} \cdot \bm{\theta}_{2}^{0} )} \, \frac{\displaystyle \re^{\ri \Delta t ( \bm{m}_{1} \cdot \bm{\Omega}_{1} (0) - \bm{m}_{2} \cdot \bm{\Omega}_{2} (0) )} - 1}{ \bm{m}_{1} \!\cdot\! \bm{\Omega}_{1} (0) - \bm{m}_{2} \!\cdot\! \bm{\Omega}_{2} (0)} \, .
\label{eq:deltaJ2}
\end{equation}
While omitting that slowly varying variables should be evaluated at ${ t \!=\! 0 }$, it follows from equation~\eqref{eq:deltaJ2} that
\begin{align}
\bigg< \frac{\Delta \bm{J}_{1} \!\otimes\! \Delta \bm{J}_{1} }{\Delta t} \bigg> = & \, \sum_{2 , 3} \mu_{2} \mu_{3} \!\sum_{\bm{m}_{1} , \bm{m}_{2}}' \! \sum_{\bm{m}_{1}' , \bm{m}_{3}}' \!\! \bm{m}_{1} \!\otimes\! \bm{m}_{1}' \, \Lambda_{\bm{m}_{1} , \bm{m}_{2}} (1 , 2 , \bm{m}_{2} \!\cdot\! \bm{\Omega}_{2}) \, \Lambda_{\bm{m}_{1}' , \bm{m}_{3}} (1 , 3 , \bm{m}_{3} \!\cdot\! \bm{\Omega}_{3}) \, \re^{ \ri (\bm{m}_{1} \cdot \bm{\theta}_{1}^{0} - \bm{m}_{2} \cdot \bm{\theta}_{2}^{0}) + \ri (\bm{m}_{1}' \cdot \bm{\theta}_{1}^{0} - \bm{m}_{3} \cdot \bm{\theta}_{3}^{0})} \nonumber
\\
& \, \times \frac{\displaystyle \big[ \re^{ \ri \Delta t ( \bm{m}_{1} \cdot \bm{\Omega}_{1} - \bm{m}_{2} \cdot \bm{\Omega}_{2} ) } \!-\! 1 \big] \big[ \re^{ \ri \Delta t ( \bm{m}_{1}' \cdot \bm{\Omega}_{1} - \bm{m}_{3} \cdot \bm{\Omega}_{3} ) } \!-\! 1 \big]}{ \Delta t \, ( \bm{m}_{1} \!\cdot\! \bm{\Omega}_{1} - \bm{m}_{2} \!\cdot\! \bm{\Omega}_{2}) ( \bm{m}_{1}' \!\cdot\! \bm{\Omega}_{1} - \bm{m}_{3} \!\cdot\! \bm{\Omega}_{3} )} \, .
\label{eq:fricdyncoef1}
\end{align}
One may now average this expression 
over the initial angles of the particles $1$, $2$, and $3$, and on the action
distribution of particles $2$ and $3$. Here, one should pay attention to the
fact that particle $1$ acts as our test star, while particles $2$ and $3$ both
run over the field stars, i.e. over all stars except particle $1$. Let us first
perform an average over the initial angles of the test and field stars. Keeping
only the dependencies w.r.t. the initial angles, equation~\eqref{eq:fricdyncoef1}
requires to consider a term generically of the form
\begin{equation}
\sum_{2, 3} \!\! \int \!\! \rd \bm{\theta}_{1}^{0} \rd \bm{\theta}_{2}^{0} \rd \bm{\theta}_{3}^{0} \, \re^{\ri (\bm{m}_{1} \cdot \bm{\theta}_{1}^{0} - \bm{m}_{2} \cdot \bm{\theta}_{2}^{0} + \bm{m}_{1}' \cdot \bm{\theta}_{1}^{0} - \bm{m}_{3} \cdot \bm{\theta}_{3}^{0})} \, ,
\label{shape_angle_average}
\end{equation}
where it is important to note that the sum on particles $2$ and $3$ is
restricted to all the field stars for particle $1$, i.e. particles $2$ and $3$
are always different from particle $1$. Because only non-zero values of
$\bm{m}_{2}$ and $\bm{m}_{3}$ contribute to the fluctuations,
equation~\eqref{shape_angle_average} immediately imposes for particle $2$ and
$3$ to be the same, so that the sum $\sum_{3}$ can straightforwardly be
executed. Averaging over $\bm{\theta}_{1}^{0}$, $\bm{\theta}_{2}^{0}$, and over
the action distribution of particle $2$, then amounts to performing in
equation~\eqref{eq:fricdyncoef1} the replacement
\begin{equation}
\sum_{2} \longrightarrow \!\! \int \!\! \frac{\rd \bm{\theta}_{1}^{0}}{(2 \pi)^{d}} \sum_{\rb} \frac{1}{\mu_{\rb}} \!\! \int \!\! \rd \bm{\theta}_{2}^{0} \rd \bm{J}_{2} \, F^{\rb} (\bm{J}_{2}) \, ,
\label{replacement_2}
\end{equation}
where the presence of the prefactor ${ 1 / \mu_{\rb} }$ is associated with the chosen normalisation of the DFs, ${ \! \int \! \rd \bm{x} \rd \bm{v} F^{\rb} \!=\! M_{\rm tot}^{\rb} \!=\! N_{\rb} \mu_{\rb} }$. In equation~\eqref{eq:fricdyncoef1}, the average over $\bm{\theta}_{1}^{0}$ yields $\delta_{\bm{m}_{1}}^{- \bm{m}_{1}'}$, while the integration over the initial angle $\bm{\theta}_{2}^{0}$ gives ${ (2 \pi)^{d} \delta_{\bm{m}_{2}}^{- \bm{m}_{3}} }$. Equation~\eqref{eq:fricdyncoef1} becomes
\begin{align}
\bigg< \frac{\displaystyle \Delta \bm{J}_{1} \!\otimes\! \Delta \bm{J}_{1} }{\Delta t} \bigg> = & \, (2 \pi)^{d} \sum_{\rb} \mu_{\rb} \!\!\sum_{\bm{m}_{1} , \bm{m}_{2}}' \!\! \bm{m}_{1} \!\otimes\! \bm{m}_{1} \!\! \int \!\! \rd \bm{J}_{2} \, F^{\rb} (\bm{J}_{2}) \, \Lambda_{\bm{m}_{1} , \bm{m}_{2}}(1 , 2 , \bm{m}_{2} \!\cdot\! \bm{\Omega}_{2}) \nonumber
\\
& \, \times \Lambda_{- \bm{m}_{1} , - \bm{m}_{2}} (1 , 2 , - \bm{m}_{2} \!\cdot\! \bm{\Omega}_{2}) \, \frac{\displaystyle \big| \re^{ \ri \Delta t ( \bm{m}_{1} \cdot \bm{\Omega}_{1} - \bm{m}_{2} \cdot \bm{\Omega}_{2} )} \!-\! 1 \big|^{2}}{\Delta t \, ( \bm{m}_{1} \!\cdot\! \bm{\Omega}_{1} - \bm{m}_{2} \!\cdot\! \bm{\Omega}_{2})^{2}} \, .
\label{eq:fricdyncoef2}
\end{align}
Now symmetries imply that (see Appendix~\ref{sec:appendixsymmetries})
\begin{equation}
\Lambda_{- \bm{m}_{1} , - \bm{m}_{2}} (1 , 2 , - \bm{m}_{2} \!\cdot\! \bm{\Omega}_{2}) = \Lambda_{\bm{m}_{1} , \bm{m}_{2}}^{*} (1 , 2 , \bm{m}_{2} \!\cdot\! \bm{\Omega}_{2}) \, .
\end{equation}
Finally, the time limit can be carried using the relation (see Appendix~\ref{sec:appendixformula})
\begin{equation}
\lim_{\Delta t \to + \infty} \frac{\displaystyle \big| \re^{ \ri \Delta t ( \bm{m}_{1} \cdot \bm{\Omega}_{1} - \bm{m}_{2} \cdot \bm{\Omega}_{2} )} \!-\! 1 \big|^{2}}{\Delta t \, (\bm{m}_{1} \!\cdot\! \bm{\Omega}_{1} - \bm{m}_{2} \!\cdot\! \bm{\Omega}_{2})^{2}} = 2 \pi \, \delta_{\rD} (\bm{m}_{1} \!\cdot\! \bm{\Omega}_{1} - \bm{m}_{2} \!\cdot\! \bm{\Omega}_{2}) \, .
\end{equation}
It follows that
\begin{equation}
\bigg< \frac{\displaystyle \Delta \bm{J}_{1} \!\otimes\! \Delta \bm{J}_{1} }{\Delta t} \bigg> = (2 \pi)^{d + 1} \sum_{\rb} \mu_{\rb} \!\!\sum_{\bm{m}_{1} , \bm{m}_{2}}' \!\! \bm{m}_{1} \!\otimes\! \bm{m}_{1} \!\!\int\!\! \rd \bm{J}_{2} \, F^{\rb} (\bm{J}_{2}) \big|\Lambda_{\bm{m}_{1} , \bm{m}_{2}} (\bm{J}_{1} , \bm{J}_{2} , \bm{m}_{2} \!\cdot\! \bm{\Omega}_{2}) \big|^{2} \delta_{\rD} (\bm{m}_{1} \!\cdot\! \bm{\Omega}_{1} \!-\! \bm{m}_{2} \!\cdot\! \bm{\Omega}_{2} ) \, .
\label{eq:fricdyncoef3}
\end{equation}
Equation~\eqref{eq:fricdyncoef3} is the final expression of the diffusion tensor acting on the action vector of the test particle.

\section{Coefficients of dynamical friction}
\label{sec:coeffdynamicalfriction}

\subsection{Mean potential and shot noise}
\label{sec:meanpotentialandshotnoise}

In section~\ref{sec:dressedpotential}, the dressed potential created by
all discrete particles was computed. This potential can be split into a
quasi-stationnary part, $U_{\rm st}$, which is essentially the mean ensemble
potential, $U_{0}$, and a fluctuation, ${ \delta\widetilde{U} }$, coming from
the discrete particles. The real potential, including the noise contributions,
which is felt by particle 1 is given by
\begin{equation}
U^{\rm di} (1 , t) = U_{\rm st} (1) + \delta \widetilde{U} (1 , t).
\end{equation}
The potential $U_{0}$ is the average, computed from the one point 
DF of the zero frequency component of the discrete potential, i.e. $U_{\rm st}$.
This averaging may differ subtly from this zero frequency part, in as much as it
might display coarse grained features induced by textures in the action
distribution which are not taken into account by the DF. Here, we will assume
that the two potentials can be identified. The power spectrum of fluctuations
caused by particle 2 is given by all the harmonics ${ \bm{m}_{2} \!\cdot\!
\bm{\Omega}_{2} }$. Its zero frequency contribution is given by the ${
\bm{m}_{2} \!=\! \bm{0} }$ component, provided the system's potential is not dynamically degenerate.
The fluctuating part induced by discrete
particles corresponds then to the non-zero values of $\bm{m}_{2}$ in
equation~\eqref{eq:Ud1short}. The components corresponding to ${ \bm{m}_{1}
\!=\! \bm{0} }$ do not impact the variation of the actions, as can be seen in
equation~\eqref{eq:rateofchangeJ1}. In contrast, they could impact the angles,
which do play a role in what follows. The potential fluctuations
corresponding to the discrete nature of the particles are accounted for in
equation~\eqref{eq:Ud1short} by the ${ \bm{m}_{2} \!\neq\! \bm{0} }$ and any
$\bm{m}_{1}$ contribution in the sum. As previously, summations over all but null vectors will
be represented by a dash. When identifying the mean potential with the zero
frequency potential, the Hamiltonian for the motion of particle 1 in the
presence of shot noise from the other particles then reads
\begin{equation}
H (1) = \mu_{1} \bm{\Omega}_{1} \!\cdot\! \bm{J}_{1} - \mu_{1} \sum_{2} \mu_{2} \sum_{\bm{m}_{1}} \sum_{\bm{m}_{2}}' \Lambda_{\bm{m}_{1} , \bm{m}_{2}} (\bm{J}_{1} , \bm{J}_{2} , \bm{m}_{2} \!\cdot\! \bm{\Omega}_{2}) \, \re^{ \ri z_{12} (t) } \, ,
\label{eq:hamiltonian1}
\end{equation}
where we relied on equation~\eqref{eq:Ud1short}.
Without any ensemble average at this stage, the dynamical evolution of particle~$1$ under the influence of the Hamiltonian~\eqref{eq:hamiltonian1} is given by the following differential equations
\begin{align}
\frac{\rd \bm{J}_{1}}{\rd t} & \, = \sum_{2} \mu_{2} \sum_{\bm{m}_{1}} \sum_{\bm{m}_{2}}' \ri \bm{m}_{1} \Lambda_{\bm{m}_{1} , \bm{m}_{2}} (\bm{J}_{1} , \bm{J}_{2} , \bm{m}_{2} \!\cdot\! \bm{\Omega}_{2} ; t) \, \re^{ \ri z_{12} (t) } \, ,
\nonumber
\\
\frac{\rd \bm{\theta}_{1}}{\rd t} & \, = \bm{\Omega}_{1} \!-\! \sum_{2} \mu_{2} \sum_{\bm{m}_{1}} \sum_{\bm{m}_{2}}' \re^{ \ri z_{12} (t) } \, \partial_{\bm{J}_{1}} \big[ \Lambda_{\bm{m}_{1} , \bm{m}_{2}} (\bm{J}_{1} , \bm{J}_{2} , \bm{m}_{2} \!\cdot\! \bm{\Omega}_{2} ; t) \big] \, .
\label{eq:dyneq1}
\end{align}

\subsection{Mean friction at second order}
\label{sec:meanfrictionsecondorder}

Let us now compute the mean drag ${ \big< \Delta \bm{J}_{1} \big> }$ applied
onto particle 1 during ${ \Delta t }$. This change in the vector action must be
computed to second order in the level of the noise. Since
equation~\eqref{eq:dyneq1} is clearly first order only, this implies that we
cannot rely on zeroth order approximation for the ${ \Lambda_{\bm{m}_{1} ,
\bm{m}_{2}} (\bm{J}_{1} , \bm{J}_{2} , \bm{m}_{2} \!\cdot\! \bm{\Omega}_{2} ; t) }$
and ${ \re^{\ri z_{12} (t)} }$ factors.\footnote{Should we do so, ${ \big< \Delta
\bm{J}_{1} \big> }$ would vanish identically.}
They must then be computed at
the next order following, e.g.,~\cite{Ecker2013}.

The calculation at second order of the change in action ${ \big< \Delta \bm{J}_{1} \big> }$ is a somewhat technical calculation, that we present in detail in Appendix~\ref{sec:appendixdrift}. In the same Appendix, we also detail how one may average this drift vector over the initial angles of the involved particles. One finally obtains in equation~\eqref{final_drift_app} that the averaged drift vector is given by
\begin{equation}
\bigg< \frac{\Delta \bm{J}_{1}}{\Delta t} \bigg> = 
\sum_{\rb} \!\!\sum_{\bm{m}_{1} , \bm{m}_{2}}' \!\! \int \!\! \rd \bm{J}_{2} \,
F^{\rb} (\bm{J}_{2}) \, \pi (2 \pi)^{d} \bm{m}_{1} \big( \mu_{\rb} \,
\bm{m}_{1} \!\cdot\! \partial_{\bm{J}_{1}} \!-\! \mu_{\ra} \, \bm{m}_{2}
\!\cdot\! \partial_{\bm{J}_{2}} \big) \delta_{\rD} (\bm{m}_{1} \!\cdot\!
\bm{\Omega}_{1} \!-\! \bm{m}_{2} \!\cdot\! \bm{\Omega}_{2}) \, \big|
\Lambda_{\bm{m}_{1} , \bm{m}_{2}} (1 , 2 , \bm{m}_{2} \!\cdot\! \bm{\Omega}_{2})
\big|^{2} \, .
\label{final_drift}
\end{equation}
Equation~\eqref{final_drift} is the final expression of the dynamical friction coefficient
acting on the action vector of the test particle.

\section{From Fokker-Planck to Balescu-Lenard}
\label{sec:fromFPtoBL}

Let us now show how the Fokker-Planck equation based on the friction and
diffusions coefficients obtained previously is in fact fully consistent with the Balescu-Lenard
equation (see also Appendix~\ref{sec:tpa} for a discussion of the test particle approach).
The Fokker-Planck equation for the system's DF can be obtained from
the Master equation of a Markov process by
using the Kramers-Moyal expansion for the transition
probability~\citep{Risken1996}.
If the expansion stops after the second
term, one gets the Fokker-Planck equation
(also called the forward Kolmogorov equation), reading
\begin{equation}
\partial_{t} F^{\ra} (\bm{J}_{1},t) = \frac{1}{2} \partial_{\bm{J}_{1}}
\!\otimes\! \partial_{\bm{J}_{1}} \!\cdot\! \bigg[ \bigg< \frac{\displaystyle
\Delta \bm{J}_{1} \!\otimes\! \Delta \bm{J}_{1} }{\Delta t} \bigg> F^{\ra}
(\bm{J}_{1},t) \bigg] - \partial_{\bm{J}_{1}} \!\cdot\! \bigg[ \bigg<
\frac{\displaystyle \Delta \bm{J}_{1} }{\Delta t} \bigg> F^{\ra} (\bm{J}_{1},t)
\bigg] \, .
\label{eq:FPgeneric}
\end{equation}
It involves only the drift and
diffusion coefficients
\begin{equation}
 \bm{D}^{(1)}=
\bigg< \frac{\displaystyle \Delta \bm{J}_{1} }{\Delta t}
\bigg> = \bm{F}_{\rm fric} , \qquad \bm{D}^{(2)}=\frac{1}{2}\bigg<
\frac{\displaystyle \Delta
\bm{J}_{1} \!\otimes\! \Delta \bm{J}_{1}
}{\Delta t} \bigg> = \bm{D} ,
\label{d1d2}
\end{equation}
where one should note that these coefficients depend on ``${\ra}$'', the considered
component. In the present context, $\bm{F}_{\rm fric}$ represents the friction
force and $\bm{D}$ is the diffusion matrix in action space. In general, for
complex systems, it is not possible to determine the Kramers-Moyal coefficients
$\bm{D}^{(n)}$ from first principles. However, in the present case, this could be
achieved, in the two previous sections, by considering an expansion of the equations of Hamiltonian
dynamics in powers of ${1/N}$ in the limit ${ N \!\rightarrow\! + \infty }$.\footnote{It can
be shown that the quasilinear approximation amounts to
neglecting terms of order ${ 1/N^{2} }$ or smaller~\citep{Chavanis2012}.}
At order ${1/N}$, the diffusion and drift coefficients were
obtained in equations~\eqref{eq:fricdyncoef3} and~\eqref{final_drift}, and read
\begin{align}
\bm{D} & \, = \pi (2 \pi)^{d} \sum_{\rb} \mu_{\rb}
\!\!\sum_{\bm{m}_{1} , \bm{m}_{2}}' \!\! \int\!\! \rd \bm{J}_{2} \, F^{\rb}
(\bm{J}_{2},t) \, \bm{m}_{1} \!\otimes\! \bm{m}_{1} \,\frac{ \delta_{\rD}
(\bm{m}_{1} \!\cdot\! \bm{\Omega}_{1} \!-\! \bm{m}_{2} \!\cdot\!
\bm{\Omega}_{2})}{ \big| \mathcal{D}_{\bm{m}_{1} , \bm{m}_{2}} (\bm{J}_{1} ,
\bm{J}_{2} , \bm{m}_{2} \!\cdot\! {\bm{\Omega}_{2}}) \big|^{2}}
\, ,
\label{final_diff_drift1}
\\
\bm{F}_{\rm fric} & \, = \pi (2
\pi)^{d} \sum_{\rb} \!\!\sum_{\bm{m}_{1} , \bm{m}_{2}}' \!\! \int\!\! \rd
\bm{J}_{2} \, F^{\rb} (\bm{J}_{2},t) \, \bm{m}_{1} \bigg[ \mu_{\rb} \,
\bm{m}_{1}
\!\cdot\! \frac{\partial}{\partial {\bm{J}_{1}}} \!-\! \mu_{\ra} \, \bm{m}_{2}
\!\cdot\!
\frac{\partial}{\partial {\bm{J}_{2}}} \bigg] \frac{ \delta_{\rD}
(\bm{m}_{1} \!\cdot\! \bm{\Omega}_{1} \!-\! \bm{m}_{2} \!\cdot\!
\bm{\Omega}_{2})}{ \big| \mathcal{D}_{\bm{m}_{1} , \bm{m}_{2}} (\bm{J}_{1} ,
\bm{J}_{2} , \bm{m}_{2} \!\cdot\! {\bm{\Omega}_{2}}) \big|^{2}} \, ,
\label{final_diff_drift2}
\end{align}
where we note that the sums ${ \sum_{\bm{m}_{1} , \bm{m}_{2}}' }$ are restricted
to non-zero values of $\bm{m}_{1}$ and $\bm{m}_{2}$. It can be shown that the
higher order Kramers-Moyal coefficients are negligible at order ${ 1/N }$. This
fully justifies the Fokker-Planck equation~\eqref{eq:FPgeneric}.

Using the notations from equation~\eqref{d1d2}, the Fokker-Planck equation~\eqref{eq:FPgeneric} can be rewritten
as
\begin{equation}
\partial_{t} F^{\ra} (\bm{J}_{1},t) = \partial_{\bm{J}_{1}}
\!\otimes\! \partial_{\bm{J}_{1}} \!\cdot\! \bigg[ \bm{D}^{(2)}(\bm{J}_{1} , t) \, F^{\ra}
(\bm{J}_{1},t) \bigg] - \partial_{\bm{J}_{1}} \!\cdot\!
\bigg[\bm{D}^{(1)}(\bm{J}_{1} , t) \, F^{\ra} (\bm{J}_{1},t)
\bigg] 
\label{fpd1d2}
\end{equation}
or, equivalently, as
\begin{equation}
\frac{\partial F^{\ra}}{\partial t} (\bm{J}_{1},t)\!=\! \frac{\partial }{\partial
\bm{J}_{1}}
\!\cdot\! \bigg[ \frac{\partial }{\partial \bm{J}_{1}} \!\cdot\! \bigg(
\bm{D} (\bm{J}_{1} , t) \, F^{\ra} (\bm{J}_{1} , t) \bigg) - \bm{F}_{\rm fric}
(\bm{J}_{1} , t) \, F^{\ra} (\bm{J}_{1} , t)
\, \bigg] \, .
\label{nice}
\end{equation}
In order to make the connection with the Balescu-Lenard equation, let us rewrite
equation~\eqref{nice} under a form in which the diffusion coefficient is
``sandwiched'' between the two action derivatives, i.e.
\begin{equation}
\frac{\partial F^{\ra} }{\partial t}(\bm{J}_{1},t) = \frac{\partial }{\partial
\bm{J}_{1}}
\!\cdot\! \bigg[ \bm{D}
(\bm{J}_{1} , t) \!\cdot\! \frac{\partial F^{\ra} }{\partial \bm{J}_{1}}
 \!-\! \bm{F}_{\rm pol} (\bm{J}_{1} , t) \, F^{\ra} (\bm{J}_{1} , t)\bigg]
\,
,
\label{sandwichedFP}
\end{equation}
where we defined
\begin{equation}
\bm{F}_{\rm pol}=\bm{F}_{\rm fric}-\frac{\partial
\bm{D}}{\partial \bm{J}_{1}}.
\label{Fpol}
\end{equation}
Here, $\bm{F}_{\rm pol}$ represents the friction force by
polarisation~\citep{Chavanis2012}. It differs from the true friction force $\bm{F}_{\rm fric}$ by a term involving the
derivatives of the diffusion tensor $\bm{D}$. Integrating equation~\eqref{final_diff_drift2} by parts and comparing the resulting expression with equations~\eqref{final_diff_drift1} and~\eqref{Fpol}, we finally get
\begin{align}
\bm{F}_{\rm pol}=\pi (2
\pi)^{d} \mu_{\ra} \sum_{\rb} \!\!\sum_{\bm{m}_{1} , \bm{m}_{2}}' \!\! \int\!\! \rd
\bm{J}_{2} \, \bm{m}_{1} \bigg[ \bm{m}_{2} \!\cdot\! \frac{\partial F^{\rb}}{\partial 
\bm{J}_{2}} \bigg] \, \frac{ \delta_{\rD}
(\bm{m}_{1} \!\cdot\! \bm{\Omega}_{1} \!-\! \bm{m}_{2} \!\cdot\!
\bm{\Omega}_{2})}{ \big| \mathcal{D}_{\bm{m}_{1} , \bm{m}_{2}} (\bm{J}_{1} ,
\bm{J}_{2} , \bm{m}_{2} \!\cdot\! {\bm{\Omega}_{2}}) \big|^{2}} \, .
\label{Fpolexp}
\end{align}
The friction force by polarisation can also be obtained from a
direct calculation based on a linear response theory (see Appendix~\ref{sec:appendixFpol}).
Substituting equations~\eqref{final_diff_drift1} and~\eqref{Fpolexp} into
equation~\eqref{sandwichedFP}, we immediately obtain
\begin{equation}
\frac{\partial F^{\ra}}{\partial t}(\bm{J}_{1},t) = \pi (2 \pi)^{d} \sum_{\rb}
\!\!\!\sum_{\bm{m}_{1} , \bm{m}_{2}}' \!\! \bm{m}_{1} \!\cdot\!
\frac{\partial}{\partial{\bm{J}_{1}}} \!\! \int\!\! \rd \bm{J}_{2} \, \frac{
\delta_{\rD}
(\bm{m}_{1} \!\cdot\! \bm{\Omega}_{1} \!-\! \bm{m}_{2} \!\cdot\!
\bm{\Omega}_{2})}{ \big| \mathcal{D}_{\bm{m}_{1} , \bm{m}_{2}} (\bm{J}_{1} ,
\bm{J}_{2} , \bm{m}_{2} \!\cdot\! {\bm{\Omega}_{2}}) \big|^{2}} \bigg[
\mu_{\rb} \, \bm{m}_{1}
\!\cdot\! \frac{\partial}{\partial {\bm{J}_{1}}} \!-\! \mu_{\ra} \, \bm{m}_{2}
\!\cdot\!
\frac{\partial}{\partial {\bm{J}_{2}}} \bigg]
\, F^{\ra} (\bm{J}_{1},t) \, F^{\rb}
(\bm{J}_{2},t) \, .
\label{eq:BL}
\end{equation}
This is the inhomogeneous Balescu-Lenard
equation~\citep{Heyvaerts2010,Chavanis2012}. Hence we have demonstrated that
the Balescu-Lenard equation is equivalent to the Fokker-Planck equation.
These are the kinetic equations describing the secular evolution of dressed particles in inhomogeneous systems via resonant binary interactions.

It is straightforward to specialise the previous expressions to the case where
collective effects are not accounted for. This amounts to replacing the dressed
potential perturbation ${ \delta U_{\rd} }$ from equation~\eqref{eq:defUd} by
the bare potential perturbation ${ \delta U_{\rm bare} }$ from equation~\eqref{eq:potbare}, while all the
following calculations remain the same. The dressed susceptibility coefficients
${ 1/\mathcal{D}_{\bm{m}_{1} , \bm{m}_{2}} }$ from
equation~\eqref{definition_1/D} then become the bare susceptibility coefficients
${ 1/\mathcal{D}_{\bm{m}_{1} , \bm{m}_{2}}^{\rm bare} }$ reading
\begin{equation}
\frac{1}{\mathcal{D}_{\bm{m}_{1} , \bm{m}_{2}}^{\rm bare} (\bm{J}_{1} ,
\bm{J}_{2}) } = \psi_{\bm{m}_{1}}^{(\alpha)} (\bm{J}_{1}) \,
\psi_{\bm{m}_{2}}^{(\alpha) *} (\bm{J}_{2}) = - A_{\bm{m}_{1} , \bm{m}_{2}}
(\bm{J}_{1} , \bm{J}_{2}) \, ,
\label{definition_1/D_bare}
\end{equation}
where the bare susceptibility coefficients ${ A_{\bm{m}_{1} , \bm{m}_{1}}
(\bm{J}_{1} , \bm{J}_{2}) }$~\citep{LyndenBell1994,Pichon1994,Chavanis2013} are
given by the Fourier transform in angles of the interaction potential $u$, so
that
\begin{align}
& \, u (\bm{x} ( \bm{\theta}_{1} , \bm{J}_{1}) \!-\! \bm{x} (\bm{\theta}_{2} ,
\bm{J}_{2}) ) = \!\! \sum_{\bm{m}_{1} , \bm{m}_{2}} \!\! A_{\bm{m}_{1} ,
\bm{m}_{2}} (\bm{J}_{1} , \bm{J}_{2}) \, \re^{\ri (\bm{m}_{1} \cdot
\bm{\theta}_{1} - \bm{m}_{2} \cdot \bm{\theta}_{2})} \, , \nonumber
\\
& \, A_{\bm{m}_{1} , \bm{m}_{2}} (\bm{J}_{1} , \bm{J}_{2}) = \frac{1}{(2
\pi)^{2d}} \!\! \int \!\! \rd \bm{\theta}_{1} \rd \bm{\theta}_{2} \, u ( \bm{x}
(\bm{\theta}_{1} , \bm{J}_{1}) \!-\! \bm{x} (\bm{\theta}_{2} , \bm{J}_{2}) ) \,
\re^{- \ri (\bm{m}_{1} \cdot \bm{\theta}_{1} - \bm{m}_{2} \cdot
\bm{\theta}_{2})} \, .
\label{definition_A_bare}
\end{align}
The detailed calculations leading to the third equality of
equation~\eqref{definition_1/D_bare} are given in Appendix B of~\cite{FouvryPichonChavanis2015}.
Because of these strong similarities, the bare analogs of the drift and
diffusion coefficients from equations~\eqref{final_diff_drift1},~\eqref{final_diff_drift2}
and~\eqref{Fpolexp} are immediately given by
\begin{align}
& \, \bm{D} = \pi (2 \pi)^{d } \sum_{\rb} \mu_{\rb}
\!\!\sum_{\bm{m}_{1} , \bm{m}_{2}}' \!\! \int\!\! \rd \bm{J}_{2} \, F^{\rb}
(\bm{J}_{2},t) \, \bm{m}_{1} \!\otimes\! \bm{m}_{1} \, \delta_{\rD}
(\bm{m}_{1} \!\cdot\! \bm{\Omega}_{1} \!-\! \bm{m}_{2} \!\cdot\!
\bm{\Omega}_{2}) \,
\big| A_{\bm{m}_{1} , \bm{m}_{2}} (\bm{J}_{1} , \bm{J}_{2}) \big|^{2} \, ,
\label{final_diff_drift_bare1}
\\
& \, \bm{F}_{\rm fric} = \pi (2
\pi)^{d} \sum_{\rb} \!\!\sum_{\bm{m}_{1} , \bm{m}_{2}}' \!\! \int\!\! \rd
\bm{J}_{2} \, F^{\rb} (\bm{J}_{2},t) \, \bm{m}_{1} \bigg[
\mu_{\rb} \, \bm{m}_{1}
\!\cdot\! \frac{\partial}{\partial {\bm{J}_{1}}} \!-\! \mu_{\ra} \, \bm{m}_{2}
\!\cdot\!
\frac{\partial}{\partial {\bm{J}_{2}}} \bigg] \, 
\delta_{\rD}
(\bm{m}_{1} \!\cdot\! \bm{\Omega}_{1} \!-\! \bm{m}_{2} \!\cdot\!
\bm{\Omega}_{2}) \, \big| A_{\bm{m}_{1}
, \bm{m}_{2}} (\bm{J}_{1} , \bm{J}_{2}) \big|^{2} \, ,
\label{final_diff_drift_bare2}
\\
& \, \bm{F}_{\rm pol} = \pi (2
\pi)^{d} \mu_{\ra} \,\sum_{\rb} \!\!\sum_{\bm{m}_{1} , \bm{m}_{2}}' \!\! \int\!\! \rd
\bm{J}_{2} \, \bm{m}_{1} \bigg[ \bm{m}_{2} \!\cdot\! \frac{\partial F^{\rb}}{\partial 
\bm{J}_{2}} \bigg] \, \delta_{\rD}
(\bm{m}_{1} \!\cdot\! \bm{\Omega}_{1} \!-\! \bm{m}_{2} \!\cdot\!
\bm{\Omega}_{2}) \, \big| A_{\bm{m}_{1}
, \bm{m}_{2}} (\bm{J}_{1} , \bm{J}_{2}) \big|^{2} \, .
\label{Fpolexpbar}
\end{align}
Similarly, the inhomogeneous Balescu-Lenard equation~\eqref{eq:BL}, when
neglecting collective effects, becomes the inhomogeneous Landau
equation~\citep{Chavanis2013} reading
\begin{equation}
\frac{\partial F^{\ra}}{\partial t} (\bm{J}_{1},t) \!=\! \pi (2 \pi)^{d}
\sum_{\rb}
\!\!\!\sum_{\bm{m}_{1} , \bm{m}_{2}}' \!\!\! \bm{m}_{1} \!\cdot\!
\frac{\partial}{\partial{\bm{J}_{1}}} \!\! \int\!\! \rd \bm{J}_{2} \,
\delta_{\rD}
(\bm{m}_{1} \!\cdot\! \bm{\Omega}_{1} \!-\! \bm{m}_{2} \!\cdot\!
\bm{\Omega}_{2}) \, \big| A_{\bm{m}_{1} , \bm{m}_{2}} (\bm{J}_{1} ,
\bm{J}_{2}) \big|^{2} \bigg[\!
\mu_{\rb} \, \bm{m}_{1}
\!\cdot\! \frac{\partial}{\partial {\bm{J}_{1}}} \!-\! \mu_{\ra} \, \bm{m}_{2}
\!\cdot\!
\frac{\partial}{\partial {\bm{J}_{2}}} \!\bigg] F^{\ra}
(\bm{J}_{1},t) \, F^{\rb} (\bm{J}_{2},t) \, .
\label{eq:Landau}
\end{equation}
One can easily check that, in the single species case, all the results presented
in this section 
agree with those obtained in~\cite{Chavanis2012} via a different method.

\section{From Fokker-Planck to Langevin}
\label{sec:langevin}

The Kramers-Moyal coefficients appearing in the Fokker-Planck equation~\eqref{eq:FPgeneric} may be derived from stochastic
Langevin equations~\citep{Risken1996}. In the present context,
such a Langevin equation describes the evolution
of the action ${ \bm{{J}} (t) }$ of a given (test) star. Let us consider a
general Langevin equation of the form
\begin{equation}
\frac{\rd \bm{{J}}}{\rd t} = \bm{h} (\bm{{J}} , t) + \bm{g}
(\bm{{J}} , t) \!\cdot\! \bm{\Gamma} (t) \, ,
\label{Langevin_equation}
\end{equation}
where $\bm{h}(\bm{{J}} , t)$ is a vector, $\bm{g}(\bm{{J}} , t)$ is a tensor,
and ${ \bm{\Gamma} (t)
}$ is a Gaussian white noise (Langevin force) whose
statistics satisfy
\begin{equation}
\big< \bm{\Gamma} (t) \big> = 0 \;\;\; ; \;\;\; \big< \bm{\Gamma} (t)
\!\otimes\! \bm{\Gamma} (t') \big> = 2 \, \bm{I} \, \delta_{\rm D} (t
\!-\! t') \, ,
\label{Langevin_forces}
\end{equation}
where $\bm{I}$ is the identity matrix. When the tensor $\bm{g}(\bm{{J}} , t)$
explicitly depends on the action $\bm{{J}}$ of the particle, we say that the
noise is multiplicative. Using the Stratonovich picture (ordinarily used by physicists),
the drift and diffusion coefficients are given by
\begin{equation}
{D}^{(1)}_{i} ={h}_{i} \!+\! \sum_{j,k} g_{kj}\frac{\partial g_{ij}}{\partial
J_k}\;\;\; ; \;\;\; D^{(2)}_{ij}= \sum_{k} g_{ik} g_{jk} \, .
\label{g_h_relation}
\end{equation}
The other Kramers-Moyal coefficients are zero. The last term in the expression
of $\bm{D}^{(1)}$ is the noise-induced drift or spurious drift.
The drift and diffusion coefficients determine the Fokker-Planck equation~\eqref{fpd1d2} which describes the evolution of the probability density. The
drift and diffusion coefficients $\bm{D}^{(1)}$ and $\bm{D}^{(2)}$ are uniquely
determined by the functions $\bm{h}$ and $\bm{g}$ of the Langevin equations as
given by equation~\eqref{g_h_relation}.

Let us now consider the inverse problem, i.e., the determination of the Langevin
equations from the Fokker-Planck equation. As discussed in~\cite{Risken1996}, in
the multidimensional case, the functions $\bm{h}$ and $\bm{g}$ are not
uniquely determined by the drift and diffusion coefficients $\bm{D}^{(1)}$ and
$\bm{D}^{(2)}$. One particular solution obtained by diagonalising the positive
definite matrix $\bm{D}^{(2)}$ is given by
\begin{equation}
{h}_{i} = {D}^{(1)}_{i} \!-\! \sum_{j,k} \big(\! \sqrt{{D}}^{(2)}
\big)_{kj} \frac{\partial \big(\! \sqrt{{D}}^{(2)} \big)_{ij}}{\partial
J_{k}} \;\;\; ; \;\;\; {g}_{ij} = \big(\! \sqrt{{D}}^{(2)} \big)_{ij} =
\big(\! \sqrt{{D}}^{(2)} \big)_{ji} \, .
\label{g_h_relation_inverse}
\end{equation}
The general solution can then be obtained by multiplying the matrix
$\big(\sqrt{{D}}^{(2)}
\big)_{ij}$ with arbitrary orthogonal matrices. However, the expression of
equation~\eqref{g_h_relation_inverse} is sufficient for our purposes.

It is not easy to numerically solve the Balescu-Lenard
equation~\eqref{eq:BL}. However, since we have established that the Balescu-Lenard
equation is equivalent to the Fokker-Planck equation, it may be more
convenient to solve numerically the stochastic Langevin equations for each
individual star
(characterised by its action $\bm{{J}}$) and make an ensemble average to
reconstruct the system's DF. In the theory of Brownian motion, this is the
so-called molecular dynamics method. The main idea is to simulate the Langevin
force on a computer, integrate the equations of motion with the simulated
Langevin force and then take the average for a large number of realisations~\citep{Risken1996}.
Of course, in the present context, the diffusion and friction
coefficients $\bm{D}^{(1)}$ and $\bm{D}^{(2)}$, and
therefore the Langevin coefficients $\bm{h}$ and $\bm{g}$, which describe the
self-induced noisy environment, must be updated
self-consistently as
the system's DF ${ F(\bm{J},t) }$ changes on secular timescales.
Even if we are led
back to a discrete ${N-}$particles system (recall that we started from a Hamiltonian system of $N$ stars), the gain of the stochastic approach is
to allow for a time discretisation of the particles' trajectories with a timestep ${ \Delta t }$ that is
orders of magnitude larger than the timestep required to solve the Hamiltonian
dynamics, since the complicated effects of collisions are encapsulated
in the stochastic force and in the drift.

In addition, this method may be more flexible for generalisation. For example, for non integrable systems,
equation~\eqref{Langevin_equation} could
be extended to account for chaotic stochasticity as 
\begin{equation}
\frac{\rd \bm{{J}}}{\rd t} = \bm{h} (\bm{{J}} , t) + \bm{g}
(\bm{{J}} , t) \!\cdot\! \bm{\Gamma} (t)+ \bm{g}_{\rc}
(\bm{{J}} , t) \!\cdot\! \bm{\Gamma}_{\rc} (t) \, ,
\label{Langevin_equation_chaos}
\end{equation}
where the stochastic Langevin force, ${ \bm{\Gamma}_{\rc} (t) }$, and its action
varying amplitude, $\bm{g}_{\rc}$, are set to match the orbital diffusion induced
by the chaotic sea within action space.

\section{The different stages in the evolution of stellar systems}
\label{sec:stages}

We are now in a position to describe accurately the different stages that occur
in the evolution of a stellar system. Fundamentally, a stellar system is a
Hamiltonian system of $N$ stars in gravitational interaction. If we are
interested in the evolution of the DF, ${ F \!=\! F(\bm{x}, \bm{v},t) }$, one can
identify different dynamical regimes, each of them characterised by a different
kinetic equation: 

(i) For sufficiently ``short'' times (that can be astronomical in practice!), the
evolution of the DF of a stellar system is governed by the 
Vlasov-Poisson equations~\citep{Jeans1915,Vlasov1938}. The Vlasov equation is a
mean
field equation which describes the ``collisionless'' evolution of the system.
Mathematically speaking, it is valid in the limit ${ N \!\rightarrow\! + \infty }$ with
${ \mu \!\sim\! 1/N }$.\footnote{For self-gravitating systems, there are
mathematical difficulties to
rigorously justify the Vlasov equation because of the $r^{-1}$ divergence of the
gravitational potential as ${ r \!\rightarrow\! 0 }$.}
A stellar system described by
the Vlasov equation that is initially in an unsteady state, or in a dynamically
unstable steady state, generically undergoes a process of violent
relaxation ~\citep{LyndenBell1967} and reaches a quasi-stationary
(virialised)
state on a coarse-grained scale. This process takes place in a few
dynamical times $t_{\rd}$. Violent relaxation is a complex process
associated with large potential fluctuations, phase mixing and nonlinear Landau damping~\citep{MouhotVillani2011}. 
The quasi-stationary state resulting from violent relaxation is difficult to
predict in general. However, it must be (close to) a stable steady state of the
Vlasov-Poisson equations, as found in numerical simulations. According to Jeans' theorem, the DF
of a stellar system trapped in a quasi-stationary state is generically a function
of the actions only, ${F \!=\! F(\bm{J}) }$. 

(ii) On a secular timescale ${ \!\sim\! N t_{\rd} }$, gravitational encounters between stars
(departures from the mean field dynamics, granularities, graininess, finite${-N}$ effects,...)
come into play and must be taken into account in the
dynamics. Because of gravitational encounters (via resonances), the system's DF slowly
changes by evolving through a succession of quasi-stationary states, ${ F \!=\! F(\bm{J},t) }$. This self-induced
``collisional'' evolution of the DF is described by the inhomogeneous
Balescu-Lenard or Fokker-Planck equation~\citep{Heyvaerts2010,Chavanis2012}, which
is a refinement of the homogeneous
Chandrasekhar~\citep{Chandrasekhar1942} and
Landau~\citep{Landau1936} equations, taking into account spatial inhomogeneity and collective effects. We note that
gravitational encounters between stars need not be local~\citep{LyndenBellKalnajs1972}
but can be distant, e.g. capturing the mechanism of resonant
relaxation~\citep{RauchTremaine1996}. An alternative description of the system's
dynamics can be written in
terms of ${N-}$body stochastic Langevin equations associated with the
inhomogeneous Fokker-Planck equation (see section~\ref{sec:langevin}).

(iii) It may happen that, during the collisional evolution, the system's DF, ${ F \!=\! F(\bm{J},t)}$,
becomes dynamically (Vlasov) unstable. In that case, one has to come back
to the Vlasov-Poisson equations to describe its evolution. This drives a dynamical phase
transition from this unstable state to a new stable state. This has been found in~\cite{Sellwood2012} and explicity
demonstrated in~\cite{FouvryPichonMagorrianChavanis2015} in the case of stellar
discs, showing the transition between a disc-phase (axisymmetric) and a bar-phase
(non-axisymmetric). A similar dynamical phase transition was evidenced
previously for a toy model of particles with long-range
interactions~\citep{CampaChavanis2008}.

(iv) It can be shown that the (inhomogeneous) Balescu-Lenard equation
conserves mass and energy and satisfies a ${H-}$theorem for the Boltzmann
entropy~\citep{Chavanis2007,Heyvaerts2010}, see Appendix~\ref{sec:prop}.
As a result, one expects that
the DF relaxes for ${ t \!\rightarrow\! + \infty }$
towards the Boltzmann DF which maximises the entropy at fixed mass and energy.
However, for self-gravitating systems, in most cases,
the Boltzmann entropy has no
maximum~\citep[e.g.][]{Padmanabhan1990}, so that there exists no
statistical equilibrium state in a strict sense. For example, the late time
evolution of globular clusters
proceeds through stellar evaporation~\citep{Spitzer1940}. According to the
virial theorem, the central density increases as the system expands. When the
system becomes sufficiently centrally condensed, an instability develops and
leads to core collapse. This instability, called the gravothermal
catastrophe~\citep{LyndenBellWood1968}, arises from the negative
specific heat of the inner part of the cluster. Core collapse can be
stopped by the formation of a binary star that can release an enormous amount of
energy able to reverse the collapse and drive a re-expansion of the whole
cluster until the next collapse takes place. This can lead to a series of 
gravothermal oscillations~\citep{BettwieserSugimoto1984}.

(v) Finally, even in the collisionless regime ${ N \!\rightarrow\! + \infty }$, the DF
may evolve under the effect of external perturbations, again passing through a
succession of quasi-stationary states, ${F \!=\! F (\bm{J} , t) }$. The
kinetic equation that governes this dynamics is the secular collisionless
diffusion equation introduced
in~\cite{BinneyLacey1988,Weinberg2001a,PichonAubert2006,FouvryPichonPrunet2015}
for inhomogeneous systems and 
in~\cite{Nardini2012,Chavanis2012epjp}
for homogeneous systems.
We note that, contrary to the Balescu-Lenard equation which has no free
parameter, this equation needs an input which is the power spectrum of the
external potential fluctuations. The effect of the external environment may or may not outrun that of the self-induced evolution:
this is the classical conundrum of ``nature'' and ``nurture'' driven secular evolutions.

\section{Conclusion}
\label{sec:conclusion}

The derivation of the coefficients 
of diffusion and dynamical friction in a stable, inhomogeneous,
multicomponent, self-gravitating system was presented.
The method followed here is based on the detailed study of the dynamics of a test particle, when perturbed by the dressed potential perturbations induced by a discrete bath of background particles.
It was shown in particular how the averaged
coefficients of diffusion and dynamical friction are fully consistent with those involved in the associated inhomogeneous
Balescu-Lenard equation.
As a result, the Balescu-Lenard equation
can be interpreted as a Fokker-Planck equation in which the diffusion and
friction coefficients evolve self-consistently (i.e. they depend on the DF
itself).

The present derivation has several advantages. First of all, it clarifies the
physical content of the Balescu-Lenard equation by showing its equivalence
with the traditional Fokker-Planck equation that was introduced initially
in the seminal work of~\cite{Chandrasekhar1943I} and that has
been adopted by
most astrophysicists. This approach confirms that the force acting on a star can
be decomposed into a smooth component due to the mean field of the whole system
and fluctuations due to finite${-N}$ effects (encounters). In turn, the
fluctuations have a completely random part that can be described by a
multiplicative Gaussian white noise and a systematic part
corresponding to the effect of
dynamical friction. This is in complete agreement with the physical picture
given by~\cite{Chandrasekhar1943I}. However, considerable progress
has been made in the
calculation of the coefficients of diffusion and dynamical friction with respect
to early approaches that focused on spatially homogeneous stellar systems
(making a local approximation) and
neglected collective effects (the dressing of a star by its gravitational wake).
We are now in a position to account for spatial inhomogeneity and collective
effects accurately. This results in
self-consistent expressions of the coefficients of diffusion and dynamical friction
(given by equations~\eqref{final_diff_drift1} and~\eqref{final_diff_drift2})
at order ${1/N}$, that encompass previous results obtained in the
literature.\footnote{Actually, there can remain a logarithmic
divergence at small
scales (in particular for ${3D}$ spherical systems) due to strong collisions. This
divergence can be solved by developing a treatment ``\`a la Boltzmann'' or by
taking into account the bending of trajectories like in the work of~\cite{Chandrasekhar1943I}.
Alternatively, one may introduce in
the diverging expression a cut-off at the Landau length which corresponds to the
impact parameter producing a deflection at $90^{\circ}$ of the particle's
trajectory. There is no such divergence in the case of stellar discs.}
We refer the reader to~\cite{Chavanis2013} for a thorough and detailed discussion of the links between self-consistent
kinetic equations such as the Balescu-Lenard and Landau equations, and other approaches,
such as the two-body encounters theory introduced
in~\cite{Chandrasekhar1943I}.
The present formalism also allows for a self-consistent description of a spectrum of masses,
with a proper accounting of the induced secular mass segregation, which should be of interest
to various astrophysical contexts, from galactic centers to protostellar discs. 
Another advantage of the present derivation is
practical. Instead of numerically solving the
Balescu-Lenard equation or the Fokker-Planck equation, it may be more
convenient to solve a system of $N$ Langevin equations describing the
stochastic trajectories of stars on an intermediate timescale with the coefficients of diffusion and
dynamical friction obtained from the Fokker-Planck approach. This procedure may
be useful in stellar dynamical simulations since it allows one to
use larger timesteps compared to the ones used in the original ${N-}$body
Hamiltonian equations, as the encounters between stars have been taken into
account in the kinetic parametrisation.
Yet, computing the diffusion flux of such kinetic equations remains a challenge. In order to deal with the system's inhomogeneity, one has to construct a set of angle-action coordinates ${ (\bm{\theta} , \bm{J}) }$. In order to characterise the self-gravitating amplification, one may rely on the matrix method to construct a biorthogonal basis of potential and density elements ${ (\psi^{(p)} , \rho^{(p)}) }$ and estimate the system's global response matrix ${ \widehat{\mathbf{M}} (\omega) }$. Finally, the secular evolution being driven by resonant encounters, one has to solve the non-local resonance condition, ${ \delta_{\rD} (\bm{m}_{1} \!\cdot\! \bm{\Omega}_{1} \!-\! \bm{m}_{2} \!\cdot\! \bm{\Omega}_{2}) }$, present in the Balescu-Lenard equation~\eqref{eq:BL}. See, e.g.,~\cite{FouvryPichonMagorrianChavanis2015} for an illustration of how these various difficulties may be solved in the context of razor-thin axisymmetric stellar discs.

Although the formalism and discussions are presented in the context of
self-gravitating systems, this approach is actually valid for arbitrary systems
with long-range interactions in any dimension of space. One just has
to introduce a proper biorthogonal basis of potentials and densities associated
with the specific interaction potential, as defined by equation~\eqref{definition_basis},
and introduce the set of angle-action variables associated with
the unperturbed Hamiltonian $H_{0}$. In addition, there are strong analogies
between two-dimensional point vortices and stellar
systems~\citep[see, e.g.][]{houches}.
In the
same manner that a test star in a star cluster has a diffusion motion in
velocity space due to the fluctuations of the gravitational force and
experiences a dynamical friction due to a polarisation process, a point vortex
evolving in a sea of field vortices has a diffusion motion in position space due to the
fluctuations of the velocity field and experiences a systematic drift~\citep{drift} due to a
polarisation process. The evolution of the probability density of
its position
is governed by a Fokker-Planck equation that can be written in the form of a
Balescu-Lenard equation~\citep{quasivortex} in complete parallel with the 
Fokker-Planck and Balescu-Lenard equations of stellar systems.
In the thermal bath approach, the friction and drift coefficients are
related to the diffusion coefficients by a form of Einstein relation expressing
the fluctuation-dissipation theorem.


\subsection*{Acknowledgements}
This paper was initially drafted by Jean Heyvaerts in 2010, independently of~\cite{Chavanis2012}, and completed posthumously by his co-authors to honour his memory. JBF thanks Scott Tremaine for insightful comments. Support for Program number HST-HF2-51374 was provided by NASA through a grant from the Space Telescope Science Institute, which is operated by the Association of Universities for Research in Astronomy, Incorporated, under NASA contract NAS5-26555. This research is part of ANR grant Spin(e) (ANR-13-BS05-0005, \url{http://cosmicorigin.org}).

\bibliography{references}

\appendix

\section{Computing the drift vector}
\label{sec:appendixdrift}

In this Appendix, we compute the mean drag ${ \big< \Delta \bm{J}_{1} \big> }$ acting on particle $1$ during the time ${ \Delta t }$. Using the shorthand notation
introduced in equation~\eqref{eq:shorthand}, equation~\eqref{eq:dyneq1} can be
rewritten as
\begin{align}
& \, \frac{\rd \bm{J}_{1}}{\rd t} = \sum_{2} \mu_{2} \sum_{\bm{m}_{1}} \sum_{\bm{m}_{2}}' \ri \bm{m}_{1} \Lambda_{\bm{m}_{1} , \bm{m}_{2}} (\bm{J}_{1} , \bm{J}_{2} , \omega_{2} ; t) \, \re^{ \ri z_{12} (t)} \, ,
\label{eq:dyneq3a}
\\
& \, \frac{\rd z_{12}}{\rd t} = \bm{m}_{1} \!\cdot \bm{\Omega}_{1} \!-\! \bm{m}_{2} \!\cdot\! \bm{\Omega}_{2} \!-\! \sum_{3} \mu_{3} \sum_{\bm{m}_{1}'} \sum_{\bm{m}_{3}'}' \re^{\ri z_{1' 3'} (t)} \bm{m}_{1} \!\cdot\! \partial_{\bm{J}_{1}} \big[ \Lambda_{\bm{m}_{1}' , \bm{m}_{3}'} (1 , 3 , \omega_{3}' ; t) \big] \nonumber
\\
& \;\;\;\;\;\;\;\;\;\;\;\; + \sum_{4} \mu_{4} \sum_{\bm{m}_{2}'} \sum_{\bm{m}_{4}'}' \re^{\ri z_{2' 4'} (t)} \bm{m}_{2} \!\cdot\! \partial_{\bm{J}_{2}} \big[ \Lambda_{\bm{m}_{2}' , \bm{m}_{4}'} (2 , 4 , \omega_{4} ' ; t) \big] \, ,
\label{eq:dyneq3b}
\\
& \, \frac{\rd \Lambda_{\bm{m}_{1} , \bm{m}_{2}} (1 , 2 , \omega_{2} ; t)}{\rd t} = \partial_{\bm{J}_{1}} \big[ \Lambda_{\bm{m}_{1} , \bm{m}_{2}} (1 , 2 , \omega_{2} ; t) \big] \!\cdot\! \frac{\rd \bm{J}_{1}}{\rd t} + \partial_{\bm{J}_{2}} \big[ \Lambda_{\bm{m}_{1} , \bm{m}_{2}} (1 , 2 , \omega_{2} ; t) \big] \!\cdot\! \frac{\rd \bm{J}_{2}}{\rd t} \, ,
\label{eq:dyneq3c}
\end{align}
where in the last equation, the gradient w.r.t. $\bm{J}_{2}$ also includes the dependency of ${ \omega_{2} \!=\! \bm{m}_{2} \!\cdot\! \bm{\Omega}_{2} (\bm{J}_{2}) }$. The solution to the previous system must be sought to second order in the noise, which is given by the number of $\Lambda$ factors. The change in ${ \Delta \bm{J}_{1} }$ during ${ \Delta t }$ is formally given through integration of equation~\eqref{eq:dyneq3a} between $0$ and ${ \Delta t }$, so that
\begin{equation}
\Delta \bm{J}_{1} = \sum_{2} \mu_{2} \sum_{\bm{m}_{1}} \sum_{\bm{m}_{2}}' \ri \bm{m}_{1} \!\! \int_{0}^{\Delta t} \!\!\!\!\!\! \rd t_{1} \, \Lambda_{\bm{m}_{1} , \bm{m}_{2}} (\bm{J}_{1} , \bm{J}_{2} , \bm{m}_{2} \!\cdot\! \bm{\Omega}_{2} ; t_{1}) \, \re^{ \ri z_{12} (t_{1})} \, ,
\label{eq:DeltaJ1}
\end{equation}
where one must note the time dependence of ${ \Lambda ( \, \cdot \, ; t) }$ which has to be accounted for at this order in the noise.

Progress can be made towards solution accurate to second order by computing ${ \Lambda_{\bm{m}_{1} , \bm{m}_{2}} (1 , 2 , \omega_{2} ; t) }$, after substituting the expressions for $\dot{\bm{J}}_{1}$ and $\dot{\bm{J}}_{2}$ given by equation~\eqref{eq:dyneq3a}. Then equation~\eqref{eq:dyneq3c} gives
\begin{align}
\frac{\rd \Lambda_{\bm{m}_{1} , \bm{m}_{2}} (1 , 2 , \omega_{2} ; t)}{\rd t} =
& \, \sum_{3} \mu_{3} \sum_{\bm{m}_{1}'} \sum_{\bm{m}_{3}'}' \ri \, \re^{\ri
z_{1'3'} (t)} \Lambda_{\bm{m}_{1}' , \bm{m}_{3}'} (1 , 3 , \omega_{3}' ; t) \,
\bm{m}_{1}' \!\cdot\! \partial_{\bm{J}_{1}} \big[ \Lambda_{\bm{m}_{1} ,
\bm{m}_{2}} (1 , 2 , \omega_{2} ; t) \big] \nonumber
\\
& \, + \sum_{4} \mu_{4} \sum_{\bm{m}_{2}'} \sum_{\bm{m}_{4}'}' \ri \, \re^{\ri z_{2'4'} (t)} \Lambda_{\bm{m}_{2}' , \bm{m}_{4}'} (2 , 4 , \omega_{4}' ; t) \, \bm{m}_{2}' \!\cdot\! \partial_{\bm{J}_{2}} \big[ \Lambda_{\bm{m}_{1} , \bm{m}_{2}} (1 , 2 , \omega_{2} ; t) \big] \, .
\label{expression_dLambda_dt}
\end{align}
Via time integration, we may then obtain an expression for ${ \Lambda_{\bm{m}_{1} , \bm{m}_{2}} (1 , 2 , \omega_{2} ; t ) }$ which is explicitly second order (recalling that $\Lambda$ is already first order). At this order, all the involved elements can be evaluated to zeroth order, so that in the r.h.s. of equation~\eqref{expression_dLambda_dt}, all the occurences of $\Lambda$ can be evaluated for ${ t \!=\! 0 }$. In the upcoming calculations, this is no more explicitly written to simplify the notations. The time integration of equation~\eqref{expression_dLambda_dt} gives
\begin{align}
\Lambda_{\bm{m}_{1} , \bm{m}_{2}} (1 , 2 , \omega_{2} ; t_{1}) = & \, \Lambda_{\bm{m}_{1} , \bm{m}_{2}} (1 , 2 , \omega_{2}) \nonumber
\\
& \, + \ri \!\! \int_{0}^{t_{1}} \!\!\!\!\!\! \rd t_{2} \, \sum_{3} \mu_{3} \sum_{\bm{m}_{1}'} \sum_{\bm{m}_{3}'}' \, \re^{\ri z_{1'3'} (t_{2})} \Lambda_{\bm{m}_{1}' , \bm{m}_{3}'} (1 , 3 , \omega_{3}') \, \bm{m}_{1}' \!\cdot\! \partial_{\bm{J}_{1}} \big[ \Lambda_{\bm{m}_{1} , \bm{m}_{2}} (1 , 2 , \omega_{2}) \big] \nonumber
\\
& \, + \ri \!\! \int_{0}^{t_{1}} \!\!\!\!\!\! \rd t_{2} \, \sum_{4} \mu_{4} \sum_{\bm{m}_{2}'} \sum_{\bm{m}_{4}'}' \, \re^{\ri z_{2'4'} (t_{2})} \Lambda_{\bm{m}_{2}' , \bm{m}_{4}'} (2 , 4 , \omega_{4}') \, \bm{m}_{2}' \!\cdot\! \partial_{\bm{J}_{2}} \big[ \Lambda_{\bm{m}_{1} , \bm{m}_{2}} (1 , 2 , \omega_{2}) \big] \, ,
\label{eq:dLambda}
\end{align}
where we insist on the fact that the occurences of ${ \Lambda }$ in the r.h.s.
of equation~\eqref{eq:dLambda} are evaluated at zeroth order, i.e. for ${ t
\!=\! 0 }$. Let us now obtain ${ z_{12} (t_{1}) }$, which enters
equation~\eqref{eq:DeltaJ1}, thanks to the time integration of
equation~\eqref{eq:dyneq3b}. One gets
\begin{align}
z_{12} (t_{1}) = & \, z_{12} (0) + \!\! \int_{0}^{t_{1}} \!\!\!\!\!\! \rd t_{2} \, (\bm{m}_{1} \!\cdot\! \bm{\Omega}_{1} (t_{2}) \!-\! \bm{m}_{2} \!\cdot\! \bm{\Omega}_{2} (t_{2}))
\label{z12_time}
\\
& \, - \!\! \int_{0}^{t_{1}} \!\!\!\!\!\! \rd t_{2} \, \sum_{3} \mu_{3} \sum_{\bm{m}_{1}'} \sum_{\bm{m}_{3}'}' \re^{\ri z_{1'3'} (t_{2})} \bm{m}_{1} \!\cdot\! \partial_{\bm{J}_{1}} \big[ \Lambda_{\bm{m}_{1}' , \bm{m}_{3}'} (1 , 3 , \omega_{3}') \big] + \!\! \int_{0}^{t_{1}} \!\!\!\!\!\! \rd t_{2} \, \sum_{4} \mu_{4} \sum_{\bm{m}_{2}'} \sum_{\bm{m}_{4}'}' \re^{\ri
z_{2'4'} (t_{2})} \bm{m}_{2} \!\cdot\! \partial_{\bm{J}_{2}} \big[ \Lambda_{\bm{m}_{2}' , \bm{m}_{4}'} (2 , 4 , \omega_{4}') \big] \, , \nonumber
\end{align}
where ${ \Lambda }$ is evaluated at zeroth-order, i.e. for ${ t \!=\! 0 }$. The frequencies ${ \bm{\Omega}_{1} (t_{2}) }$ and ${ \bm{\Omega}_{2} (t_{2}) }$
follow by time integration of equation~\eqref{eq:dyneq1} and read
\begin{equation}
\bm{\Omega}_{1} (t_{2}) = \bm{\Omega}_{1} (0) + \!\! \int_{0}^{t_{2}} \!\!\!\!\!\! \rd t_{3} \, \big[ \partial_{\bm{J}_{1}} \!\otimes\! \bm{\Omega}_{1} \big] \!\cdot\! \dot{\bm{J}}_{1} (t_{3}) \, .
\end{equation}
Then, the relative angular velocity $g_{12}$ at time $t_{2}$ reads
\begin{equation}
g_{12} (t_{2}) \equiv \bm{m}_{1} \!\cdot\! \bm{\Omega}_{1} (t_{2}) \!-\!
\bm{m}_{2} \!\cdot\! \bm{\Omega}_{2} (t_{2}) = g_{12} (0) + \!\!
\int_{0}^{t_{2}} \!\!\!\!\!\! \rd t_{3} \, \bigg[ \big[ (\bm{m}_{1} \!\cdot\!
\partial_{\bm{J}_{1}}) \, \bm{\Omega}_{1} (t_{3}) \big] \!\cdot\!
\dot{\bm{J}}_{1} (t_{3}) \!-\! \big[ (\bm{m}_{2} \!\cdot\!
\partial_{\bm{J}_{2}}) \, \bm{\Omega}_{2} (t_{3}) \big] \!\cdot\!
\dot{\bm{J}}_{2} (t_{3})\bigg] \, .
\end{equation}
Equation~\eqref{z12_time} finally becomes
\begin{align}
z_{12} (t_{1}) = & \, z_{12} (0) + g_{12} (0) \, t_{1} \nonumber
\\
& \, + \!\! \int_{0}^{t_{1}} \!\!\!\!\! \rd t_{2} \!\! \int_{0}^{t_{2}} \!\!\!\!\!\! \rd t_{3} \, \bigg[ \big[ (\bm{m}_{1} \!\cdot\! \partial_{\bm{J}_{1}}) \, \bm{\Omega}_{1} (t_{3}) \big] \!\cdot\! \dot{\bm{J}}_{1} (t_{3}) - \big[ (\bm{m}_{2} \!\cdot\! \partial_{\bm{J}_{2}}) \, \bm{\Omega}_{2} (t_{3}) \big] \!\cdot\! \dot{\bm{J}}_{2} (t_{3}) \bigg]
\label{eq:complicated}
\\
& \, - \!\! \int_{0}^{t_{1}} \!\!\!\!\!\! \rd t_{2} \, \sum_{3} \mu_{3} \sum_{\bm{m}_{1}'} \sum_{\bm{m}_{3}'}' \re^{\ri z_{1'3'} (t_{2})} \bm{m}_{1} \!\cdot\! \partial_{\bm{J}_{1}} \big[ \Lambda_{\bm{m}_{1}' , \bm{m}_{3}'} (1 , 3 , \omega_{3}') \big] + \!\! \int_{0}^{t_{1}} \!\!\!\!\!\! \rd t_{2} \,\sum_{4} \mu_{4} \sum_{\bm{m}_{2}'} \sum_{\bm{m}_{4}'}' \re^{\ri z_{2'4'} (t_{2})} \bm{m}_{2} \!\cdot\! \partial_{\bm{J}_{2}} \big[ \Lambda_{\bm{m}_{2}' , \bm{m}_{4}'} (2 , 4 , \omega_{4}') \big] \, . \nonumber
\end{align}
The lines 2 and 3 of equation~\eqref{eq:complicated} correspond to the
first order correction relative to the zeroth expression on the first line.
Hence the complex exponential ${ \re^{\ri z_{12} (t_{1})} }$ can be expanded as
\begin{align}
& \, \re^{\ri z_{12} (t_{1})} = \re^{\ri ( z_{12} (0) + g_{12} (0) t_{1})} \bigg\{ 1 + \ri \!\! \int_{0}^{t_{1}} \!\!\!\!\!\! \rd t_{2} \!\! \int_{0}^{t_{2}} \!\!\!\!\!\! \rd t_{3} \, \bigg[ \big[ ( \bm{m}_{1} \!\cdot\! \partial_{\bm{J}_{1}}) \, \bm{\Omega}_{1} (t_{3}) \big] \!\cdot\! \dot{\bm{J}}_{1} (t_{3}) - \big[ (\bm{m}_{2} \!\cdot\! \partial_{\bm{J}_{2}}) \, \bm{\Omega}_{2} (t_{3}) \big] \!\cdot\! \dot{\bm{J}}_{2} (t_{3}) \bigg] \nonumber
\\
& \, - \ri \!\! \int_{0}^{t_{1}} \!\!\!\!\!\! \rd t_{2} \, \! \sum_{3} \mu_{3} \sum_{\bm{m}_{1}'} \sum_{\bm{m}_{3}'}' \re^{\ri z_{1'3'} (t_{2})} \bm{m}_{1} \!\cdot\! \partial_{\bm{J}_{1}} \big[ \Lambda_{\bm{m}_{1}' , \bm{m}_{3}'} (1 , 3 , \omega_{3}') \big] + \ri \!\! \int_{0}^{t_{1}} \!\!\!\!\!\! \rd t_{2} \, \! \sum_{4} \mu_{4} \sum_{\bm{m}_{2}'} \sum_{\bm{m}_{4}'}' \re^{\ri z_{2'4'} (t_{2})} \bm{m}_{2} \!\cdot\! \partial_{\bm{J}_{2}} \big[ \Lambda_{\bm{m}_{2}' , \bm{m}_{4}'} (2 , 4 , \omega_{4}') \big] \bigg\} \, ,
\label{eq:expcomplex}
\end{align}
where we relied on the usual development ${ \re^{\ri \varepsilon} \!\simeq\! 1
\!+\! \ri \varepsilon }$ at first order in $\varepsilon$. In
equation~\eqref{eq:expcomplex}, the expressions of ${ \dot{\bm{J}}_{1} (t_{3})
}$ and ${ \dot{\bm{J}}_{2} (t_{3}) }$ are given by Hamilton's
equation~\eqref{eq:dyneq3a}. The expressions~\eqref{eq:dLambda} and~\eqref{eq:expcomplex}
are then replaced in equation~\eqref{eq:DeltaJ1}. It
yields the expression of ${ \Delta \bm{J}_{1} / \Delta t }$ which now reaches
the required second order level, so that
\begin{align}
\frac{\Delta \bm{J}_{1} }{\Delta t} = & \, \sum_{2} \mu_{2} \sum_{\bm{m}_{1}} \sum_{\bm{m}_{2}}' \ri \bm{m}_{1} \!\! \int_{0}^{\Delta t} \!\! \frac{\rd t_{1}}{\Delta t} \, \re^{\ri (z_{12} (0) + g_{12} (0) t_{1})} \nonumber
\\
& \, \times \bigg[ \Lambda_{\bm{m}_{1} , \bm{m}_{2}} (1 , 2 , \omega_{2}) + \ri \!\! \int_{0}^{t_{1}} \!\!\!\!\!\! \rd t_{2} \, \sum_{3} \mu_{3} \sum_{\bm{m}_{1}'} \sum_{\bm{m}_{3}'}' \re^{\ri z_{1'3'} (t_{2})} \Lambda_{\bm{m}_{1}' , \bm{m}_{3}'} (1 , 3 , \omega_{3}') \, \bm{m}_{1}' \!\cdot\! \partial_{\bm{J}_{1}} \big[ \Lambda_{\bm{m}_{1} , \bm{m}_{2}} (1 , 2 , \omega_{2}) \big] \nonumber 
\\
& \, + \ri \!\! \int_{0}^{t_{1}} \!\!\!\!\!\! \rd t_{2} \, \sum_{4} \mu_{4} \sum_{\bm{m}_{2}'} \sum_{\bm{m}_{4}'}' \re^{\ri z_{2'4'} (t_{2})} \Lambda_{\bm{m}_{2}' , \bm{m}_{4}'} (2 , 4 , \omega_{4}') \, \bm{m}_{2}' \!\cdot\! \partial_{\bm{J}_{2}} \big[ \Lambda_{\bm{m}_{1} , \bm{m}_{2}} (1 , 2 , \omega_{2}) \big] \bigg] \nonumber
\\
& \, \times \bigg[ 1 + \ri \sum_{3} \mu_{3} \sum_{\bm{m}_{1}'} \sum_{\bm{m}_{3}'}' \!\! \int_{0}^{t_{1}} \!\!\!\!\!\! \rd t_{2} \!\! \int_{0}^{t_{2}} \!\!\!\!\!\! \rd t_{3} \, \ri \, \bm{m}_{1} \!\cdot\! \partial_{\bm{J}_{1}} \big[ \bm{m}_{1}' \!\cdot\! \bm{\Omega}_{1} \big] \, \re^{\ri z_{1'3'} (t_{3})} \, \Lambda_{\bm{m}_{1}' , \bm{m}_{3}'} (1 , 3 , \omega_{3}') \nonumber
\\
& \, - \ri \sum_{4} \mu_{4} \sum_{\bm{m}_{2}'} \sum_{\bm{m}_{4}'}' \!\! \int_{0}^{t_{1}} \!\!\!\!\!\! \rd t_{2} \!\! \int_{0}^{t_{2}} \!\!\!\!\!\! \rd t_{3} \, \ri \, \bm{m}_{2} \!\cdot\! \partial_{\bm{J}_{2}} \big[ \bm{m}_{2}' \!\cdot\! \bm{\Omega}_{2} \big] \, \re^{\ri z_{2'4'} (t_{3})} \, \Lambda_{\bm{m}_{2}' , \bm{m}_{4}'} (2 , 4 , \omega_{4}') \label{eq:verycomplicated}
\\
& \, - \ri \!\! \int_{0}^{t_{1}} \!\!\!\!\!\! \rd t_{2} \, \sum_{3} \mu_{3} \sum_{\bm{m}_{1}'} \sum_{\bm{m}_{3}'}' \re^{\ri z_{1'3'} (t_{2})} \, \bm{m}_{1} \!\cdot\! \partial_{\bm{J}_{1}} \big[ \Lambda_{\bm{m}_{1}' , \bm{m}_{3}'} (1 , 3 , \omega_{3}') \big] + \ri \!\! \int_{0}^{t_{1}} \!\!\!\!\!\! \rd t_{2} \, \sum_{4} \mu_{4} \sum_{\bm{m}_{2}'} \sum_{\bm{m}_{4}'}' \re^{\ri z_{2'4'} (t_{2})} \, \bm{m}_{2} \!\cdot\! \partial_{\bm{J}_{2}} \big[ \Lambda_{\bm{m}_{2}' , \bm{m}_{4}'} (2 , 4 , \omega_{4}') \big] \bigg] \, . \nonumber
\end{align}
Let us first insist on the fact that equation~\eqref{eq:verycomplicated} is explicitly second order in the noise. Indeed, the only linear term in equation~\eqref{eq:verycomplicated}, arising from ${ \re^{\ri z_{12} (0)} \Lambda_{\bm{m}_{1} , \bm{m}_{2}} (1 , 2 , \omega_{2}) }$ will vanish when averaged over the initial angle $\bm{\theta}_{2}^{0}$. We also recall that at the present level of approximation, the complex exponentials $\re^{\ri z (t)}$ should be evaluated to zeroth order, that corresponding to the uniform angular motion at fixed frequency. For example, in equation~\eqref{eq:verycomplicated}, one should read ${ \re^{\ri z_{1' 3'} (t_{2})} }$ as
\begin{equation}
\re^{\ri z_{1' 3'} (t_{2})} = \re^{\ri (\bm{m}_{1}' \cdot \bm{\theta}_{1}^{0} - \bm{m}_{3}' \cdot \bm{\theta}_{3}^{0})} \re^{\ri t_{2} (\bm{m}_{1}' \cdot \bm{\Omega}_{1} (0) - \bm{m}_{3}' \cdot \bm{\Omega}_{3} (0))} \, .
\label{example_zero_order}
\end{equation}
Similarly, at the order considered here, the susceptibility coefficients ${ \Lambda (... ; t) }$ should be evaluated at ${ t \!=\! 0 }$.

One may then follow the same method as the one presented in equation~\eqref{eq:fricdyncoef1} to obtain the averaged action diffusion tensor. Indeed, one can average equation~\eqref{eq:verycomplicated} over the initial angles of the particles $1$, $2$, $3$, and $4$, as well as on the action distribution of particles $2$, $3$, and $4$. Here, one should pay attention to the fact that particle $1$ acts as our test star, while particles $2$ and $3$ both run over the field stars associated with particle $1$, i.e. over all stars except particle $1$. The situation is slightly different for particle $4$ which runs over the field stars associated with particle $2$, i.e. over all stars except particle $2$. Let us first perform an average over the initial angles of all stars. Considering only second order terms (i.e. involving two factors $\Lambda$) and keeping only the dependencies w.r.t. the initial angles, equation~\eqref{eq:verycomplicated} requires to study two different generic terms
\begin{equation}
\sum_{2,3} \!\! \int \!\! \rd \bm{\theta}_{1}^{0} \rd \bm{\theta}_{2}^{0} \rd \bm{\theta}_{3}^{0} \, \re^{\ri (\bm{m}_{1} \cdot \bm{\theta}_{1}^{0} - \bm{m}_{2} \cdot \bm{\theta}_{2}^{0} + \bm{m}_{1}' \cdot \bm{\theta}_{1}^{0} - \bm{m}_{3}' \cdot \bm{\theta}_{3}^{0})} \;\;\; ; \;\;\; \sum_{2,4} \!\! \int \!\! \rd \bm{\theta}_{1}^{0} \rd \bm{\theta}_{2}^{0} \rd \bm{\theta}_{4}^{0} \, \re^{\ri (\bm{m}_{1} \cdot \bm{\theta}_{1}^{0} - \bm{m}_{2} \cdot \bm{\theta}_{2}^{0} + \bm{m}_{2}' \cdot \bm{\theta}_{2}^{0} - \bm{m}_{4}' \cdot \bm{\theta}_{4}^{0})} \, ,
\label{shape_angle_dependence_drift}
\end{equation}
where it is important to note that the sums on particles $2$ and $3$ are
restricted to all stars 
except particle $1$, while the sum on particle $4$ is restricted to all stars
except particle $2$. Because only non-zero values of $\bm{m}_{2}$, ${
\bm{m}_{3}' }$, and ${ \bm{m}_{4}' }$ contribute to the fluctuations,
equation~\eqref{shape_angle_dependence_drift} therefore immediately imposes
for particles $3$ and $2$ to be the same and for particles $4$ and
$1$ to be the same. As a consequence, the sums $\sum_{3}$ and $\sum_{4}$ can be
straightforwarldy executed. We may then average
equation~\eqref{eq:verycomplicated} over $\bm{\theta}_{1}^{0}$,
$\bm{\theta}_{2}^{0}$, and over the action distribution of particle $2$,
following the same substitution as in equation~\eqref{replacement_2}. All in
all, equation~\eqref{eq:verycomplicated}, when averaged and restricted to second
order terms, becomes
\begin{align}
\bigg< \frac{\Delta \bm{J}_{1}}{\Delta t} \bigg> = & \, \sum_{\rb} \!\!\sum_{\bm{m}_{1} , \bm{m}_{1}'} \! \sum_{\bm{m}_{2} , \bm{m}_{2}'}' \!\! \int \!\! \frac{\rd \bm{\theta}_{1}^{0}}{(2 \pi)^{d}} \!\! \int \!\! \rd \bm{\theta}_{2}^{0} \rd \bm{J}_{2} \, F^{\rb} (\bm{J}_{2}) \, \ri \bm{m}_{1} \!\! \int_{0}^{\Delta t} \!\! \frac{\rd t_{1}}{\Delta t} \nonumber
\\
& \, \times \bigg[ \ri \!\! \int_{0}^{t_{1}} \!\!\!\!\!\! \rd t_{2} \, \re^{ \ri (z_{12} (t_{1}) + z_{1'2'} (t_{2}))} \Lambda_{\bm{m}_{1}' , \bm{m}_{2}'} (1 , 2 , \omega_{2}') \, \mu_{\rb} \, \bm{m}_{1}' \!\cdot\! \partial_{\bm{J}_{1}} \big[ \Lambda_{\bm{m}_{1} , \bm{m}_{2}} (1 , 2 , \omega_{2}) \big] \nonumber
\\
& \, + \ri \!\! \int_{0}^{t_{1}} \!\!\!\!\!\! \rd t_{2} \, \re^{\ri (z_{12} (t_{1}) + z_{2'1'} (t_{2}))} \Lambda_{\bm{m}_{2}' , \bm{m}_{1}'} (2 , 1 , \omega_{1}') \, \mu_{\ra} \, \bm{m}_{2}' \!\cdot\! \partial_{\bm{J}_{2}} \big[ \Lambda_{\bm{m}_{1} , \bm{m}_{2}} (1 , 2 , \omega_{2}) \big] \nonumber
\\
& \, - \!\! \int_{0}^{t_{1}} \!\!\!\!\!\! \rd t_{2} \!\! \int_{0}^{t_{2}} \!\!\!\!\!\! \rd t_{3} \, \re^{ \ri (z_{12} (t_{1}) + z_{1'2'} (t_{3}))} \Lambda_{\bm{m}_{1} , \bm{m}_{2}} (1 , 2 , \omega_{2}) \, \mu_{\rb} \, \bm{m}_{1} \!\cdot\! \partial_{\bm{J}_{1}} \big[ \bm{m}_{1}' \!\cdot\! \bm{\Omega}_{1} \big] \, \Lambda_{\bm{m}_{1}' , \bm{m}_{2}'} (1 , 2 , \omega_{2}') \nonumber
\\
& \, + \!\! \int_{0}^{t_{1}} \!\!\!\!\!\! \rd t_{2} \!\! \int_{0}^{t_{2}} \!\!\!\!\!\! \rd t_{3} \, \re^{\ri (z_{12} (t_{1}) + z_{2'1'}(t_{3}))} \Lambda_{\bm{m}_{1} , \bm{m}_{2}} (1 , 2 , \omega_{2}) \, \mu_{\ra} \, \bm{m}_{2} \!\cdot\! \partial_{\bm{J}_{2}} \big[ \bm{m}_{2}' \!\cdot\! \bm{\Omega}_{2} \big] \, \Lambda_{\bm{m}_{2}' , \bm{m}_{1}'} (2 , 1 , \omega'_{1}) \nonumber
\\
& \, - \ri \!\! \int_{0}^{t_{1}} \!\!\!\!\!\! \rd t_{2} \, \re^{\ri (z_{12} (t_{1}) + z_{1'2'} (t_{2}))} \Lambda_{\bm{m}_{1} , \bm{m}_{2}} (1 , 2 , \omega_{2}) \, \mu_{\rb} \, \bm{m}_{1} \!\cdot\! \partial_{\bm{J}_{1}} \big[ \Lambda_{\bm{m}_{1}' , \bm{m}_{2}'} (1 , 2 , \omega_{2}') \big]\nonumber
\\
& \, + \ri \!\! \int_{0}^{t_{1}} \!\!\!\!\!\! \rd t_{2} \, \re^{\ri (z_{12} (t_{1}) + z_{2'1'} (t_{2}))} \Lambda_{\bm{m}_{1} , \bm{m}_{2}} (1 , 2 , \omega_{2}) \, \mu_{\ra} \, \bm{m}_{2} \!\cdot\! \partial_{\bm{J}_{2}} \big[ \Lambda_{\bm{m}_{2}' , \bm{m}_{1}'} (2 , 1 , \omega_{1}') \big] \bigg] \, ,
\label{calculation_drift}
\end{align}
where one should pay attention to the different mass prefactors ${ \mu_{\ra}
\!=\! \mu_{1} }$ and $\mu_{\rb}$, whose role is essential to induce mass
segregation in multicomponent systems. In equation~\eqref{calculation_drift},
we also performed the change of notations ${ \bm{m}_{3}' \!\to\! \bm{m}_{2}' }$
and ${ \bm{m}_{4}' \!\to\! \bm{m}_{1}' }$. The averaging process over the
initial angles $\bm{\theta}_{1}^{0}$ and $\bm{\theta}_{2}^{0}$ yields
\begin{equation}
\!\! \int \!\! \frac{\rd \bm{\theta}_{1}^{0}}{(2 \pi)^{d}} \rd \bm{\theta}_{2}^{0} \, \re^{\ri (z_{12} (t_{1}) + z_{1'2'} (t_{2}))} \!=\! (2 \pi)^{d} \, \delta_{\bm{m}_{1}}^{-\bm{m}_{1}'} \, \delta_{\bm{m}_{2}}^{- \bm{m}_{2}'} \, \re^{\ri g_{12} (t_{1} - t_{2})} \; ; \; \!\! \int \!\! \frac{\rd \bm{\theta}_{1}^{0}}{(2 \pi)^{d}} \bm{\theta}_{2}^{0} \, \re^{\ri (z_{12} (t_{1}) + z_{2'1'} (t_{2}))} \!=\! (2 \pi)^{d} \, \delta_{\bm{m}_{1}}^{\bm{m}_{1}'} \, \delta_{\bm{m}_{2}}^{\bm{m}_{2}'} \, \re^{\ri g_{12} (t_{1} - t_{2})} \, ,
\end{equation}
where we recall that ${ g_{12} \!\equiv\! \bm{m}_{1} \!\cdot\! \bm{\Omega}_{1} \!-\! \bm{m}_{2} \!\cdot\! \bm{\Omega}_{2} }$, and is evaluated at zeroth order. Equation~\eqref{calculation_drift} then becomes
\begin{align}
\bigg< \frac{\Delta \bm{J}_{1}}{\Delta t} \bigg> = & \, \sum_{\rb} \sum_{\bm{m}_{1}} \sum_{\bm{m}_{2}}' \!\! \int \!\! \rd \bm{J}_{2} \, F^{\rb} (\bm{J}_{2}) (2 \pi)^{d} \, \bm{m}_{1} \!\! \int_{0}^{\Delta t} \! \frac{\rd t_{1}}{\Delta t} \nonumber
\\
& \, \times \bigg[ \! \int_{0}^{t_{1}} \!\!\!\!\!\! \rd t_{2} \, \re^{\ri g_{12} (t_{1} - t_{2})} \Lambda_{- \bm{m}_{1} , - \bm{m}_{2}} (1 , 2 , - \omega_{2}) \, \mu_{\rb} \, \bm{m}_{1} \!\cdot\! \partial_{\bm{J}_{1}} \big[ \Lambda_{\bm{m}_{1} , \bm{m}_{2}} (1 , 2 , \omega_{2}) \big] \nonumber
\\
& \, - \!\! \int_{0}^{t_{1}} \!\!\!\!\!\! \rd t_{2} \, \re^{\ri g_{12} (t_{1} - t_{2})} \Lambda_{\bm{m}_{2} , \bm{m}_{1}} (2 , 1 , \omega_{1}) \, \mu_{\ra} \, \bm{m}_{2} \!\cdot\! \partial_{\bm{J}_{2}} \big[ \Lambda_{\bm{m}_{1} , \bm{m}_{2}} (1 , 2 , \omega_{2}) \big] \nonumber
\\
& \, + \ri \!\! \int_{0}^{t_{1}} \!\!\!\!\!\! \rd t_{2} \!\! \int_{0}^{t_{2}} \!\!\!\!\!\! \rd t_{3} \, \re^{\ri g_{12} (t_{1} - t_{3})} \Lambda_{\bm{m}_{1} , \bm{m}_{2}} (1 , 2 , \omega_{2}) \, \mu_{\rb} \, \bm{m}_{1} \!\cdot\! \partial_{\bm{J}_{1}} \big[ \bm{m}_{1} \!\cdot\! \bm{\Omega}_{1} \big] \, \Lambda_{ - \bm{m}_{1} , - \bm{m}_{2}} (1 , 2 , - \omega_{2}) \nonumber
\\
& \, + \ri \!\! \int_{0}^{t_{1}} \!\!\!\!\!\! \rd t_{2} \!\! \int_{0}^{t_{2}} \!\!\!\!\!\! \rd t_{3} \, \re^{ \ri g_{12} (t_{1} - t_{3})} \Lambda_{\bm{m}_{1} , \bm{m}_{2}} (1 , 2 , \omega_{2}) \, \mu_{\ra} \, \bm{m}_{2} \!\cdot\! \partial_{\bm{J}_{2}} \big[ \bm{m}_{2} \!\cdot\! \bm{\Omega}_{2} \big] \, \Lambda_{\bm{m}_{2} , \bm{m}_{1}} (2 , 1 , \omega_{1}) \nonumber
\\
& \, + \!\! \int_{0}^{t_{1}} \!\!\!\!\!\! \rd t_{2} \, \re^{\ri g_{12} (t_{1} - t_{2})} \Lambda_{\bm{m}_{1} , \bm{m}_{2}} (1 , 2 , \omega_{2}) \, \mu_{\rb} \, \bm{m}_{1} \!\cdot\! \partial_{\bm{J}_{1}} \big[ \Lambda_{ - \bm{m}_{1} , - \bm{m}_{2}} (1 , 2 , - \omega_{2}) \big] \nonumber
\\
& \, - \!\! \int_{0}^{t_{1}} \!\!\!\!\!\! \rd t_{2} \, \re^{\ri g_{12} (t_{1} - t_{2})} \Lambda_{\bm{m}_{1} , \bm{m}_{2}} (1 , 2 , \omega_{2}) \, \mu_{\ra} \, \bm{m}_{2} \!\cdot\! \partial_{\bm{J}_{2}} \big[ \Lambda_{\bm{m}_{2} , \bm{m}_{1}} (2 , 1 , \omega_{1}) \big] \bigg] \, .
\label{eq:dj1dtaverage}
\end{align}
The various double and triple time integrals of complex exponentials
occuring in equation~\eqref{eq:dj1dtaverage} can be replaced by distributions
in the limit ${ \Delta t \!\to\! + \infty }$. They are computed in
Appendix~\ref{sec:appendixformula}, and one has
\begin{equation}
\lim_{\Delta t \to + \infty} \! \int_{0}^{\Delta t} \!\! \frac{\rd t_{1}}{\Delta
t} \!\! \int_{0}^{t_{1}} \!\!\!\!\!\! \rd t_{2} \, \re^{ \ri x (t_{1} - t_{2})}
= \pi \delta_{\rD} (x) \;\;\; ; \;\;\; \lim_{\Delta t \to + \infty} \! \int_{0}^{\Delta t} \!\! \frac{\rd t_{1}}{\Delta
t} \!\! \int_{0}^{t_{1}} \!\!\!\!\!\! \rd t_{2} \!\! \int_{0}^{t_{2}}
\!\!\!\!\!\! \rd t_{3} \, \re^{ \ri x (t_{1} - t_{3})} = - \ri \pi
\frac{\rd}{\rd x} \big( \delta_{\rD} (x) \big) \, .
\end{equation}
These relations allow us to rewrite equation~\eqref{eq:dj1dtaverage} as
\begin{align}
\bigg< \frac{\Delta \bm{J}_{1} }{\Delta t} \bigg> = & \, \sum_{\rb} \sum_{\bm{m}_{1}} \sum_{\bm{m}_{2}}' \!\! \int \!\! \rd \bm{J}_{2} \, F^{\rb} (\bm{J}_{2}) \, \pi (2 \pi)^{d} \, \bm{m}_{1} \nonumber
\\
& \, \times \bigg[ \, \delta_{\rD} (g_{12}) \, \Lambda_{ - \bm{m}_{1} , - \bm{m}_{2}} (1 , 2 , - \omega_{2}) \, \mu_{\rb} \, \bm{m}_{1} \!\cdot\! \partial_{\bm{J}_{1}} \big[ \Lambda_{\bm{m}_{1} , \bm{m}_{2}} (1 , 2 , \omega_{2}) \big] 
\nonumber
\\
& \,
 - \delta_{\rD} (g_{12}) \, \Lambda_{\bm{m}_{2} , \bm{m}_{1}} (2 , 1 , \omega_{1}) \, \mu_{\ra} \, \bm{m}_{2} \!\cdot\! \partial_{\bm{J}_{2}} \big[ \Lambda_{\bm{m}_{1} , \bm{m}_{2}} (1 , 2 , \omega_{2}) \big] \nonumber
\\
& \, + \frac{\rd \delta_{\rD} (g_{12})}{\rd g_{12}} \, \Lambda_{\bm{m}_{1} , \bm{m}_{2}} (1 , 2 , \omega_{2}) \, \mu_{\rb} \, \bm{m}_{1} \!\cdot\! \partial_{\bm{J}_{1}} \big[ \bm{m}_{1} \!\cdot\! \bm{\Omega}_{1} \big] \, \Lambda_{ - \bm{m}_{1} , - \bm{m}_{2}} (1 , 2 , - \omega_{2}) \nonumber
\\
& \, + \frac{\rd \delta_{\rD} (g_{12})}{\rd g_{12}} \, \Lambda_{\bm{m}_{1} , \bm{m}_{2}} (1 , 2 , \omega_{2}) \, \mu_{\ra} \, \bm{m}_{2} \!\cdot\! \partial_{\bm{J}_{2}} \big[ \bm{m}_{2} \!\cdot\! \bm{\Omega}_{2} \big] \, \Lambda_{\bm{m}_{2} , \bm{m}_{1}} (2 , 1 , \omega_{1}) \nonumber
\\
& \, + \delta_{\rD} (g_{12}) \, \Lambda_{\bm{m}_{1} , \bm{m}_{2}} (1 , 2 , \omega_{2}) \, \mu_{\rb} \, \bm{m}_{1} \!\cdot\! \partial_{\bm{J}_{1}} \big[ \Lambda_{ - \bm{m}_{1} , - \bm{m}_{2}} (1 , 2 , - \omega_{2}) \big] \nonumber
\\
& \, - \delta_{\rD} (g_{12}) \, \Lambda_{\bm{m}_{1} , \bm{m}_{2}} (1 , 2 , \omega_{2}) \, \mu_{\ra} \, \bm{m}_{2} \!\cdot\! \partial_{\bm{J}_{2}} \big[ \Lambda_{\bm{m}_{2} , \bm{m}_{1}} (2 , 1 , \omega_{1} ) \big] \bigg] \, .
\label{eq:dj1dtaverage2}
\end{align}
One can straightforwardly show that for any function ${ G (g_{12}) }$, one has
\begin{equation}
\bm{m}_{1} \!\cdot\! \partial_{\bm{J}_{1}} \big[ G (g_{12}) \big] = \bm{m}_{1} \!\cdot\! \partial_{\bm{J}_{1}} \big[ \bm{m}_{1} \!\cdot\! \bm{\Omega}_{1} \big] \, \frac{\rd G}{\rd g_{12}} \;\;\; ; \;\;\; \bm{m}_{2} \!\cdot\! \partial_{\bm{J}_{2}} \big[ G (g_{12}) \big] = -\bm{m}_{2} \!\cdot\! \partial_{\bm{J}_{2}} \big[ \bm{m}_{2} \!\cdot\! \bm{\Omega}_{2} \big] \, \frac{\rd G}{\rd g_{12}} \, .
\end{equation}
Applying these formulae to equation~\eqref{eq:dj1dtaverage2}, one can collect quite a few terms in this equation, to get
\begin{align}
\bigg< \frac{\Delta \bm{J}_{1} }{\Delta t} \bigg> = & \, \sum_{\rb} \sum_{\bm{m}_{1}} \sum_{\bm{m}_{2}}' \pi (2 \pi)^{d} \bm{m}_{1} \!\! \int \!\! \rd \bm{J}_{2} \, F^{\rb} (\bm{J}_{2}) \nonumber
\\
\times \bigg\{ & \, \mu_{\rb} \, \bm{m}_{1} \!\cdot\! \partial_{\bm{J}_{1}} \bigg[ \delta_{\rD} (\bm{m}_{1} \!\cdot\! \bm{\Omega}_{1} \!-\! \bm{m}_{2} \!\cdot\! \bm{\Omega}_{2}) \, \Lambda_{ - \bm{m}_{1} , - \bm{m}_{2}} (1 , 2 , - \bm{m}_{2} \!\cdot\! \bm{\Omega}_{2}) \, \Lambda_{\bm{m}_{1} , \bm{m}_{2}} (1 , 2 , \bm{m}_{2} \!\cdot\! \bm{\Omega}_{2}) \bigg] \nonumber
\\
& \, - \mu_{\ra} \, \bm{m}_{2} \!\cdot\! \partial_{\bm{J}_{2}} \bigg[ \delta_{\rD} (\bm{m}_{1} \!\cdot\! \bm{\Omega}_{1} \!-\! \bm{m}_{2} \!\cdot\! \bm{\Omega}_{2}) \, \Lambda_{\bm{m}_{2} , \bm{m}_{1}} (2 , 1 , \bm{m}_{1} \!\cdot\! \bm{\Omega}_{1}) \, \Lambda_{\bm{m}_{1} , \bm{m}_{2}} (1 , 2 , \bm{m}_{2} \!\cdot\! \bm{\Omega}_{2}) \bigg] \bigg\} \, .
\label{eq:dJ1dtsemifinal}
\end{align}
The lack of symmetry of this equation 
might look troublesome. In fact at the second order level, one can identify ${
\Lambda_{\bm{m}_{2} , \bm{m}_{1}} (2 , 1 , \omega_{1}) }$ with the complex
conjugate of ${ \Lambda_{\bm{m}_{2} , \bm{m}_{1}} (2 , 1 , \omega_{1}) }$, as
demonstrated in Appendix~\ref{sec:appendixsymmetries}. Similarly, ${ \Lambda_{ -
\bm{m}_{1} , - \bm{m}_{2}} (2 , 1 , - \omega_{2}) }$ is the complex conjugate 
of ${ \Lambda_{\bm{m}_{1} , \bm{m}_{2}} (1 , 2 , \omega_{2}) }$ (see Appendix~\ref{sec:appendixsymmetries}). Equation~\eqref{eq:dJ1dtsemifinal} can therefore be rewritten as
\begin{equation}
\bigg< \frac{\Delta \bm{J}_{1}}{\Delta t} \bigg> = 
\sum_{\rb} \!\!\sum_{\bm{m}_{1} , \bm{m}_{2}}' \!\! \int \!\! \rd \bm{J}_{2} \,
F^{\rb} (\bm{J}_{2}) \, \pi (2 \pi)^{d} \bm{m}_{1} \big( \mu_{\rb} \,
\bm{m}_{1} \!\cdot\! \partial_{\bm{J}_{1}} \!-\! \mu_{\ra} \, \bm{m}_{2}
\!\cdot\! \partial_{\bm{J}_{2}} \big) \delta_{\rD} (\bm{m}_{1} \!\cdot\!
\bm{\Omega}_{1} \!-\! \bm{m}_{2} \!\cdot\! \bm{\Omega}_{2}) \, \big|
\Lambda_{\bm{m}_{1} , \bm{m}_{2}} (1 , 2 , \bm{m}_{2} \!\cdot\! \bm{\Omega}_{2})
\big|^{2} \, .
\label{final_drift_app}
\end{equation}

\section{Direct calculation of the friction force by polarisation}
\label{sec:appendixFpol}

The friction force $\bm{F}_{\rm pol}$ that appears in the
Fokker-Planck equation~\eqref{sandwichedFP} is called the ``friction by
polarisation''~\citep{Chavanis2013}. This is just one component
of the total friction
force $\bm{F}_{\rm fric}$ that appears in the Fokker-Planck equation~\eqref{nice}.
Physically, $\bm{F}_{\rm pol}$ is the force resulting directly from the
retroaction of the field stars to the perturbation caused by the test star, like
in a polarisation process. Some particular, or formal, expressions of the
friction force by polarisation have been derived
in~\cite{Marochnik1968,Kalnajs1971,Kandrup1983,BekensteinZamir1990,Chavanis2008}
from a linear response theory based on the Liouville equation or on the Klimontovich equation.
 Let us derive here its general expression in
angle-action
variables taking spatial inhomogeneity
and collective effects into account and check that its expression agrees with
equation~\eqref{Fpolexp}. This calculation is inspired by section 3.3 of~\cite{Chavanis2012}.
The perturbation induced by the test particle is determined by the
coefficients $\widehat{a}^{\alpha}_{\rp} (\omega)$.
Following equation~\eqref{eq:calc_ad}, they can be rewritten as
\begin{equation}
\widehat{a}^{\alpha}_{\rp} (\omega)= (\varepsilon_{\alpha \beta}^{-1}
(\omega)-\delta_{\alpha\beta}) \, \widehat{a}^{\beta}_{\rt} (\omega) \, .
\end{equation}
The temporal Fourier transform of the potential perturbation ${ \delta U_{\rp}
}$ associated with the gravitational wake reads
\begin{equation}
\delta \widehat{U}_{\rp} (\bm{x} , \omega) = \psi^{(\alpha)} (\bm{x}) \,
(\varepsilon_{\alpha \beta}^{-1} (\omega)-\delta_{\alpha\beta}) \,
\widehat{a}^{\beta}_{\rt} (\omega) \, ,
\end{equation}
where the coefficients ${ \widehat{a}^{\beta}_{\rt} (\omega) }$ of the 
test particle potential are given by equation~\eqref{eq:aexpt3}.
Therefore
\begin{equation}
\delta \widehat{U}_{\rp} (\bm{x} , \omega) = - 2\pi \mu_{\rt} \sum_{\bm{m}'}
\psi^{(\alpha)} (\bm{x})
\, (\varepsilon_{\alpha \beta}^{-1} (\omega)-\delta_{\alpha\beta}) \,
\psi^{(\beta) *}_{\bm{m}'} (\bm{J}_{\rt}) \, \re^{- \ri \bm{m}' \cdot
\bm{\theta}_{\rt}^{0}} \, \delta_{D}(\omega-\bm{m}' \!\cdot\!
\bm{\Omega}_{\rt}) \, .
\label{fp1}
\end{equation}
Written as a function of time, equation~\eqref{fp1} becomes
\begin{equation}
\delta U_{\rp} (\bm{x} , t) = -\mu_{\rt} \sum_{\bm{m}'}
\psi^{(\alpha)} (\bm{x}) \, (\varepsilon_{\alpha \beta}^{-1} (\bm{m}' \!\cdot\!
\bm{\Omega}_{\rt})-\delta_{\alpha\beta}) \, \psi^{(\beta) *}_{\bm{m}'}
(\bm{J}_{\rt}) \, \re^{- \ri \bm{m}' \cdot (\bm{\theta}_{\rt}^{0} +
\bm{\Omega}_{\rt} t)} \, .
\end{equation}
Replacing ${ \psi^{(\alpha)}
(\bm{x}) }$ by the sum
\begin{equation}
\psi^{(\alpha)} (\bm{x}) = \sum_{\bm{m}} \psi^{(\alpha)}_{\bm{m}}
(\bm{J}) \, \re^{ \ri \bm{m} \cdot \bm{\theta}} \, ,
\end{equation}
we obtain
\begin{equation}
\delta U_{\rp} (\bm{x} , t) = - \mu_{\rt} \!\!\sum_{\bm{m} ,
\bm{m}'}\!\! \re^{\ri (\bm{m} \cdot \bm{\theta} - \bm{m}' \cdot
\bm{\theta}_{\rt})} \, \psi^{(\alpha)}_{\bm{m}} (\bm{J}) \,
(\varepsilon_{\alpha \beta}^{-1} (\bm{m}' \!\cdot\!
\bm{\Omega}_{\rt})-\delta_{\alpha\beta}) \,
\psi^{(\beta) *}_{\bm{m}'} (\bm{J}_{\rt}) \, .
\end{equation}
This formula completely specifies the polarisation cloud. The corresponding force in action, ${ \delta \bm{F}_{\rp} (\bm{x} , t) \!=\! -\partial_{\bm{\theta}} \delta U_{\rp} (\bm{x} , t) }$, is
\begin{equation}
 \delta \bm{F}_{\rp} (\bm{x} , t) = \mu_{\rt} \!\!\sum_{\bm{m} ,
\bm{m}'}\!\! \ri \bm{m} \, \re^{\ri (\bm{m} \cdot \bm{\theta} - \bm{m}' \cdot
\bm{\theta}_{\rt})} \, \psi^{(\alpha)}_{\bm{m}} (\bm{J}) \,
(\varepsilon_{\alpha \beta}^{-1} (\bm{m}' \!\cdot\!
\bm{\Omega}_{\rt})-\delta_{\alpha\beta}) \,
\psi^{(\beta) *}_{\bm{m}'} (\bm{J}_{\rt}) \, .
\end{equation}
One may then evaluate this force at the location of the test particle, and average it over the orbit of the test particle. One obtains the friction force by polarisation, ${ \bm{F}_{\rm pol} \!=\! \! \int\! \rd \bm{\theta}_{\rt} / (2 \pi)^{d} \delta F_{\rp} (\bm{x} (\bm{\theta}_{\rt} , \bm{J}_{\rt})) }$, reading
\begin{equation}
\bm{F}_{\rm pol} = \mu_{\rt} \sum_{\bm{m}} \ri\bm{m} 
\, \psi^{(\alpha)}_{\bm{m}} (\bm{J}_{\rt}) \,
(\varepsilon_{\alpha \beta}^{-1} (\bm{m} \!\cdot\!
\bm{\Omega}_{\rt})-\delta_{\alpha\beta}) \,
\psi^{(\beta) *}_{\bm{m}} (\bm{J}_{\rt}) \, .
\end{equation}
Introducing the system's dressed and bare susceptibility coefficients ${ 1
/ \mathcal{D}_{\bm{m}_{1} , \bm{m}_{2}} (\bm{J}_{1} , \bm{J}_{2} , \omega) }$
and ${ 1/\mathcal{D}_{\bm{m}_{1} , \bm{m}_{2}}^{\rm bare}(\bm{J}_{1} ,
\bm{J}_{2}) } \!=\! -{ A_{\bm{m}_{1} , \bm{m}_{1}}
(\bm{J}_{1} , \bm{J}_{2}) }$ from equations
\eqref{definition_1/D} and \eqref{definition_1/D_bare}, one gets 
\begin{equation}
\bm{F}_{\rm pol} = \mu_{\rt} \sum_{\bm{m}} \ri \bm{m} 
\, \bigg[ \frac{1}{\mathcal{D}_{\bm{m} , \bm{m}} (\bm{J}_{\rt} ,
\bm{J}_{\rt}, \bm{m} \!\cdot\!
\bm{\Omega}_{\rt})}+A_{\bm{m} , \bm{m}}
(\bm{J}_{\rt} , \bm{J}_{\rt}) \bigg] = - \mu_{\rt} \sum_{\bm{m}} \bm{m} 
\, \text{Im} \bigg[ \frac{1}{\mathcal{D}_{\bm{m} , \bm{m}} (\bm{J}_{\rt} ,
\bm{J}_{\rt}, \bm{m} \!\cdot\!
\bm{\Omega}_{\rt})} \bigg] \, .
\end{equation}
Finally, using identity (54) of~\cite{Chavanis2012}, i.e.
\begin{equation}
\text{Im} \bigg[ \frac{1}{\mathcal{D}_{\bm{m} , \bm{m}} (\bm{J} ,
\bm{J}, \bm{m} \!\cdot\!
\bm{\Omega})} \bigg] = - \pi (2\pi)^{d} \sum_{\rb} \sum_{\bm{m}'} \!\! \int \!\!
\rd \bm{J}' \, \frac{1}{|\mathcal{D}_{\bm{m} , \bm{m}'} (\bm{J} ,
\bm{J}', \bm{m} \!\cdot\! \bm{\Omega})|^{2}} \, \delta_{\rD}
(\bm{m} \!\cdot\! \bm{\Omega} \!-\! \bm{m}' \!\cdot\!
\bm{\Omega}') \bigg( \bm{m}' \!\cdot\! \frac{\partial F^{\rb}}{\partial
\bm{J}'} \bigg) \, , 
\end{equation}
we obtain
\begin{equation}
\bm{F}_{\rm pol} = \pi (2\pi)^{d} \mu_{\rt} \sum_{\rb} \sum_{\bm{m} , \bm{m}'} \bm{m} \!\! \int \!\!
\rd \bm{J}' \, \frac{1}{|\mathcal{D}_{\bm{m} , \bm{m}'} (\bm{J}_{\rt} ,
\bm{J}', \bm{m} \!\cdot\! \bm{\Omega}_{\rt})|^{2}} \, \delta_{\rD}
(\bm{m} \!\cdot\! \bm{\Omega}_{\rt} \!-\! \bm{m}' \!\cdot\!
\bm{\Omega}') \bigg( \bm{m}' \!\cdot\! \frac{\partial F^{\rb}}{\partial
\bm{J}'} \bigg) \, , 
\end{equation}
which coincides with equation~\eqref{Fpolexp}. If we neglect collective
effects, we recover equation~\eqref{Fpolexpbar}. The fact that $\bm{F}_{\rm
pol}$ is just one component of the true friction
force $\bm{F}_{\rm fric}$ is clear from equation~\eqref{Fpol} (see also
section~3 of~\cite{Chavanis2012}).

Note that this calculation remains valid if the test particle is of
different nature from the field particles. In particular, it could be a
satellite of mass $\mu_{\rt}$ moving in a collisionless fluid
of stars with mass ${ \mu_{\rb} \!\sim\! 1/N_{\rb} }$
governed by the Vlasov equation when ${ N_{\rb} \!\rightarrow\! + \infty }$.
In the limit ${ \mu_{\rt} \!\gg\! \mu_{\rb} }$, the friction by polarisation is the only force that
acts on the test particle, and the satellite sinks at the center of the system (see
Appendix~\ref{sec:sink}).

\section{Properties of the multicomponent Balescu-Lenard equation}
\label{sec:prop}

In this Appendix, we derive the main properties of the
inhomogeneous Balescu-Lenard equation~\eqref{eq:BL} following the works of~\cite{Chavanis2007,Heyvaerts2010,Chavanis2012} and references therein.
This extends their results to the important class of astrophysical systems containing different components. 

\subsection{Conservative form}

The Balescu-Lenard equation~\eqref{eq:BL} can be written in the conservative
form
\begin{equation}
\frac{\partial F^{\ra}}{\partial t} (\bm{J}_{1},t) = \frac{\partial}{\partial
\bm{J}_{1}} \!\cdot\! \bm{\mathcal{F}}^{\ra} \, ,
\label{prop1}
\end{equation}
where
\begin{align}
\bm{\mathcal{F}}^{\ra} & \, = \bm{D}
(\bm{J}_{1} , t) \!\cdot\! \frac{\partial F^{\ra} }{\partial
\bm{J}_{1}}(\bm{J}_{1} , t)
 \!-\! \bm{F}_{\rm pol} (\bm{J}_{1} , t) \, F^{\ra} (\bm{J}_{1} , t) \nonumber
\\
& \, = \pi (2 \pi)^{d} \sum_{\rb}
\!\!\!\sum_{\bm{m}_{1} , \bm{m}_{2}}' \!\!\! \bm{m}_{1} 
\!\! \int\!\! \rd \bm{J}_{2} \, \frac{ \delta_{\rD}
(\bm{m}_{1} \!\cdot\! \bm{\Omega}_{1} \!-\! \bm{m}_{2} \!\cdot\!
\bm{\Omega}_{2})}{ \big| \mathcal{D}_{\bm{m}_{1} , \bm{m}_{2}} (\bm{J}_{1} ,
\bm{J}_{2} , \bm{m}_{2} \!\cdot\! {\bm{\Omega}_{2}}) \big|^{2}} \,
 ( \mu_{\rb}
\, \bm{m}_{1} \!\cdot\! \partial_{\bm{J}_{1}} \!-\! \mu_{\ra} \, \bm{m}_{2}
\!\cdot\! \partial_{\bm{J}_{2}} )\, F^{\ra}
(\bm{J}_{1},t) \, F^{\rb}
(\bm{J}_{2},t)
\label{prop2}
\end{align}
is the flux of particles of component ``$\ra$''.
Under that form, it is immediately
clear that the Balescu-Lenard equation conserves the total mass of each component.

\subsection{Equilibrium state: Boltzmann distribution}
\label{sec_bd}

It is straightforward to check that the Boltzmann distribution 
\begin{equation}
F^{\ra} (\bm{J}) = A_{\ra} \, \re^{-\beta \mu_{\ra} \epsilon(\bm{J})} \, , 
\label{prop3}
\end{equation}
where ${ \epsilon(\bm{J}) }$ is the energy of a star by unit of mass, ${ \beta \!=\! 1/T }$
is the inverse ``temperature'' and $A_{\ra}$ is a normalisation constant, is a steady
state of the Balescu-Lenard equation. Indeed, using
\begin{equation}
\frac{\partial\epsilon}{\partial \bm{J}} = \bm{\Omega}(\bm{J}) \, ,
\label{prop4}
\end{equation} 
we get
\begin{equation}
\frac{\partial F^{\ra}}{\partial \bm{J}} = - \beta \mu_{\ra} \, F^{\ra} (\bm{J}) \, \bm{\Omega}(\bm{J}) \, .
\label{prop5}
\end{equation} 
Substituting this relation into the Balescu-Lenard flux from equation~\eqref{prop2}, we find
\begin{equation}
\bm{\mathcal{F}}^{\ra} = -\beta \pi (2 \pi)^{d} \sum_{\rb}
\!\!\!\sum_{\bm{m}_{1} , \bm{m}_{2}}' \!\! \bm{m}_{1} 
\!\! \int\!\! \rd \bm{J}_{2} \, \frac{ \delta_{\rD}
(\bm{m}_{1} \!\cdot\! \bm{\Omega}_{1} \!-\! \bm{m}_{2} \!\cdot\!
\bm{\Omega}_{2})}{ \big| \mathcal{D}_{\bm{m}_{1} , \bm{m}_{2}} (\bm{J}_{1} ,
\bm{J}_{2} , \bm{m}_{2} \!\cdot\! {\bm{\Omega}_{2}}) \big|^{2}} \,
 \mu_{\ra}\mu_{\rb} (\bm{m}_{1} \!\cdot\! \bm{\Omega}_{1} \!-\! \bm{m}_{2}
\!\cdot\!
\bm{\Omega}_{2}) \, F^{\ra}
(\bm{J}_{1}) \, F^{\rb}
(\bm{J}_{2}).
\label{prop2b}
\end{equation}
The integrand involves the term ${ \delta_{\rD} (\bm{m}_{1} \!\cdot\! \bm{\Omega}_{1} \!-\! \bm{m}_{2} \!\cdot\! \bm{\Omega}_{2})(\bm{m}_{1} \!\cdot\! \bm{\Omega}_{1} \!-\! \bm{m}_{2} \!\cdot\! \bm{\Omega}_{2}) }$, which is obviously equal to zero, so that one has ${ \partial_{t} F^{\ra} \!=\! 0 }$ for each component.
 Note that the temperature in the Boltzmann distribution
of equation~\eqref{prop3} is the same for all the components. This corresponds to
an equipartition of energy, and usually implies that heavy particles sink at
the center of the system while light particles wander around.

The Boltzmann distribution can be obtained by maximising
the Boltzmann entropy $S_{\rB}$ defined by equation~\eqref{prop14} while conserving
the total energy $E$ and the total mass $M_{\ra}$ of each species of particles.
Writing the variational principle as 
\begin{equation}
\delta S_{\rB} - \beta \delta E - \sum_{\ra} \alpha_{\ra} \delta M_{\ra} = 0 \, ,
\label{prop7b}
\end{equation}
where $\beta$ (inverse temperature) and $\alpha_{\ra}$ (chemical potentials) are the
Lagrange multipliers associated with energy and mass conservation, we obtain
\begin{equation}
\sum_{\ra} \bigg\{\! -\frac{1}{\mu_{\ra}} \bigg[ \ln \!\bigg(\! \frac{F^{\ra}}{\mu_{\ra}} \!\bigg) \!+\! 1 \bigg] - \beta\epsilon - \alpha_{\ra} \bigg\} \delta F^{\ra} = 0 \, .
\label{prop7c}
\end{equation}
Since this condition must be satisfied for arbitrary variations ${ \delta F^{\ra} }$ the
term in braces must vanish, leading to the Boltzmann distribution~\eqref{prop3}.
The Boltzmann distribution is therefore a critical
point of entropy at fixed mass and energy. However, let us stress once again that a
statistical equilibrium state does
not always exist for self-gravitating systems (notably for ${3D}$ spherical
systems). The Boltzmann entropy may not have a (global or local) maximum. Even
worse, the Boltzmann distribution from equation~\eqref{prop3} may not be normalisable (i.e.
the Boltzmann entropy may not have any critical point).

\subsection{Energy conservation}

The total energy of the system is
\begin{equation}
E = \sum_{\ra} \!\! \int \!\! \rd \bm{J}_{1} \, F^{\ra} (\bm{J}_{1},t)
\epsilon(\bm{J}_{1}) = \sum_{\ra} E_{\ra} \, .
\label{prop8}
\end{equation}
Taking its time derivative and using equation~\eqref{prop1}, we get
\begin{equation}
\dot{E} = \sum_{\ra} \!\! \int \!\! \rd \bm{J}_{1} \bigg( \frac{\partial}{\partial
\bm{J}_{1}} \!\cdot\! \bm{\mathcal{F}}^{\ra} \bigg) \epsilon(\bm{J}_{1}) \, .
\label{prop9}
\end{equation}
Integrating by parts, assuming that boundary terms do not
contribute, and using equation~\eqref{prop4}, we obtain 
\begin{equation}
\dot{E} = - \sum_{\ra} \!\! \int \!\! \rd \bm{J}_{1} \, \bm{\mathcal{F}}^{\ra} \!\cdot\! \bm{\Omega} (\bm{J}_{1}) \, .
\label{prop10}
\end{equation}
Substituting equation~\eqref{prop2} into equation~\eqref{prop10}, we get
\begin{equation}
\dot{E} = -\pi (2 \pi)^{d} \sum_{\ra,\rb}
\!\sum_{\bm{m}_{1}, \bm{m}_{2}}' 
 \!\! \int\!\! \rd \bm{J}_{1} \rd \bm{J}_{2} \, \frac{ \delta_{\rD}
(\bm{m}_{1} \!\cdot\! \bm{\Omega}_{1} \!-\! \bm{m}_{2} \!\cdot\!
\bm{\Omega}_{2})}{ \big| \mathcal{D}_{\bm{m}_{1} , \bm{m}_{2}} (\bm{J}_{1} ,
\bm{J}_{2} , \bm{m}_{2} \!\cdot\! {\bm{\Omega}_{2}}) \big|^{2}} \,(\bm{m}_{1}
\!\cdot\!
\bm{\Omega}_{1}) 
 ( \mu_{\rb}
\, \bm{m}_{1} \!\cdot\! \partial_{\bm{J}_{1}} \!-\! \mu_{\ra} \, \bm{m}_{2}
\!\cdot\! \partial_{\bm{J}_{2}} )
\, F^{\ra} (\bm{J}_{1}) \, F^{\rb}
(\bm{J}_{2}) \, .
\label{prop11}
\end{equation}
Interchanging the dummy variables ${ (\ra , \rb) }$, ${ (\bm{m}_{1} , \bm{m}_{2}) }$ and 
${ (\bm{J}_{1} , \bm{J}_{2}) }$, and using the property ${ \mathcal{D}_{\bm{m}_{2} ,
\bm{m}_{1}} (\bm{J}_{2} ,
\bm{J}_{1} , \omega) \!=\! \mathcal{D}_{\bm{m}_{1} ,
\bm{m}_{2}} (\bm{J}_{1} ,
\bm{J}_{2} , \omega)^{*} }$ (see Appendix~\ref{sec:LinkLambdatwo}), we obtain
\begin{equation}
\dot{E} = \pi (2 \pi)^{d} \sum_{\ra,\rb}
\!\sum_{\bm{m}_{1}, \bm{m}_{2}}' 
 \!\! \int\!\! \rd \bm{J}_{1}\rd \bm{J}_{2} \, \frac{ \delta_{\rD}
(\bm{m}_{1} \!\cdot\! \bm{\Omega}_{1} \!-\! \bm{m}_{2} \!\cdot\!
\bm{\Omega}_{2})}{ \big| \mathcal{D}_{\bm{m}_{1} , \bm{m}_{2}} (\bm{J}_{1} ,
\bm{J}_{2} , \bm{m}_{2} \!\cdot\! {\bm{\Omega}_{2}}) \big|^{2}} \,(\bm{m}_{2}
\!\cdot\!
\bm{\Omega}_{2})
( \mu_{\rb}
\, \bm{m}_{1} \!\cdot\! \partial_{\bm{J}_{1}} \!-\! \mu_{\ra} \, \bm{m}_{2}
\!\cdot\! \partial_{\bm{J}_{2}} )
\, F^{\ra} (\bm{J}_{1}) \, F^{\rb}
(\bm{J}_{2}) \, .
\label{prop12}
\end{equation}
Taking the half-sum of these equations, we get
\begin{equation}
\dot{E} = -\frac{\pi (2 \pi)^{d}}{2} \sum_{\ra,\rb}
\!\sum_{\bm{m}_{1} , \bm{m}_{2}}' 
 \!\! \int\!\! \rd \bm{J}_{1}\rd \bm{J}_{2} \, \frac{\delta_{\rD} (\bm{m}_{1} \!\cdot\! \bm{\Omega}_{1} \!-\! \bm{m}_{2} \!\cdot\!
\bm{\Omega}_{2}) \, (\bm{m}_{1} \!\cdot\! \bm{\Omega}_{1}-\bm{m}_{2} \!\cdot\! \bm{\Omega}_{2}) }{ \big| \mathcal{D}_{\bm{m}_{1} , \bm{m}_{2}} (\bm{J}_{1} ,
\bm{J}_{2} , \bm{m}_{2} \!\cdot\! {\bm{\Omega}_{2}}) \big|^{2}} \, ( \mu_{\rb}
\, \bm{m}_{1} \!\cdot\! \partial_{\bm{J}_{1}} \!-\! \mu_{\ra} \, \bm{m}_{2}
\!\cdot\! \partial_{\bm{J}_{2}} )
\, F^{\ra} (\bm{J}_{1}) \, F^{\rb}
(\bm{J}_{2}) \, .
\label{prop11b}
\end{equation}
As in equation~\eqref{prop2b}, the integrand involves the 
term ${ \delta_{\rD} (\bm{m}_{1} \!\cdot\! \bm{\Omega}_{1} \!-\! \bm{m_{2}}
\!\cdot\! \bm{\Omega}_{2}) (\bm{m}_{1} \!\cdot\! \bm{\Omega}_{1} \!-\!
\bm{m}_{2} \!\cdot\! \bm{\Omega}_{2}) }$, which is identically zero. One
therefore has ${ \dot{E} \!=\! 0 }$, the total energy of the system is
conserved. Note, however, that the energy of each species is not
individually conserved. Using equations~\eqref{prop2} and~\eqref{prop10}, one has
\begin{equation}
\dot{E}_{\ra} = - \!\! \int \!\! \rd \bm{J}_{1} \, 
\bm{D}(\bm{J}_{1} , t) \!\cdot\! \bm{\Omega}(\bm{J}_{1} , t ) \!\otimes\! \frac{\partial
F^{\ra}}{\partial \bm{J}_{1}} (\bm{J}_{1} , t ) + \!\! \int \!\! \rd \bm{J}_{1} \, 
\bm{F}_{\rm pol} (\bm{J}_{1} , t ) \!\cdot\!
\bm{\Omega} (\bm{J}_{1} , t ) F^{\ra} (\bm{J}_{1} , t ) \, .
\label{mod2}
\end{equation}

\subsection{${H-}$theorem}
\label{sec_ht}

The multicomponent Boltzmann entropy is 
\begin{equation}
S_{\rB} = - \sum_{\ra} \!\! \int \!\! \rd \bm{J}_{1} \, \frac{F^{\ra}}{\mu_{\ra}} \ln \!\bigg(\! \frac{F^{\ra}}{\mu_{\ra}} \!\bigg) = \sum_{\ra} S_{\ra} \, .
\label{prop14}
\end{equation}
Taking its time derivative, and using equation~\eqref{prop1}, we get
\begin{equation}
\dot{S}_{\rB} = - \sum_{\ra} \!\! \int \!\! \rd \bm{J}_{1} \, \frac{1}{\mu_{\ra}} \bigg[ 1 \!+\! \ln \!\bigg(\! \frac{F^{\ra}}{\mu_{\ra}} \bigg) \bigg] \bigg( \frac{\partial}{\partial \bm{J}_{1}} \!\cdot\! \bm{\mathcal{F}}^{\ra} \bigg) \, .
\label{prop15}
\end{equation}
Integrating by parts and assuming that boundary terms do not contribute, we obtain
\begin{equation}
\dot{S}_{\rB} = \sum_{\ra} \!\! \int \!\! \rd \bm{J}_{1} \, \frac{1}{\mu_{\ra} F^{\ra} (\bm{J}_{1})} \frac{\partial F^{\ra}}{\partial
\bm{J}_{1}} \!\cdot\! \bm{\mathcal{F}}^{\ra} \, .
\label{prop16}
\end{equation}
Substituting equation~\eqref{prop2} into equation~\eqref{prop16}, we get
\begin{equation}
\dot{S}_{\rB} \!=\! \pi (2 \pi)^{d} \sum_{\ra,\rb}
\!\sum_{\bm{m}_{1} , \bm{m}_{2}}'
\!\!\int\!\! \rd \bm{J}_{1}\rd \bm{J}_{2} \, \frac{ \delta_{\rD}
(\bm{m}_{1} \!\cdot\! \bm{\Omega}_{1} \!-\! \bm{m}_{2} \!\cdot\!
\bm{\Omega}_{2})}{ \big| \mathcal{D}_{\bm{m}_{1} , \bm{m}_{2}} (\bm{J}_{1} ,
\bm{J}_{2} , \bm{m}_{2} \!\cdot\! {\bm{\Omega}_{2}}) \big|^{2}}
\bigg[ \frac{1}{\mu_{\ra} F^{\ra} (\bm{J}_{1})} \, \bm{m}_{1} \!\cdot\! \partial_{\bm{J}_{1}} F^{\ra} \bigg] ( \mu_{\rb}
\, \bm{m}_{1} \!\cdot\! \partial_{\bm{J}_{1}} \!-\! \mu_{\ra} \, \bm{m}_{2}
\!\cdot\! \partial_{\bm{J}_{2}} ) \, F^{\ra} (\bm{J}_{1}) \, F^{\rb}
(\bm{J}_{2}) \, .
\label{prop17}
\end{equation}
Interchanging the dummy variables ${ (\ra , \rb) }$, ${ (\bm{m}_{1} , \bm{m}_{2}) }$ and 
${ (\bm{J}_{1},\bm{J}_{2}) }$, and
using the property $\mathcal{D}_{\bm{m}_{2} ,
\bm{m}_{1}} (\bm{J}_{2} ,
\bm{J}_{1} , \omega)=\mathcal{D}_{\bm{m}_{1} ,
\bm{m}_{2}} (\bm{J}_{1} ,
\bm{J}_{2} , \omega)^*$ (see Appendix~\ref{sec:LinkLambdatwo}), we
obtain
\begin{equation}
\dot{S}_{\rB} \!=\! - \pi (2 \pi)^{d} \sum_{\ra,\rb}
\!\sum_{\bm{m}_{1} , \bm{m}_{2}}' 
 \!\! \int\!\! \rd \bm{J}_{1}\rd \bm{J}_{2} \, \frac{ \delta_{\rD}
(\bm{m}_{1} \!\cdot\! \bm{\Omega}_{1} \!-\! \bm{m}_{2} \!\cdot\!
\bm{\Omega}_{2})}{ \big| \mathcal{D}_{\bm{m}_{1} , \bm{m}_{2}} (\bm{J}_{1} ,
\bm{J}_{2} , \bm{m}_{2} \!\cdot\! {\bm{\Omega}_{2}}) \big|^{2}}
\bigg[ \frac{1}{\mu_{\rb} F^{\rb} (\bm{J}_{2})} \, \bm{m}_{2} \!\cdot\! \partial_{\bm{J}_{2}} F^{\rb} \bigg] ( \mu_{\rb}
\, \bm{m}_{1} \!\cdot\! \partial_{\bm{J}_{1}} \!-\! \mu_{\ra} \, \bm{m}_{2}
\!\cdot\! \partial_{\bm{J}_{2}} ) \, F^{\ra} (\bm{J}_{1}) \, F^{\rb}
(\bm{J}_{2}) \, .
\label{prop18}
\end{equation}
Taking the half-sum of these equations, we get
\begin{equation}
\dot{S}_{\rB} \!=\! \frac{1}{2}\pi (2 \pi)^{d} \sum_{\ra,\rb}
\!\sum_{\bm{m}_{1} , \bm{m}_{2}}'
 \!\! \int\!\! \rd \bm{J}_{1}\rd \bm{J}_{2} \, \frac{ \delta_{\rD}
(\bm{m}_{1} \!\cdot\! \bm{\Omega}_{1} \!-\! \bm{m}_{2} \!\cdot\!
\bm{\Omega}_{2})}{ \big| \mathcal{D}_{\bm{m}_{1} , \bm{m}_{2}} (\bm{J}_{1} ,
\bm{J}_{2} , \bm{m}_{2} \!\cdot\! {\bm{\Omega}_{2}}) \big|^{2}}
\frac{1}{\mu_{\ra} F^{\ra} (\bm{J}_{1}) \mu_{\rb} F^{\rb} (\bm{J}_{2})}
\bigg[ ( \mu_{\rb}
\, \bm{m}_{1} \!\cdot\! \partial_{\bm{J}_{1}} \!-\! \mu_{\ra} \, \bm{m}_{2}
\!\cdot\! \partial_{\bm{J}_{2}} ) \, F^{\ra} (\bm{J}_{1}) \, F^{\rb}
(\bm{J}_{2}) \bigg]^{2} \!\! ,
\label{prop19}
\end{equation}
from which we obtain the ${H-}$theorem
\begin{equation}
\dot{S}_{\rB} \ge 0 \, .
\label{prop20}
\end{equation}
It establishes that the Boltzmann entropy is monotonically
increasing. Note however that the Boltzmann entropy of each
species does not individually satisfy a ${H-}$theorem. Using equations~\eqref{prop2} and~\eqref{prop16}, one has
\begin{equation}
\dot{S}_{\ra} = \!\! \int \!\! \rd \bm{J}_{1} \, 
\frac{1}{\mu_{\ra} F^{\ra} (\bm{J}_{1} , t)} \bm{D} (\bm{J}_{1} , t ) \!\cdot\!
\frac{\partial F^{\ra} }{\partial\bm{J}_{1}} (\bm{J}_{1} , t) \!\otimes\! \frac{\partial
F^{\ra} }{\partial \bm{J}_{1}} (\bm{J}_{1} , t ) - \!\! \int \!\! \rd \bm{J}_{1}
\frac{1}{\mu_{\ra}} \bm{F}_{\rm pol} (\bm{J}_{1} , t ) \!\cdot\!
\frac{\partial F^{\ra}}{\partial\bm{J}_{1}} (\bm{J}_{1} , t ) \, .
\label{mod3}
\end{equation}

For neutral plasmas, that are spatially homogeneous, the Boltzmann entropy
is bounded from above. In that case, one can show from the ${H-}$theorem that the
homogeneous Balescu-Lenard relaxes, for ${ t \!\rightarrow\! + \infty}$, towards the
Boltzmann distribution~\eqref{prop3}. This is the maximum entropy state at
fixed mass and energy. We note that the Balescu-Lenard equation singles out the
Boltzmann distribution
among all possible steady states of the Vlasov equation. For self-gravitating
systems, that are spatially
inhomogeneous, the Boltzmann entropy is typically not bounded from above (this
is notably the case for ${3D}$ spherical or ${2D}$ flat systems). 
In that case, the inhomogeneous Balescu-Lenard equation does not
relax towards an equilibrium state. It can describe stellar evaporation and core
collapse (gravothermal catastrophe) as discussed in section~\ref{sec:stages}.
It cannot, however, account for the formation of binary stars
and for gravothermal oscillations since the formation of binaries results from
three-body collisions that are neglected in the Balescu-Lenard treatment. For
other
systems with long-range interactions for which the Boltzmann entropy
is bounded from above, the Boltzmann distribution is always a steady state of
the inhomogeneous Balescu-Lenard equation (see section~\ref{sec_bd}). However,
the inhomogeneous Balescu-Lenard equation does not necessarily relax towards this
distribution. In very specific situations, the system may remain blocked in
another state if the resonance condition cannot be fulfilled. This ``kinetic
blocking'' is illustrated in the case of ${2D}$ point vortices in~\cite{ChavanisLemou2007}
when the profile of angular velocity is monotonic.
In that case, no resonance is possible whatever the distribution function and the
system remains frozen in a distribution that is different from the Boltzmann
distribution. Only higher order correlations (three-body, four-body...) may
unblock the system. Such correlations have their own higher order (e.g. ${ 1/N^{2} }$) kinetic equations.

\subsection{Initial flux in non-thermalised mix}
\label{sec_if}

Let us consider an initial condition in which all the particles have
a Boltzmann distribution of the form 
\begin{equation}
F^{\ra} (\bm{J}) = A_{\ra} \, \re^{- \beta_{\ra} \mu_{\ra} \epsilon(\bm{J})} \, .
\label{prop3q}
\end{equation}
However, we do not assume equipartition of energy: the temperature
${ T_{\ra} \!=\! 1/\beta_{\ra} }$ of each species may be different. This may correspond to the situation in a stellar disc 
where old and young populations of stars co-exist.
The collisions, described by the Balescu-Lenard equation, change the energy of each
component so that their temperature is the same at equilibrium (equipartition of
energy). The initial value of the flux defined by equation~\eqref{prop2} is
\begin{equation}
\bm{\mathcal{F}}^{\ra} = -\pi (2 \pi)^{d} \sum_{\rb}
\!\!\!\sum_{\bm{m}_{1} , \bm{m}_{2}}' \!\! \bm{m}_{1} 
\!\! \int\!\! \rd \bm{J}_{2} \, \frac{ \delta_{\rD}
(\bm{m}_{1} \!\cdot\! \bm{\Omega}_{1} \!-\! \bm{m}_{2} \!\cdot\!
\bm{\Omega}_{2})}{ \big| \mathcal{D}_{\bm{m}_{1} , \bm{m}_{2}} (\bm{J}_{1} ,
\bm{J}_{2} , \bm{m}_{2} \!\cdot\! {\bm{\Omega}_{2}}) \big|^{2}} \,
 \mu_{\ra} \mu_{\rb}(\bm{m}_{1} \!\cdot\! \bm{\Omega}_{1})(\beta_{\ra} \!-\! \beta_{\rb}) \,
F^{\ra}
(\bm{J}_{1}) \, F^{\rb}
(\bm{J}_{2}) \, . 
\label{prop2q}
\end{equation}
It can be written as
\begin{equation}
\bm{\mathcal{F}}^{\ra} = -\sum_{\rb} \mu_{\ra} (\beta_{\ra} \!-\! \beta_{\rb}) \bm{D}^{\ra \rb} \!\cdot
\bm{\Omega}_{1} F^{\ra} (\bm{J}_{1}) \, ,
\label{prop2qa}
\end{equation}
where $\bm{D}^{\ra \rb}$ is the diffusion tensor of component ``$\ra$'' caused by
the collisions with the particles of component ``$\rb$''. The total diffusion tensor
defined by equation~\eqref{final_diff_drift1} can be written as ${ \bm{D} \!=\! \sum_{\rb}
\bm{D}^{\ra \rb} }$. The initial rate of change of energy, obtained from equations~\eqref{prop10} and~\eqref{prop2qa} is given by
\begin{equation}
\dot{E}_{\ra} = \sum_{\rb} \!\! \int \!\! \rd \bm{J}_{1} \, \mu_{\ra}
(\beta_{\ra} \!-\! \beta_{\rb}) \bm{D}^{\ra \rb} \!\cdot
\bm{\Omega}_{1} \!\otimes\! \bm{\Omega}_{1} F^{\ra} (\bm{J}_{1}).
\label{prop2qb}
\end{equation}
It can be written more explicitly as
\begin{equation}
 \dot{E}_{\ra} = \pi (2 \pi)^{d} \sum_{\rb}
\!\!\!\sum_{\bm{m}_{1} , \bm{m}_{2}}' 
\!\! \int\!\! \rd \bm{J}_{1}\rd \bm{J}_{2} \, \frac{ \delta_{\rD}
(\bm{m}_{1} \!\cdot\! \bm{\Omega}_{1} \!-\! \bm{m}_{2} \!\cdot\!
\bm{\Omega}_{2})}{ \big| \mathcal{D}_{\bm{m}_{1} , \bm{m}_{2}} (\bm{J}_{1} ,
\bm{J}_{2} , \bm{m}_{2} \!\cdot\! {\bm{\Omega}_{2}}) \big|^{2}} \,
 \mu_{\ra} \mu_{\rb}(\bm{m}_{1} \!\cdot\! \bm{\Omega}_{1})^{2} (\beta_{\ra} \!-\! \beta_{\rb}) \,
F^{\ra} (\bm{J}_{1}) \, F^{\rb} (\bm{J}_{2}) \, . 
\label{prop2qc}
\end{equation}
The rate of change of energy of species ``$\ra$'' scales like the mass weighted sum of inverse temperature differences.
We can similarly compute the initial rate of change of entropy from
equations~\eqref{prop16} and~\eqref{prop2qa}.
We find that
\begin{equation}
\dot{S}_{\ra} = \beta_{\ra} \dot{E}_{\ra} \, .
\label{prop2qd}
\end{equation}

\section{Test particle approach}
\label{sec:tpa}

Different interpretations can be given to the Balescu-Lenard
equation. The Balescu-Lenard equation~\eqref{eq:BL} is an
integro-differential
equation that describes the evolution
of an ensemble of particles (e.g. stars) in interaction. In this interpretation,
all the particles are treated on the same footing and their distribution
function ${ F^{\ra} (\bm{J}_{1} , t) }$ evolves self-consistently according to equation~\eqref{eq:BL}.
In a second interpretation, one can select a particular test particle
with mass $\mu_{\rt}$ and study the
evolution of its probability density ${ P(\bm{J}_{1} , t) }$ in a cloud of field
particles with a static distribution function ${ F^{\rb} (\bm{J}_{2}) }$. In
this interpretation,\footnote{As discussed in Appendix F
of~\cite{Chavanis2013}, this interpretation is valid either for a single
component system or for a multicomponent system. The test particle may
represent just one particle or an ensemble of non-interacting particles of the
same species. In all cases, one has to assume that the collisions between the
test particle(s) and the field particles do not alter the distribution of the
field particles.}
the Balescu-Lenard
equation~\eqref{eq:BL} is transformed into a differential equation 
\begin{equation}
\frac{\partial P}{\partial t}(\bm{J}_{1},t) = \pi (2 \pi)^{d} \sum_{\rb}
\!\!\!\sum_{\bm{m}_{1} , \bm{m}_{2}}' \!\! \bm{m}_{1} \!\cdot\!
\frac{\partial}{\partial{\bm{J}_{1}}} \!\! \int\!\! \rd \bm{J}_{2} \, \frac{
\delta_{\rD}
(\bm{m}_{1} \!\cdot\! \bm{\Omega}_{1} \!-\! \bm{m}_{2} \!\cdot\!
\bm{\Omega}_{2})}{ \big| \mathcal{D}_{\bm{m}_{1} , \bm{m}_{2}} (\bm{J}_{1} ,
\bm{J}_{2} , \bm{m}_{2} \!\cdot\! {\bm{\Omega}_{2}}) \big|^{2}} \, \bigg(
\mu_{\rb} \, \bm{m}_{1}
\!\cdot\! \frac{\partial}{\partial {\bm{J}_{1}}} \!-\! \mu_{\rt} \, \bm{m}_{2}
\!\cdot\!
\frac{\partial}{\partial {\bm{J}_{2}}} \bigg)
\, P (\bm{J}_{1},t) \, F^{\rb}
(\bm{J}_{2}) \, ,
\label{prop21}
\end{equation}
usually referred to as the Fokker-Planck equation. The diffusion and friction
coefficients are given by
\begin{align}
& \, \bm{D} = \pi (2 \pi)^{d} \sum_{\rb} \mu_{\rb}
\!\!\sum_{\bm{m}_{1} , \bm{m}_{2}}' \!\! \int\!\! \rd \bm{J}_{2} \, F^{\rb}
(\bm{J}_{2}) \, \bm{m}_{1} \!\otimes\! \bm{m}_{1} \,\frac{ \delta_{\rD}
(\bm{m}_{1} \!\cdot\! \bm{\Omega}_{1} \!-\! \bm{m}_{2} \!\cdot\!
\bm{\Omega}_{2})}{ \big| \mathcal{D}_{\bm{m}_{1} , \bm{m}_{2}} (\bm{J}_{1} ,
\bm{J}_{2} , \bm{m}_{2} \!\cdot\! {\bm{\Omega}_{2}}) \big|^{2}}
\, ,
\label{prop22}
\\
& \, \bm{F}_{\rm fric} = \pi (2
\pi)^{d} \sum_{\rb} \!\!\sum_{\bm{m}_{1} , \bm{m}_{2}}' \!\! \int\!\! \rd
\bm{J}_{2} \, F^{\rb} (\bm{J}_{2}) \, \bm{m}_{1} \bigg[ \mu_{\rb} \,
\bm{m}_{1}
\!\cdot\! \frac{\partial}{\partial {\bm{J}_{1}}} \!-\! \mu_{\rt} \, \bm{m}_{2}
\!\cdot\!
\frac{\partial}{\partial {\bm{J}_{2}}} \bigg] \frac{ \delta_{\rD}
(\bm{m}_{1} \!\cdot\! \bm{\Omega}_{1} \!-\! \bm{m}_{2} \!\cdot\!
\bm{\Omega}_{2})}{ \big| \mathcal{D}_{\bm{m}_{1} , \bm{m}_{2}} (\bm{J}_{1} ,
\bm{J}_{2} , \bm{m}_{2} \!\cdot\! {\bm{\Omega}_{2}}) \big|^{2}} \, ,
\label{prop23}
\\
& \, \bm{F}_{\rm pol} = \pi (2
\pi)^{d} \mu_{\rt} \sum_{\rb} \!\!\sum_{\bm{m}_{1} , \bm{m}_{2}}' \!\! \int\!\!
\rd
\bm{J}_{2} \, \bm{m}_{1} \bigg[ \bm{m}_{2} \!\cdot\! \frac{\partial F^{\rb}}{\partial 
\bm{J}_{2}}(\bm{J}_{2}) \bigg] \, \frac{ \delta_{\rD}
(\bm{m}_{1} \!\cdot\! \bm{\Omega}_{1} \!-\! \bm{m}_{2} \!\cdot\!
\bm{\Omega}_{2})}{ \big| \mathcal{D}_{\bm{m}_{1} , \bm{m}_{2}} (\bm{J}_{1} ,
\bm{J}_{2} , \bm{m}_{2} \!\cdot\! {\bm{\Omega}_{2}}) \big|^{2}} \, .
\label{prop24}
\end{align}
The Fokker-Planck equation~\eqref{prop21} can be
rewritten as
\begin{equation}
\frac{\partial P }{\partial t}(\bm{J}_{1} , t) = \frac{\partial }{\partial
\bm{J}_{1}}
\!\cdot\! \bigg[ \bm{D}
(\bm{J}_{1} ) \!\cdot\! \frac{\partial P }{\partial \bm{J}_{1}}(\bm{J}_{1},t)
 \!-\! \bm{F}_{\rm pol} (\bm{J}_{1} ) \, P (\bm{J}_{1} , t) \bigg]
\, .
\label{sandwichedFPb}
\end{equation}
The rates of change of energy ${ E_{\rt}(t) \!=\! \!
\int \! \rd \bm{J}_{1} \, P(\bm{J}_{1} , t) \mu_{\rt} \epsilon(\bm{J}_{1}) }$ and
Boltzmann entropy ${ S_{\rt}(t)
\!=\! - \! \int \! \rd \bm{J}_{1} P(\bm{J}_{1} , t) \ln
P(\bm{J}_{1} , t ) }$ of the test particle are
\begin{equation}
\dot{E}_{\rt} = -\mu_{\rt}\!\! \int \!\! \rd \bm{J}_{1} \, \bm{D}(\bm{J}_{1})
\!\cdot\!
 \frac{\partial P}{\partial \bm{J}_{1}}(\bm{J}_{1} , t) \!\otimes\!
\bm{\Omega}(\bm{J}_{1}) +\mu_{\rt} \!\!
\int \!\! \rd \bm{J}_{1} \, \bm{F}_{\rm pol} ( \bm{J}_{1} ) \!\cdot\!
\bm{\Omega}(\bm{J}_{1}) P(\bm{J}_{1} , t) \, ,
\label{ha2new}
\end{equation}
\begin{equation}
\dot{S}_{\rt} = \!\! \int \!\! \rd \bm{J}_{1}
\frac{1}{P(\bm{J}_{1} , t)} \bm{D} (\bm{J}_{1}) \!\cdot\!
\frac{\partial P}{\partial\bm{J}_{1}} (\bm{J}_{1} , t) \!\otimes\! \frac{\partial
P}{\partial \bm{J}_{1}} (\bm{J}_{1} , t) - \!\! \int \!\! \rd \bm{J}_{1}
 \bm{F}_{\rm pol}(\bm{J}_{1}) \!\cdot\!
\frac{\partial P}{\partial\bm{J}_{1}} (\bm{J}_{1} , t) \, .
\label{mod4}
\end{equation}
For the self-consistency of this interpretation, the field stars must be
at statistical equilibrium with the Boltzmann distribution (see section~\ref{sec_tb}), in a ``blocked state''
(see section~\ref{sec_ht}), or have a very long relaxation time (see
section~\ref{sec:sink}) so that their distribution function ${ F^{\rb} (\bm{J}_{2}) }$ does
not change under the effect of collisions among themselves on the timescale over which the test
particle evolves.
A possible astrophysical situation of relevance is a hot halo embedding a cold galactic disc.
At zeroth order, the halo can be taken to remain unresponsive to its own fluctuations and to that of its disc.

\subsection{Thermal bath: Einstein relation and fluctuation-dissipation theorem}
\label{sec_tb}

In this subsection, let us consider the relaxation of a test star described by a
probability density ${ P(\bm{J}_{1} , t) }$ in a cloud of field stars at
statistical equilibrium with the Boltzmann distribution given by
equation~\eqref{prop3}. Using equation~\eqref{prop5}, we find that the friction
by polarisation~\eqref{prop24} experienced by the test star becomes
\begin{equation}
\bm{F}_{\rm pol} = - \pi (2\pi)^d \beta \mu_{\rt} \sum_{\rb} \sum_{\bm{m}_{1} , \bm{m}_{2}}
\bm{m}_{1} \!\! \int \!\! \rd \bm{J}_{2} \, \frac{\delta_{\rD}
(\bm{m}_{1} \!\cdot\! \bm{\Omega}_{1} \!-\! \bm{m}_{2} \!\cdot\!
\bm{\Omega}_{2})}{|\mathcal{D}_{\bm{m}_{1} , \bm{m}_{2}} (\bm{J}_{1} ,
\bm{J}_{2} , \bm{m}_{1} \!\cdot\! \bm{\Omega}_{1})|^{2}} \, ( \bm{m}_{2} \!\cdot\! \bm{\Omega}_{2} ) \, \mu_{\rb} F^{\rb} (\bm{J}_{2}) \, .
\label{prop27}
\end{equation}
Using the Dirac delta function to replace ${ \bm{m}_{2} \!\cdot\! \bm{\Omega}_{2} }$ by
${ \bm{m}_{1} \!\cdot\! \bm{\Omega}_{1} }$ in the last parenthesis, we find that
\begin{equation}
\bm{F}_{\rm pol} = - \beta \mu_{\rt} \bm{D} (\bm{J}_{1}) \!\cdot\! \bm{\Omega} (\bm{J}_{1}) \, , 
\label{prop28}
\end{equation}
where $\bm{D}$ is the diffusion tensor given by equation~\eqref{prop22},
in which the field stars have the
Boltzmann distribution~\eqref{prop3}.
We may then define the friction tensor, $\bm{\xi}$, as 
\begin{equation}
\bm{\xi} = \beta \mu_{\rt} \bm{D} \, .
\label{prop29}
\end{equation}
Equation~\eqref{prop29} is the
appropriate form of the Einstein relation for this problem. It relates the friction
coefficient $\bm{\xi}$ to the diffusion coefficient $\bm{D}$, the inverse
temperature ${ \beta \!=\! 1/T }$, and the mass of the test particle $\mu_{\rt}$. This
is a manifestation
of the fluctuation-dissipation theorem. As previously emphasised~\citep{Chavanis2012}, the
Einstein relation is valid for the friction by polarisation, not for the true
friction. The Fokker-Planck equation~\eqref{prop21}, when written similarly to
equation~\eqref{sandwichedFPb},
takes the form of a generalised Kramers
equation
\begin{equation}
\frac{\partial P}{\partial t} (\bm{J}_{1} , t) = \frac{\partial}{\partial
\bm{J}_{1}} \!\cdot\! \bigg[ \bm{D}(\bm{J}_{1}) \!\cdot\! \bigg( \frac{\partial
P}{\partial \bm{J}_{1}}+\beta \mu_{\rt} P(\bm{J}_{1},t) \, \bm{\Omega} (\bm{J}_{1}) \bigg) \bigg] \, .
\label{prop30}
\end{equation}
At equilibrium, the probability density of the
test particle relaxes towards
the Boltzmann distribution
\begin{equation}
P(\bm{J}_{1}) = A_{\ra} \, \re^{-\beta \mu_{\rt} \epsilon(\bm{J}_{1})} \, , 
\label{prop31}
\end{equation}
which is the steady state of equation~\eqref{prop30}. These results generalise those of~\cite{Chandrasekhar1943I}.

For a thermal bath, the rates of energy and entropy of
the test particle are
\begin{equation}
\dot{E}_{\rt} = -\mu_{\rt} \!\! \int \!\! \rd \bm{J}_{1} \, \bm{D} (\bm{J}_{1})
\!\cdot\!
\frac{\partial P}{\partial \bm{J}_{1}} (\bm{J}_{1} , t) \!\otimes\!
\bm{\Omega} (\bm{J}_{1}) \,
- \beta \mu_{\rt}^{2} \!\! \int \!\! \rd \bm{J}_{1} \, \bm{D}(\bm{J}_{1})
\cdot\bm{\Omega}(\bm{J}_{1}) \!\otimes\!
\bm{\Omega}(\bm{J}_{1}) \, P(\bm{J}_{1} , t) \, ,
\label{ha2newb}
\end{equation}
\begin{equation}
\dot{S}_{\rt} = \!\! \int \!\! \rd \bm{J}_{1}
\frac{1}{P(\bm{J}_{1} , t)} \bm{D} (\bm{J}_{1}) \!\cdot\!
\frac{\partial P}{\partial\bm{J}_{1}} (\bm{J}_{1} , t) \!\otimes\!
\frac{\partial P}{\partial \bm{J}_{1}} (\bm{J}_{1} , t ) + \beta\mu_{\rt} \!\! \int \!\! \rd \bm{J}_{1}
\bm{D} (\bm{J}_{1}) \!\cdot\! \bm{\Omega} (\bm{J}_{1}) \!\otimes\!
\frac{\partial P}{\partial\bm{J}_{1}} (\bm{J}_{1} , t) \, .
\label{mod5}
\end{equation}
We have the general relation ${ \dot{S}_{\rm pol} \!=\! -\beta \dot{E}_{\rm diff} }$ between the rate of entropy due to the polarisation and the
rate of energy due to the diffusion.
We note that the
Fokker-Planck equation~\eqref{prop30} does not conserve the
energy and does not satisfy a ${H-}$theorem for the
Boltzmann entropy contrary to the Balescu-Lenard equation~\eqref{eq:BL}. This
is
because it describes the evolution of a test particle in a thermal bath with
a fixed temperature $T$. This corresponds to a canonical description
while the Balescu-Lenard equation corresponds to a microcanonical one.
We can obtain a form of ${H-}$theorem for the Fokker-Planck equation~\eqref{prop30}
by introducing the free energy ${ F_{\rt} \!=\! E_{\rt} \!-\! T S_{\rt} }$
which is the Legendre transform of the entropy $S_{\rt}$ w.r.t. the energy $E_{\rt}$,
with
conjugate parameter $T$. Taking the time derivative of $F_{\rt}$, substituting
the Fokker-Planck equation~\eqref{prop30}, and integrating by parts, we obtain
the (canonical) ${H-}$theorem 
\begin{equation}
\dot{F}_{\rt} = - \!\! \int \!\! \rd \bm{J}_{1} \, \frac{\bm{D}(\bm{J}_{1})}{\beta P(\bm{J}_{1})} \!\cdot\! \bigg( \frac{\partial
P}{\partial \bm{J}_{1}}+\beta \mu_{\rt} P(\bm{J}_{1},t) \bm{\Omega}(\bm{J}_{1}) \bigg) \!\otimes\! \bigg( \frac{\partial
P}{\partial \bm{J}_{1}}+\beta \mu_{\rt} P(\bm{J}_{1},t) \bm{\Omega}(\bm{J}_{1}) \bigg) \, \le 0 \, .
\label{prop31b}
\end{equation}
It establishes that the Boltzmann free energy is monotonically decreasing. At
equilibrium, ${ \dot{F}_{\rt} \!=\! 0 }$, leading to the
Boltzmann distribution
from equation~\eqref{prop31}.
The Boltzmann distribution minimises the Boltzmann
free energy $F_{\rt}$ while accounting for the normalisation
condition ${ \! \int \! \rd \bm{J}_{1}
P(\bm{J}_{1}) \!=\! 1 }$. The first variations write ${ \delta
F_{\rt} \!-\! \alpha \, \delta \! \int \! \rd \bm{J}_{1} P(\bm{J}_{1}) \!=\! 0
}$, leading to the equilibrium state from equation~\eqref{prop31}.

Let us now consider an initial condition in which the test particle has
a Boltzmann
distribution of the form 
\begin{equation}
P(\bm{J}_{1}) = A \, \re^{-\beta_{\rt} \mu_{\rt} \epsilon (\bm{J}_{1})} \, , 
\label{ha1}
\end{equation}
where ${ T_{\rt} \!=\! 1 / \beta_{\rt} }$ may be different
from the temperature ${ T \!=\! 1/\beta }$ of the bath. 
The initial rate of change of the test particle energy, $E_{\rt}$, is given by
\begin{equation}
\dot{E}_{\rt} = \mu_{\rt}^{2}
(\beta_{\rt} \!-\! \beta) \!\! \int \!\! \rd \bm{J}_{1} \,
\bm{D}_{1} \!\cdot\!
\bm{\Omega}_{1} \!\otimes\! \bm{\Omega}_{1} P(\bm{J}_{1}) \, .
\label{ha2}
\end{equation}
It can be written more explicitly as
\begin{equation}
\dot{E}_{\rt} = \mu_{\rt}^{2} (\beta_{\rt} \!-\! \beta) \pi (2 \pi)^{d} \sum_{\rb}
\!\!\!\sum_{\bm{m}_{1} , \bm{m}_{2}}'
\!\! \int\!\! \rd \bm{J}_{1} \rd \bm{J}_{2} \, \frac{ \delta_{\rD}
(\bm{m}_{1} \!\cdot\! \bm{\Omega}_{1} \!-\! \bm{m}_{2} \!\cdot\!
\bm{\Omega}_{2})}{ \big| \mathcal{D}_{\bm{m}_{1} , \bm{m}_{2}} (\bm{J}_{1} ,
\bm{J}_{2} , \bm{m}_{2} \!\cdot\! {\bm{\Omega}_{2}}) \big|^{2}} \,
\mu_{\rb}(\bm{m}_{1} \!\cdot\! \bm{\Omega}_{1})^{2} \, P(\bm{J}_{1}) \, F^{\rb} (\bm{J}_{2}) \, . 
\label{ha3}
\end{equation}
The initial rate of change of entropy is
\begin{equation}
\dot{S}_{\rt} = \beta_{\rt} \dot{E}_{\rt} \, .
\label{ha4}
\end{equation}
The initial rate of change of free energy ${ F_{\rt} \!=\! E_{\rt} \!-\! T S_{\rt} }$ can be written as
${ \dot{F}_{\rt} \!=\! (1 \!-\! T/T_{\rt}) \dot{E}_{\rt} }$.

\subsection{Sinking satellite}
\label{sec:sink}

Let us now assume that the test particle has a
mass $\mu_{\rt}$ much larger than the mass $\mu_{\rb}$ of the field particles. More
precisely, we assume that ${ \mu_{\rt} \!\sim\! 1 }$ while ${ \mu_{\rb} \!\sim\! 1/N_{\rb} }$ with
${ N_{\rb} \!\gg\! 1}$. For
${ N_{\rb} \!\rightarrow\! + \infty }$, ${ \mu_{\rb} \!\rightarrow\! 0 }$ and the field particles form a
collisionless fluid of stars. Since their distribution function
${ F^{\rb} (\bm{J}_{2}) }$ does
not evolve under the effect of collisions (the relaxation time of the field
particles scales like ${ N_{\rb} t_{\rd} \!\rightarrow\! + \infty }$), it can have an arbitrary
shape provided that it is Vlasov stable. In this fluid limit for which ${ N_{\rb} \!\rightarrow\!
+ \infty }$, the diffusion coefficient from equation~\eqref{prop22} vanishes and the friction
force from equation~\eqref{prop23} reduces to the component proportional to $\mu_{\rt}$, which corresponds to the friction by polarisation given by equation~\eqref{prop24}.
Finally, the Fokker-Planck equation~\eqref{prop21} reduces to
\begin{equation}
\frac{\partial P}{\partial t}(\bm{J}_{1},t) = -\pi (2 \pi)^{d} \mu_{\rt}
\!\!\!\sum_{\bm{m}_{1} , \bm{m}_{2}}' \!\! \bm{m}_{1} \!\cdot\!
\frac{\partial}{\partial{\bm{J}_{1}}} \!\! \int\!\! \rd \bm{J}_{2} \, \frac{
\delta_{\rD}
(\bm{m}_{1} \!\cdot\! \bm{\Omega}_{1} \!-\! \bm{m}_{2} \!\cdot\!
\bm{\Omega}_{2})}{ \big| \mathcal{D}_{\bm{m}_{1} , \bm{m}_{2}} (\bm{J}_{1} ,
\bm{J}_{2} , \bm{m}_{2} \!\cdot\! {\bm{\Omega}_{2}}) \big|^{2}} \, \bigg[ \bm{m}_{2}
\!\cdot\! \frac{\partial F}{\partial {\bm{J}_{2}}}(\bm{J}_{2}) \bigg]
\, P (\bm{J}_{1},t) \, ,
\label{prop25}
\end{equation}
where ${ F \!=\! \sum_{\rb} F^{\rb} }$ is the total distribution function of the field
particles. Equation~\eqref{prop25} can be rewritten as
\begin{equation}
\frac{\partial P }{\partial t}(\bm{J}_{1} , t) = \frac{\partial }{\partial
\bm{J}_{1}}
\!\cdot\! \bigg[\!-\! \bm{F}_{\rm
pol} (\bm{J}_{1} ) \, P (\bm{J}_{1} , t) \bigg]
\, .
\label{sandwichedFPbnew}
\end{equation}
This is just the continuity
equation corresponding to the deterministic (not stochastic) equation of motion
\begin{align}
\frac{\rd \bm{J}}{\rd t} = \bm{F}_{\rm pol} = \pi (2
\pi)^{d} \mu_{\rt} \!\!\sum_{\bm{m}_{1} , \bm{m}_{2}}' \!\! \int\!\!
\rd
\bm{J}_{2} \, \bm{m}_{1} \bigg[ \bm{m}_{2} \!\cdot\! \frac{\partial F}{\partial 
\bm{J}_{2}}(\bm{J}_{2}) \bigg] \, \frac{ \delta_{\rD}
(\bm{m}_{1} \!\cdot\! \bm{\Omega}_{1} \!-\! \bm{m}_{2} \!\cdot\!
\bm{\Omega}_{2})}{ \big| \mathcal{D}_{\bm{m}_{1} , \bm{m}_{2}} (\bm{J}_{1} ,
\bm{J}_{2} , \bm{m}_{2} \!\cdot\! {\bm{\Omega}_{2}}) \big|^{2}} \, .
\label{prop26}
\end{align}
We note that the mass $\mu_{\rb}$ of the field particles does not
appear explicitly in this expression since they form a collisionless ``fluid of
stars'' entirely determined by its total distribution function $F$. As a
result, the friction force given by equation~\eqref{prop26} can be obtained from
the calculation of Appendix~\ref{sec:appendixFpol}, which is solely 
based on the Vlasov equation (when ${ N_{\rb} \!\rightarrow\! + \infty}$) 
without reference to a discrete Hamiltonian
system. Equation~\eqref{prop26} can be viewed as the correct generalisation of
the famous Chandrasekhar's formula of dynamical friction when the mass of the test
particle is much larger than the mass of the field particles~\citep{Chavanis2013}. One can easily compute the average energy lost by the test particle.
We have
\begin{align}
\dot{E}_{\rt} = \mu_{\rt} \!\! \int \!\! \rd \bm{J}_{1} \, P (\bm{J}_{1} , t) \, \bm{F}_{\rm pol} (\bm{J}_{1} , t) \!\cdot\!
\bm{\Omega}(\bm{J}_{1} , t) \, .
\label{prop32}
\end{align}
Substituting equation~\eqref{prop26} into equation~\eqref{prop32}, we get
\begin{align}
\dot{E}_{\rt} = \pi (2
\pi)^{d} \mu_{\rt}^{2} \!\!\sum_{\bm{m}_{1} , \bm{m}_{2}}' \!\! \int\!\!
\rd
\bm{J}_{1}\, \rd
\bm{J}_{2} \, (\bm{m}_{1} \!\cdot\! \bm{\Omega}_{1}) \bigg[ \bm{m}_{2} \!\cdot\!
\frac{\partial F}{\partial 
\bm{J}_{2}}(\bm{J}_{2}) \bigg] \, \frac{ \delta_{\rD}
(\bm{m}_{1} \!\cdot\! \bm{\Omega}_{1} \!-\! \bm{m}_{2} \!\cdot\!
\bm{\Omega}_{2})}{ \big| \mathcal{D}_{\bm{m}_{1} , \bm{m}_{2}} (\bm{J}_{1} ,
\bm{J}_{2} , \bm{m}_{2} \!\cdot\! {\bm{\Omega}_{2}}) \big|^{2}} \, P
(\bm{J}_{1},t) \, .
\label{prop33}
\end{align}
When ${ F \!=\! F (\epsilon) }$, we obtain
\begin{align}
\dot{E}_{\rt} = \pi (2
\pi)^{d} \mu_{\rt}^{2} \!\!\sum_{\bm{m}_{1} , \bm{m}_{2}}' \!\! \int\!\!
\rd
\bm{J}_{1}\, \rd
\bm{J}_{2} \, (\bm{m}_{1} \!\cdot\! \bm{\Omega}_{1})^{2}
\frac{dF}{d\epsilon}[\epsilon(\bm{J}_{2})]\, \frac{ \delta_{\rD}
(\bm{m}_{1} \!\cdot\! \bm{\Omega}_{1} \!-\! \bm{m}_{2} \!\cdot\!
\bm{\Omega}_{2})}{ \big| \mathcal{D}_{\bm{m}_{1} , \bm{m}_{2}} (\bm{J}_{1} ,
\bm{J}_{2} , \bm{m}_{2} \!\cdot\! {\bm{\Omega}_{2}}) \big|^{2}} \, P
(\bm{J}_{1},t) \, .
\label{prop34}
\end{align}
For a thermal bath, considering the limit ${ \mu_{\rb} \!\sim\!
1/N_{\rb} \!\rightarrow\! 0 }$ and ${ \beta \!\rightarrow\! + \infty }$ such that ${ \beta \mu_{\rb} \!\sim\! 1 }$, we may finally write
\begin{align}
\dot{E}_{\rt} = - \pi (2
\pi)^{d} \beta \mu_{\rt}^{2} \sum_{\rb} \!\!\sum_{\bm{m}_{1} , \bm{m}_{2}}' \!\! \int\!\!
\rd
\bm{J}_{1}\, \rd
\bm{J}_{2} \, (\bm{m}_{1} \!\cdot\! \bm{\Omega}_{1})^{2} \mu_{\rb} F^{\rb} (\bm{J}_{2}) \,
\frac{
\delta_{\rD}
(\bm{m}_{1} \!\cdot\! \bm{\Omega}_{1} \!-\! \bm{m}_{2} \!\cdot\!
\bm{\Omega}_{2})}{ \big| \mathcal{D}_{\bm{m}_{1} , \bm{m}_{2}} (\bm{J}_{1} ,
\bm{J}_{2} , \bm{m}_{2} \!\cdot\! {\bm{\Omega}_{2}}) \big|^{2}} \, P
(\bm{J}_{1},t) \, .
\label{prop33w}
\end{align}
This expression is consistent with equation~\eqref{ha3}
when ${ \beta \!\rightarrow\! + \infty }$. It shows the somehow
paradoxical resurgence of the diffusion
coefficient from equation~\eqref{prop22} in the friction
term from equation~\eqref{prop26} due to the fluctuation-dissipation theorem from equation~\eqref{prop28}, although diffusion caused by
finite${-N}$ effects is neglected (at leading order) in the present approach.

The Balescu-Lenard equation~\eqref{eq:BL}, the Fokker-Planck equation~\eqref{prop21} and the generalised Kramers equation~\eqref{prop30} describe a
competition between diffusion and friction. As a result, the probability density
of the test particle relaxes
towards a statistical equilibrium state in which the two effects balance
each other establishing the Boltzmann distribution from equations~\eqref{prop3} or~\eqref{prop31}.\footnote{This is the usual picture of Brownian
theory. As we previously indicated, the situation is more complicated for
self-gravitating systems
since a statistical equilibrium state does not always exist.}
In the situation
described by the deterministic
equation~\eqref{prop26}, the test particle just feels a friction
force and sinks towards the center of the system.
In astrophysics, this is traditionally referred
to as the ``sinking satellite'' problem. For small but non-zero values of
$\mu_{\rb}$, this result is consistent with the multicomponent Balescu-Lenard equation~\eqref{eq:BL} and with the multicomponent Boltzmann distribution~\eqref{prop3} that account for a segregation of mass.
Heavy particles have the tendency to sink at the center of the system while light particles move around.
Various descriptions of this collisionless dynamical friction have already been proposed in astrophysics~\citep{Kalnajs1971,TremaineWeinberg1984,Weinberg1986,Weinberg1989}.
For a thorough illustration of this process, we refer in particular to~\cite{Weinberg1989}, which provides a detailed numerical analysis of the sinking of satellites in spherical galaxies,
when accounting for or neglecting collective effects.

Note that the diffusion coefficient~\eqref{prop22} is proportional to the mass
${ \mu_{\rb} \!\sim\! 1/N_{\rb} }$ of the
field particles because it is due to the fluctuations of the force produced by
these particles (see section~\ref{sec:diffusioncoefficients}). When ${ N_{\rb} \!\rightarrow\!
+ \infty }$, the fluctuations vanish
(the field particles form a collisionless fluid) and the diffusion
coefficient tends to zero. On the other hand, the friction force~\eqref{prop23}
involves
a term
proportional to ${ \mu_{\rb} \!\sim\! 1/N_{\rb} }$ related to
the derivative of the diffusion coefficient
${ \partial \bm{D} / \partial \bm{J} }$ 
and a term proportional to ${ \mu_{\rt} \!\sim\! 1 }$, which corresponds
to the friction by
polarisation $\bm{F}_{\rm pol}$ from equation~\eqref{prop24} (see the decomposition from equation~\eqref{Fpol}). When ${ N_{\rb} \!\rightarrow\! + \infty }$, the first
term tends to zero while
the second term remains finite. The term in ${ \partial \bm{D}
/ \partial \bm{J} }$ is proportional to the mass $\mu_{\rb}$ of the field
particles because it is due to the fluctuations of the force
produced by
these particles.
The force by polarisation from equation~\eqref{prop24} is proportional to the mass $\mu_{\rt}$ of the test
particle because it is due to the retroaction of the fluid of stars to the
perturbation caused by the test particle (see Appendix~\ref{sec:appendixFpol}).
Therefore, this force remains finite even when
the field particles have no fluctuation (${ N_{\rb} \!\rightarrow\! + \infty }$), i.e. even if the test particle is evolving within a collisionless fluid. This is the collisionless resonant dynamical friction, which accounts for both the system's inhomogeneity and collective effects.
We refer to~\cite{Chavanis2013} for a detailed discussion of the links between this formalism and other approaches, in particular the two-body encounters theory pioneered by~\cite{Chandrasekhar1943I}.

\subsection{An ilustration of dynamical friction in a Mestel disc}

The calculation of the force of dynamical friction acting on a star is a problem of considerable interest, initiated  by the seminal work of~\cite{Chandrasekhar1943I}. \cite{ChandrasekharVonNeumann1943} attempted to derive this force from a purely stochastic formalism in the case where the system is infinite and homogeneous and the stars have a Maxwellian velocity distribution. Their calculations were extended by~\cite{DelPopoloGambera1999,DelPopolo2003} to an inhomogeneous medium with a density decaying as ${ \rho \!\sim\! r^{-p} }$, assuming again that the velocity distribution of the stars is Maxwellian. In order to treat in a self-consistent manner more general situations of spatial inhomogeneity, and take into account collective effects, the formalism developed in the present paper is necessary.\footnote{We emphasise that this formalism recovers in the appropriate limit the two-body encounters theory of~\cite{Chandrasekhar1943I} for a ${3D}$ homogeneous medium (this is discussed in detail in~\cite{Chavanis2013}). Conversely, the situation presented in e.g. ~\cite{DelPopoloGambera1999,DelPopolo2003} cannot be treated with this formalism because the corresponding distribution (albeit physically interesting) is not a quasi-stationary state of the Vlasov equation and, as such, does not possess suitable angle-action coordinates.}
Explicit applications were  carried for razor-thin~\citep{FouvryPichonChavanis2015,FouvryPichonMagorrianChavanis2015}, thickened~\citep{FouvryPichonChavanisMonk2017} stellar discs, and Keplerian discs~\citep{FouvryPichonMagorrian2017}. Let us illustrate this kinetic theory 
with the calculation of the friction force acting on a sinking satellite in a collisionless fluid of stars.

Computing the resonant collisionless dynamical friction acting on a massive perturber requires the construction of the angle-action coordinates ${ (\bm{\theta} , \bm{J}) }$, the specification of the biorthogonal basis ${ (\psi^{(p)} , \rho^{(p)}) }$, the computation of the system's response matrix ${ \widehat{\mathbf{M}} (\omega) }$, and the resolution of the non-local resonance condition ${ \delta_{\rD} (\bm{m}_{1} \!\cdot\! \bm{\Omega}_{1} \!-\! \bm{m}_{2} \!\cdot\! \bm{\Omega}_{2}) }$. \cite{FouvryPichonMagorrianChavanis2015} presented this 
calculation for razor-thin axisymmetric stellar discs and recovered the spontaneous self-consistent formation on secular timescales of a narrow resonant ridge of orbits in action space, as first observed in the numerical simulations of~\cite{Sellwood2012}. 

\begin{figure}
\begin{center}
\epsfig{file=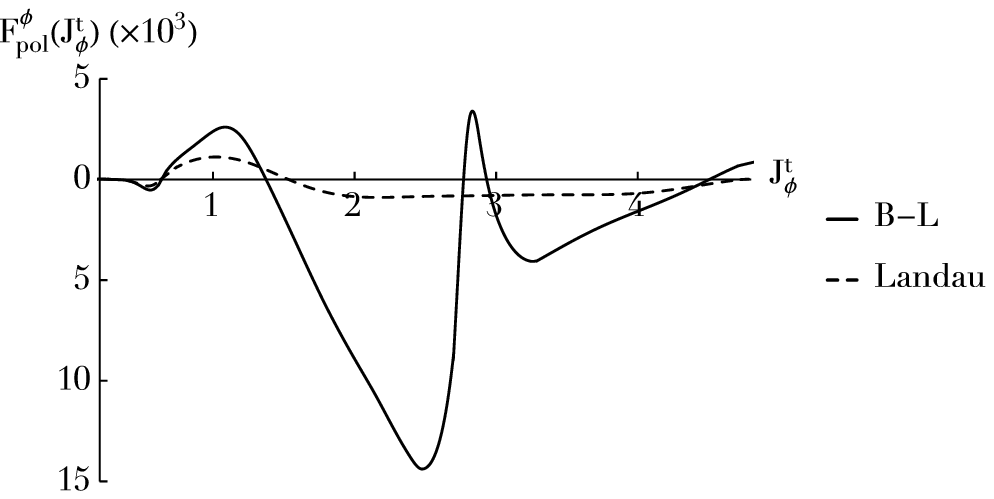,angle=-00,width=0.60\textwidth}
\caption{\small{Illustration of the friction force by polarisation acting on a massive perturber of individual mass ${ \mu_{\rt} \!=\! 10^{-3} M_{\rm tot} \!=\! 5.4 \!\times\! 10^{-3} }$, embedded in a collisionless razor-thin Mestel disc considered in~\protect\cite{Sellwood2012,FouvryPichonMagorrianChavanis2015}. Here, the perturber is assumed to remain on circular orbits, characterised by the angular momentum $J_{\phi}^{\rt}$. The two different lines correspond to the respective predictions of the Balescu-Lenard and Landau formalisms, i.e. with or without collective effects. Negative values of $F_{\rm pol}^{\phi}$ corresponds to a diffusion towards smaller angular momenta.
}}
\label{figCircularDynamicalFrictionMestelDisc}
\end{center}
\end{figure}

Relying on the calculations performed in~\cite{FouvryPichonMagorrianChavanis2015}, let us   estimate the friction force by polarisation that a massive perturber  undergoes when embedded in a collisionless disc (i.e. in the limit of an infinite number of bath particles in the disc).
Let us specifically consider an infinitely thin Mestel disc for which the circular speed is a constant $V_{0}$ independent of the radius. The stationary background potential $\psi_{\rm M}$ and its associated surface density $\Sigma_{\rm M}$ are given by
\begin{equation}
\psi_{\rm M} (R) = V_{0}^{2} \log \!\left[\! \frac{R}{R_{\rm max}} \!\right] \;\;\; ; \;\;\; \Sigma_{\rm M} (R) = \frac{V_{0}^{2}}{2 \pi G R} \, ,
\label{psi_Mestel_Sigma_Mestel}
\end{equation}
where $R_{\rm max}$ is a scale parameter of the disc. Following~\cite{BinneyTremaine2008}, a self-consistent DF for this system is given by
\begin{equation}
F_{\rm M} (E , J_{\phi}) = C_{\rm M} \, J_{\phi}^{q} \, \exp [ - E / \sigma_{r}^{2}] \, ,
\label{DF_Toomre}
\end{equation}
where the exponent $q$ is given by
$q = {V_{0}^{2}}/{\sigma_{r}^{2}} \!-\! 1 \, ,$
with $\sigma_{r}$ being the constant radial velocities spread within the disc. In equation~\eqref{DF_Toomre}, $C_{\rm M}$ is a normalisation constant. In order to ensure linear stability, the DF from equation~\eqref{DF_Toomre} is additionally tapered in the inner and outer regions of the disc. We refer to~\cite{FouvryPichonMagorrianChavanis2015} for further details on the physical system considered. Razor-thin axisymmetric discs are explicitly integrable, so that one naturally introduces the action coordinates ${ \bm{J} \!=\! (J_{\phi} , J_{r}) }$. Here, $J_{\phi}$ is the azimuthal action of the particle, its angular momentum, and encodes the typical distance of the particle to the center. The second action is $J_{r}$, the radial action, which captures the amplitude of the particle's radial libration. The larger $J_{r}$, the wider the radial oscillations and the hotter the orbit. Because the radial action satisfies ${ J_{r} \!\geq\! 0 }$, exactly circular orbits (i.e. orbits with ${ J_{r} \!=\! 0 }$) remain circular during the secular evolution and can therefore only diffuse along the ${ J_{\phi}-}$direction. For simplicity, the calculation will be restricted to massive perturbers on circular orbits in the collisionless limit, assuming that the razor-thin disc is made of an infinite number of particles. See also~\cite{Weinberg1989} for an example of calculation of the dressed dynamical friction acting on circular orbits in the case of ${3D}$ spherical systems. Following equation~\eqref{prop26}, the evolution of a massive perturber on a circular orbit of angular momentum $J_{\phi}^{\rt}$ is given by the one-dimensional differential equation
\begin{equation}
\frac{\rd J_{\phi}^{\rt}}{\rd t} = F_{\rm pol}^{\phi} (J_{\phi}^{\rt}) \, ,
\label{friction_circular}
\end{equation}
where $F_{\rm pol}^{\phi}$ stands for the component along the ${J_{\phi}-}$direction of the ${2D}$ friction force vector $\bm{F}_{\rm pol}$. Through equation~\eqref{Fpol}, such a friction force is straightforwardly estimated once the disc's self-consistent first- and second-order diffusion coefficients are computed. These coefficients were computed in~\cite{FouvryPichonMagorrianChavanis2015} for a razor-thin tapered Mestel disc. In the collisionless limit (i.e. ${ \mu_{\star} \!\to\! 0 }$, with $\mu_{\star}$ the individual mass of the particles forming the disc), the friction force ${ F_{\rm pol}^{\phi} (J_{\phi}^{\rt}) }$ acting on a massive perturber of mass $\mu_{\rt}$ on the circular orbit $J_{\phi}^{\rt}$ follows. This is illustrated in figure~\ref{figCircularDynamicalFrictionMestelDisc}, which performs the calculations respectively within the Balescu-Lenard and Landau frameworks, i.e. with or without collective effects.

Figure~\ref{figCircularDynamicalFrictionMestelDisc} exhibits a complex behaviour for the friction force by polarisation. First note the importance of collective effects, which significantly hasten the dynamical friction. Around ${ J_{\phi}^{\rt} \!\simeq\! 2.8 }$ the Balescu-Lenard formalism predicts ${ F_{\rm pol}^{\phi} \!>\! 0 }$, i.e. a diffusion towards higher angular momentum. This abnormal diffusion is also found in the lower panel of figure~8 in~\cite{Sellwood2012}, where the individual diffusion of stars also follows this trend. This can be interpreted as a consequence of the corotation resonance, whose inner Lindblad resonance is at the central ridge (${ J_{\phi}^{\rt} \!\simeq\! 1.2 }$). When collective effects are neglected, this region of positive friction vanishes. One should also keep in mind that the geometrical constraints associated with the razor-thin geometry (compared to ${3D}$ spherical systems) impact the expected properties of the friction force, leading in some regimes to its complete cancellation~\citep{Kalnajs1971}. Understanding the properties of the resonant collisionless dynamical friction in razor-thin discs in more details could be the subject of future works.

\section{Some distribution formulae}
\label{sec:appendixformula}

Let us compute here the few distributions encountered in the main text.

\subsection{Calculation of ${ \eta (x) }$}
\label{sec:formulaeta}

Let us first see how the distribution
\begin{equation}
\eta(x) = \lim_{\Delta t \to + \infty} 
\bigg[ \frac{ | \re^{\ri x \Delta t} \!-\! 1 |^{2} }{x^{2} \Delta t} \bigg] =
\lim_{\Delta t \to + \infty} \bigg[ \frac{2 \, ( 1 \!-\! \cos (x \Delta t)
)}{x^{2} \Delta t} \bigg] \, 
\end{equation}
operates on a given function ${ f(x) }$. Writing ${ u \!=\! x \Delta t }$, one has
\begin{equation}
\lim_{\Delta t \to + \infty} \!\! \int_{- \infty}^{+ \infty} \!\!\!\!\!\! \rd x \, f (x) \frac{2 \, ( 1 \!-\! \cos (x \Delta t) ) }{x^{2} \Delta t} = \lim_{\Delta t \to + \infty} \!\! \int_{- \infty}^{+ \infty} \!\!\!\!\!\! \rd u \, f ( u / \Delta t) \, \frac{2 \, ( 1 \!-\! \cos (u) ) }{u^{2}} = f (0) \!\! \int_{- \infty}^{+ \infty} \!\!\!\!\!\! \rd u \, \frac{2 \, ( 1 \!-\! \cos (u) )}{u^{2}} \, .
\end{equation}
The integral on $u$ converges and can be computed noting that
\begin{equation}
\!\! \int_{- \infty}^{+ \infty} \!\!\!\!\!\! \rd u \, \frac{2 \, ( 1 \!-\! \cos (u) )}{u^{2}} = \lim_{\varepsilon \to 0} \bigg[ \!\! \int_{- \infty}^{+ \infty} \!\!\!\!\!\! \rd u \, \frac{2 \, ( 1 \!-\! \cos(u) )}{u^{2} \!+\! \varepsilon^{2}} \bigg] = \lim_{\varepsilon \to 0} \frac{2 \pi}{\varepsilon} \big[ 1 \!-\! \re^{-\varepsilon} \big] = 2 \pi \, .
\end{equation}
Hence
\begin{equation}
\eta(x) = 2\pi \, \delta_{\rD} (x) \, .
\end{equation}

\subsection{Calculation of ${\kappa (x)}$}
\label{sec:formulakappa}

We now compute
\begin{equation}
\kappa (x) = \lim_{\Delta t \to + \infty} \!\! \int_{0}^{\Delta t} \!\! \frac{\rd t_{1}}{\Delta t} \!\! \int_{0}^{t_{1}} \!\!\!\!\!\! \rd t_{2} \, \re^{ \ri x (t_{1} - t_{2})} \, .
\end{equation}
Let us introduce ${ \tau \!=\! t_{1} \!-\! t_{2} }$ and reverse integration between $t_{1}$ and $\tau$ while keeping track of the triangular shape of the ${ (t_{1} , \tau) }$ integration domain. One gets
\begin{align}
\kappa (x) & \, = \lim_{\Delta t \to + \infty} \!\! \int_{0}^{\Delta t} \!\!\!\!\!\! \rd \tau \, \re^{ \ri x \tau} \!\! \int_{\tau}^{\Delta t} \!\! \frac{\rd t_{1}}{\Delta t} = \lim_{\Delta t \to + \infty} \!\! \int_{0}^{\Delta t} \!\!\!\!\!\! \rd \tau \, \re^{ \ri x \tau} \bigg[ 1 \!-\! \frac{\tau}{\Delta t} \bigg] \nonumber
\\
& \, = \lim_{\Delta t \to + \infty} \bigg[ \frac{\re^{\ri x \Delta t} \!-\! 1}{\ri x} \!+\! \frac{\ri}{\Delta t} \frac{\rd}{\rd x} \!\! \int_{0}^{\Delta t} \!\!\!\!\!\! \rd \tau \, \re^{ \ri x \tau} \bigg] \nonumber
\\
& \, = - \ri \lim_{\Delta t \to + \infty} \bigg[ \frac{\re^{\ri x \Delta t} \!-\! 1}{\ri x} \bigg] \!+\! \frac{1}{\Delta t} \frac{\rd}{\rd x} \bigg[ \lim_{\Delta t \to + \infty} \bigg[ \frac{\re^{\ri x \Delta t} \!-\! 1}{x} \bigg] \bigg] \, .
\end{align}
We recall the identity
\begin{equation}
\lim_{\Delta t \to + \infty} \bigg[ \frac{\re^{\ri x \Delta t} \!-\!1}{x} \bigg] = \ri \pi \delta_{\rD} (x) \, .
\end{equation}
Hence
\begin{align}
\kappa (x) = \lim_{\Delta t \to + \infty} \bigg[ \pi \delta_{\rD} (x) \!+\! \frac{\ri \pi}{\Delta t} \frac{\rd \delta_{\rD} (x) }{\rd x} \bigg] = \pi \delta_{\rD} (x) \, .
\end{align}

\subsection{Calculation of ${ \gamma (x) }$}
\label{sec:formulagamma}

Let us finally compute
\begin{equation}
\gamma (x) = \lim_{\Delta t \to + \infty} \!\! \int_{0}^{\Delta t} \!\! \frac{\rd t_{1}}{\Delta t} \!\! \int_{0}^{t_{1}} \!\!\!\!\!\! \rd t_{2} \!\! \int_{0}^{t_{2}} \!\!\!\!\!\! \rd t_{3} \, \re^{\ri x (t_{1} - t_{3})} \, .
\end{equation}
Reversing integration between $t_{2}$ and $t_{3}$ yields
\begin{align}
\gamma (x) & \, = \lim_{\Delta t \to + \infty} \!\! \int_{0}^{\Delta t} \!\! \frac{\rd t_{1}}{\Delta t} \!\! \int_{0}^{t_{1}} \!\!\!\!\!\! \rd t_{3} \, \re^{ \ri x (t_{1} - t_{3})} (t_{1} \!-\! t_{3}) = - \ri \lim_{\Delta t \to + \infty} \bigg[ \frac{\rd}{\rd x} \!\! \int_{0}^{\Delta t} \!\! \frac{\rd t_{1}}{\Delta t} \!\! \int_{0}^{t_{1}} \!\!\!\!\!\! \rd \tau \, \re^{ \ri x \tau} \bigg] \nonumber
\\
& \, = - \ri \lim_{\Delta t \to + \infty} \bigg[ \frac{\rd}{\rd x} \!\! \int_{0}^{\Delta t} \!\!\!\!\!\! \rd \tau \, \re^{ \ri x \tau} \!\! \int_{\tau}^{\Delta t} \!\! \frac{\rd t_{1}}{\Delta t} \bigg] = - \ri \frac{\rd \kappa (x)}{\rd x} = - \ri \pi \frac{\rd \delta_{\rD} (x)}{\rd x} \, .
\end{align}

\section{Symmetries}
\label{sec:appendixsymmetries}

\subsection{Relation between \texorpdfstring{$\Lambda_{- \bm{m}_{1} , - \bm{m}_{2}} (\bm{J}_{1} , \bm{J}_{2} , - \omega)$}{Lambda-m1m2} and \texorpdfstring{$\Lambda_{\bm{m}_{1} , \bm{m}_{2}} (\bm{J}_{1} , \bm{J}_{2} , \omega)$}{Lambdam1m2}}
\label{sec:linkLambdaone}

Let us first show that
\begin{equation}
\Lambda_{- \bm{m}_{1} , - \bm{m}_{2}} (\bm{J}_{1} , \bm{J}_{2} , - \omega) = \Lambda_{\bm{m}_{1} , \bm{m}_{2}}^{*} (\bm{J}_{1} , \bm{J}_{2} , \omega) \, .
\end{equation}
Recall that
\begin{equation}
\Lambda_{ - \bm{m}_{1} , - \bm{m}_{2}} (\bm{J}_{1} , \bm{J}_{2} , - \omega) = \psi^{(\alpha)}_{- \bm{m}_{1}} (\bm{J}_{1}) \, \varepsilon_{\alpha \beta}^{-1} ( - \omega) \, \psi^{(\beta) *}_{- \bm{m}_{2}} (\bm{J}_{2}) \, .
\end{equation}
Following equations~(25) and~(A25) from~\cite{Heyvaerts2010}, one can write
\begin{equation}
\psi_{- \bm{m}}^{(\alpha)} (\bm{J}) = \psi_{\bm{m}}^{(\widehat{\alpha}) *} (\bm{J}) \;\;\; ; \;\;\; \varepsilon^{-1}_{\alpha \beta} (- \omega) = \varepsilon^{-1 *}_{\widehat{\alpha} \widehat{\beta}} (\omega) \, ,
\label{definition_alpha_hat}
\end{equation}
where ${ \widehat{\alpha} }$ is an element of the basis, which is in general different from $\alpha$ (see~\cite{Heyvaerts2010}). This immediately gives
\begin{equation}
\Lambda_{- \bm{m}_{1} , - \bm{m}_{2}} (\bm{J}_{1} , \bm{J}_{2} , - \omega) = \psi_{\bm{m}_{1}}^{\widehat{\alpha} *} (\bm{J}_{1}) \, \varepsilon^{-1 *}_{\widehat{\alpha} \widehat{\beta}} (\omega) \, \psi_{\bm{m}_{2}}^{\widehat{\beta}} (\bm{J}_{2}) = \bigg[ \psi_{\bm{m}_{1}}^{\widehat{\alpha}} (\bm{J}_{1}) \, \varepsilon^{-1}_{\widehat{\alpha} \widehat{\beta}} (\omega) \, \psi_{\bm{m}_{2}}^{\widehat{\beta} *} (\bm{J}_{2}) \bigg]^{*} = \Lambda_{\bm{m}_{1} , \bm{m}_{2}}^{*} (\bm{J}_{1} , \bm{J}_{2} , \omega) \, .
\label{symmetry_Lambda_1}
\end{equation}

\subsection{Relation between \texorpdfstring{$\Lambda_{\bm{m}_{2} , \bm{m}_{1}} (\bm{J}_{2} , \bm{J}_{1} , \omega)$}{Lambdam2m1} and \texorpdfstring{$\Lambda_{\bm{m}_{1} , \bm{m}_{2}} (\bm{J}_{1} , \bm{J}_{2} , \omega)$}{Lambdam1m2}}
\label{sec:LinkLambdatwo}

Let us now demonstrate that to second order in the noise level, one can assume that ${ \Lambda_{\bm{m}_{2} , \bm{m}_{1}} (\bm{J}_{2} , \bm{J}_{1} , \omega) \!=\! \Lambda_{\bm{m}_{1} , \bm{m}_{2}}^{*} (\bm{J}_{1} , \bm{J}_{2} , \omega) }$. First note that in equation~\eqref{eq:dJ1dtsemifinal} the two frequencies $\omega_{1}$ and $\omega_{2}$ coincide because of the Dirac Delta factor ${ \delta_{\rD} (\omega_{1} \!-\! \omega_{2}) }$. Now we have
\begin{align}
\Lambda_{\bm{m}_{2} , \bm{m}_{1}} (\bm{J}_{2} , \bm{J}_{1} , \omega) & \, = \psi^{(\alpha)}_{\bm{m}_{2}} (\bm{J}_{2}) \, \varepsilon_{\alpha \beta}^{-1} (\omega) \, \psi^{(\beta) *}_{\bm{m}_{1}} (\bm{J}_{1}) \nonumber
\\
 & \, = \psi^{(\alpha)}_{\bm{m}_{2}} (\bm{J}_{2}) \big\{ \varepsilon_{\beta \alpha}^{-1 *} (\omega) \, + \big[ \varepsilon_{\beta \alpha}^{-1} (\omega) - \varepsilon_{\beta \alpha}^{-1 *} (\omega) \big] \big\} \psi^{(\beta) *}_{\bm{m}_{1}} (\bm{J}_{1}) \nonumber
 \\
& \, = \bigg[ \psi^{(\beta)}_{\bm{m}_{1}} (\bm{J}_{2}) \, \varepsilon_{\beta \alpha}^{-1} (\omega) \, \psi^{(\alpha) *}_{\bm{m}_{2}} (\bm{J}_{2}) \bigg]^{*} + \psi^{(\alpha)}_{\bm{m}_{2}} (\bm{J}_{2}) \big[ \varepsilon_{\alpha \beta}^{-1} (\omega) - \varepsilon_{\beta \alpha}^{-1 *} (\omega) \big] \psi^{(\beta) *}_{\bm{m}_{1}} (\bm{J}_{1}) \nonumber
\\
& \, = \Lambda_{\bm{m}_{1} , \bm{m}_{2}}^{*} (\bm{J}_{1} , \bm{J}_{2} , \omega) + \psi^{(\alpha)}_{\bm{m}_{2}} (\bm{J}_{2}) \, \big[ \varepsilon^{-1} (\omega) - \varepsilon^{-1 \dagger} (\omega) \big]_{\alpha \beta} \, \psi^{(\beta) *}_{\bm{m}_{1}} (\bm{J}_{1}) \, ,
\label{LambdaCalculation}
\end{align}
where, in the last line, we introduced $\varepsilon^{-1 \dagger}$ as the hermitian conjugate of $\varepsilon^{-1}$. The anti-hermitic part of $\varepsilon^{-1}$, which appears in the square bracket, was computed by~\cite{Heyvaerts2010} and can be written as
\begin{equation}
\big[ \varepsilon^{-1} - \varepsilon^{-1 \dagger} \big]_{\alpha \beta} = \varepsilon_{\alpha \lambda}^{-1} \, \big[ \varepsilon^{\dagger} - \varepsilon \big]_{\lambda \mu} \, \varepsilon_{\mu \beta}^{-1 \dagger} \, .
\end{equation}
Now we have (see equation~(25) in~\cite{Heyvaerts2010})
\begin{equation}
\big[ \varepsilon^{\dagger} (\omega) - \varepsilon (\omega) \big]_{\lambda \mu} = - \ri \sum_{\rc} \sum_{\bm{m}_{3}}' (2 \pi)^{d + 1} \!\!\! \int \!\! \rd \bm{J}_{3} \, \delta_{\rD} (\omega \!-\! \bm{m}_{3} \!\cdot\! \bm{\Omega}_{3}) \, \bm{m}_{3} \!\cdot\! \partial_{\bm{J}_{3}} \big[ F^{\rc} (\bm{J}_{3}) \big] \, \psi_{\bm{m}_{3}}^{(\lambda) *} (\bm{J}_{3}) \, \psi_{\bm{m}_{3}}^{(\mu)} (\bm{J}_{3}) \, .
\end{equation}
Injecting this relation into equation~\eqref{LambdaCalculation} yields
\begin{align}
\Lambda_{\bm{m}_{2} , \bm{m}_{1}} (\bm{J}_{2} , \bm{J}_{1} , \omega) & \, = \Lambda_{\bm{m}_{1} , \bm{m}_{2}}^{*} (\bm{J}_{1} , \bm{J}_{2} , \omega) -\ri \sum_{\rc} \sum_{\bm{m}_{3}}' (2 \pi)^{d + 1} \!\!\! \int \!\! \rd \bm{J}_{3} \, \delta_{\rD} (\omega \!-\! \omega_{3}) \, \bm{m}_{3} \!\cdot\! \partial_{\bm{J}_{3}} \big[ F^{\rc} (\bm{J}_{3}) \big] \nonumber
\\
& \quad\quad\quad\quad\quad\quad\quad\quad\quad\quad \times \, \psi^{(\alpha)}_{\bm{m}_{2}} (\bm{J}_{2}) \, \varepsilon_{\alpha \lambda}^{-1} (\omega) \, \psi_{\bm{m}_{3}}^{(\lambda) *} (\bm{J}_{3}) \, \psi_{\bm{m}_{3}}^{(\mu)} (\bm{J}_{3}) \, \varepsilon_{\beta \mu}^{-1 *} (\omega) \, \psi^{(\beta) *}_{\bm{m}_{1}} (\bm{J}_{1}) \nonumber
\\
 & \hspace{-20mm} \, = \Lambda_{\bm{m}_{1} , \bm{m}_{2}}^{*} (\bm{J}_{1} , \bm{J}_{2} , \omega) - \ri \sum_{\rc} \sum_{\bm{m}_{3}}' (2 \pi)^{d + 1} \!\!\! \int \!\! \rd \bm{J}_{3} \, \delta_{\rD} (\omega - \omega_{3}) \, \bm{m}_{3} \!\cdot\! \partial_{\bm{J}_{3}} \big[ F^{\rc} (\bm{J}_{3}) \big] \nonumber
\\
& \hspace{-20mm} \quad\quad\quad\quad\quad\quad\quad\quad\quad\quad \times \, \bigg[ \psi^{(\alpha)}_{\bm{m}_{2}} (\bm{J}_{2}) \, \varepsilon_{\alpha \lambda}^{-1} (\omega) \, \psi_{\bm{m}_{3}}^{(\lambda) *} (\bm{J}_{3}) \bigg] \bigg[ \psi_{\bm{m}_{1}}^{(\beta)} (\bm{J}_{1}) \, \varepsilon_{\beta \mu}^{-1} (\omega) \, \psi_{\bm{m}_{3}}^{\mu *} (\bm{J}_{3}) \bigg]^{*} \nonumber
\\
& \hspace{-20mm} \, = \Lambda_{\bm{m}_{1} , \bm{m}_{2}}^{*} (\bm{J}_{1} , \bm{J}_{2} , \omega) - \ri \sum_{\rc} \sum_{\bm{m}_{3}}' (2 \pi)^{d + 1} \!\! \int \!\! \rd \bm{J}_{3} \, \delta_{\rD} (\omega \!-\! \omega_{3}) \, \bm{m}_{3} \!\cdot\! \partial_{\bm{J}_{3}} \big[ F^{\rc} (\bm{J}_{3}) \big] \, \Lambda_{\bm{m}_{2} , \bm{m}_{3}} (\bm{J}_{2} , \bm{J}_{3} , \omega) \, \Lambda_{\bm{m}_{1} , \bm{m}_{3}}^{*} (\bm{J}_{1} , \bm{J}_{3} , \omega) \, .
\end{align}
This relation shows that the difference between ${ \Lambda_{\bm{m}_{2} , \bm{m}_{1}} (\bm{J}_{2} , \bm{J}_{1} , \omega) }$ and ${ \Lambda_{\bm{m}_{1} , \bm{m}_{2}}^{*} (\bm{J}_{1} , \bm{J}_{2} , \omega) }$ is a term involving two $\Lambda$ factors, which implies that it is of higher order with respect to the noise, as discussed in the end of section~\ref{sec:dressedpotential}. This difference is proportional to the anti-hermitic part of $\varepsilon$, and therefore corresponds to a particular form of the fluctuation-dissipation theorem. Since equation~\eqref{eq:dJ1dtsemifinal} aims for a second order expression in the noise, the difference ${ \delta_{\rD} (\omega_{1} \!-\! \omega_{2}) \big[ \Lambda_{\bm{m}_{2} , \bm{m}_{1}} (\bm{J}_{2} , \bm{J}_{1} , \omega_{1}) \!-\! \Lambda_{\bm{m}_{1} , \bm{m}_{2}}^{*} (\bm{J}_{1} , \bm{J}_{2} , \omega_{2}) \big] }$, because it would introduce a third order correction, can be neglected here.

\label{lastpage}
\end{document}